\documentclass[preprint,12pt]{elsarticle}

\usepackage{amssymb}
\usepackage{amsmath}

\usepackage{float}
\usepackage[utf8]{inputenc}
\usepackage[T1]{fontenc}
\usepackage{bm}
\usepackage{tabularx} 
\usepackage{amsmath} 
\usepackage{ragged2e} 
\usepackage{array}
\usepackage{graphicx}
\usepackage{multirow}

\usepackage{hyperref}
\usepackage{booktabs}

\journal{Physics Reports}

\begin{document}

\begin{frontmatter}

\newcommand{\revise}[1]{\textcolor{red}{#1}}

\title{Integrating Artificial Intelligence and Geophysical Insights for Earthquake Forecasting: A Cross-Disciplinary Review}

\author[1,2]{Zhang Ying}
\author[3]{Wen Congcong}
\author[4]{Sornette Didier\corref{cor1}}
\author[5]{Zhan Chengxiang}

\cortext[cor1]{Corresponding Author. E-mail: dsornette@ethz.ch}


\affiliation[1]{
organization={School of Automation and Electrical Engineering, University of Science and Technology Beijing}, 
city={Beijing}, 
country={China}}

\affiliation[2]{
organization={Beijing Engineering Research Center of Industrial Spectrum Imaging}, 
city={Beijing}, 
country={China}}

\affiliation[3]{
organization={Department of Computer Engineering, New York University Abu Dhabi}, 
city={Abu Dhabi}, 
country={UAE}}

\affiliation[4]{
organization={Institute of Risk Analysis, Prediction and Management, Southern University of Science and Technology}, 
city={Shenzhen}, 
country={China}}

\affiliation[5]{
organization={School of Science, China University of Geosciences (Beijing)}, 
city={Beijing}, 
country={China}}


\begin{abstract}
 Earthquake forecasting remains a significant scientific challenge, with current methods falling short of achieving the performance necessary for meaningful societal benefits. Traditional models, primarily based on past seismicity and geomechanical data, struggle to capture the complexity of seismic patterns and often overlook valuable non-seismic precursors such as geophysical, geochemical, and atmospheric anomalies. The integration of such diverse data sources into forecasting models, combined with advancements in AI technologies, offers a promising path forward. AI methods, particularly deep learning, excel at processing complex, large-scale datasets, identifying subtle patterns, and handling multidimensional relationships, making them well-suited for overcoming the limitations of conventional approaches.

This review highlights the importance of combining AI with geophysical knowledge to create robust, physics-informed forecasting models. It explores current AI methods, input data types, loss functions, and practical considerations for model development, offering guidance to both geophysicists and AI researchers. While many AI-based studies oversimplify earthquake prediction, neglecting critical features such as data imbalance and spatio-temporal clustering, the integration of specialized geophysical insights into AI models can address these shortcomings.

We emphasize the importance of interdisciplinary collaboration, urging geophysicists to experiment with AI architectures thoughtfully and encouraging AI experts to deepen their understanding of seismology. By bridging these disciplines, we can develop more accurate, reliable, and societally impactful earthquake forecasting tools. 
\end{abstract}










\end{frontmatter}












\tableofcontents

\section{Introduction and Motivation}

Despite decades of research, earthquake forecasting remains in an exploratory stage, with current methods still falling short of achieving performance levels that would provide meaningful benefits for society. Forecasting models based primarily on past seismicity and other geomechanical measurements have shown limited progress, appearing to exhibit diminishing returns on the large deployed research efforts, likely because the available techniques are insufficient to extract useful information from the complex seismicity patterns. Moreover, the information they rely on may be too narrow. These methods, while valuable, focus predominantly on geo-mechanical information (mostly on seismic source data and space-time seismicity patterns, and sometimes on geodetic and deformation information, fault zone properties, subsurface structure data, plate kinematics), which may overlook critical aspects of earthquake precursors. 

To overcome these limitations, there is a growing recognition of the need to integrate non-seismic and non-mechanical information into earthquake forecasting models. Traditional approaches, which rely heavily on seismic and geomechanical data, struggle to capture the full complexity of earthquake precursors. Expanding the scope of data inputs to include geophysical and geochemical anomalies--such as atmospheric and ionospheric precursors (total electron content, low-frequency electromagnetic emissions, geomagnetic field variations, atmospheric electric field), thermal infrared anomalies, gas and hydrochemical emissions,  biological and ecological indicators, gravitational and tidal anomalies and even satellite-based remote sensing data\cite{freund2021earthquake,MearnsSor2021}--can provide a more comprehensive picture of the conditions leading to earthquakes. Incorporating diverse data sources allows for a holistic view that bridges gaps in traditional methods and leverages insights from a broader range of disciplines.

However, the heterogeneity and sheer volume of these data, combined with their intricate interdependencies, pose a significant analytical challenge. Extracting meaningful, multivariate, and multidimensional information from such complex datasets exceeds the capabilities of conventional statistical and computational techniques. This is where AI becomes indispensable. With its capacity to process large-scale, diverse data, identify subtle patterns, and handle multidimensional relationships, AI offers the tools necessary to transform earthquake forecasting. By integrating advanced AI methods with data from many varied sources augmented by geophysical knowledge, it will become possible to not only enhance predictive accuracy but also uncover new insights into the precursors and mechanisms of seismic events. This review highlights the potential of such an approach to push the boundaries of current forecasting capabilities and deliver meaningful societal benefits.
 
For seismologists and geophysicists, this comprehensive review offers an in-depth exploration of current AI methods applied to earthquake forecasting. We detail the types of input data utilized, the different machine learning models, the loss functions commonly employed, and key practical considerations in model development. Additionally, we examine the future integration of AI technologies with geophysical expertise, highlighting how this synergy can significantly improve the accuracy and reliability of earthquake forecasting.

Our aim is to equip domain experts with limited AI experience with the foundational knowledge needed to choose appropriate models, select relevant data inputs, and implement effective training strategies for creating high-quality forecasting systems tailored to their specific needs. For those already proficient in both geophysics and AI, we present cutting-edge techniques, future directions, and advanced methodologies that can help push the limits of current models, addressing existing challenges and paving the way for more powerful forecasting tools.

While many AI researchers have shown interest in earthquake forecasting and proposed various models, our review reveals that many of these approaches oversimplify the problem, often treating it as a straightforward binary classification or regression task, with little attention to the underlying physical processes. Critical features of earthquake data—such as the severe imbalance between positive and negative samples and the spatio-temporal clustering of seismic events—are frequently overlooked. Moreover, without a deep understanding of geophysics, performance evaluations in many studies are often overly optimistic, failing to reflect real-world conditions.

We collected 142 relevant papers on earthquake prediction and forecasting using machine learning from 1994 to the present (Q4-2025), and the information about these papers is available at https://github.com/AI-earthquake/citations-for-Predicting-future- Californian-earthquakes-with-deep-neural-networks/tree/main. This review addresses these common oversights and provides guidance on building specialized models that are both practically effective and grounded in physical principles. By doing so, we aim to bridge the gap between the AI and geophysics communities, helping AI experts gain the geophysical insights necessary to develop more robust and credible earthquake forecasting models. Our goal is to ensure that these models are not only technically sound but also recognized and validated by the geophysics community, fostering true interdisciplinary collaboration.

\section{Fundamentals of earthquake forecasting }\label{funda}

In this section, we aim to assist non-expert earthquake forecasting researchers in locating valuable learning resources and swiftly establishing a professional earthquake forecasting system. We also highlight various organizations and platforms currently involved in earthquake forecasting, detailing the models they employ. This approach is intended to enable future AI experts to conduct prospective testing after developing their own earthquake forecasting models and to identify robust benchmarks for comparison. Those already proficient in the field can feel free to skip this part.

\subsection{Differences between an earthquake forecast and an earthquake prediction}

The terms ``earthquake forecast'' and ``earthquake prediction'' are often confused, even among seismologists, and especially by non-experts.

Earthquake prediction refers to the precise determination of when, where, and with what magnitude an earthquake will occur. This approach aims to provide exact details about specific earthquake parameters, including the date, location, and intensity. An example of this would be an attempt to predict that a magnitude 6.5 earthquake will strike San Francisco on June 15th, 2025, at 10:30 AM. Such predictions require an exactness that is extremely difficult to achieve due to the complexity and unpredictability of earthquake-generating processes within the Earth's crust. Historically, earthquake prediction has proven to be highly elusive, and most claims have turned out to fail or to be incomplete.

In contrast, earthquake forecasting focuses on estimating the likelihood of seismic activity over well-defined time frames, spatial regions and magnitude range, generating probabilistic estimates rather than predicting specific events. In other words, while a prediction is more like a point prediction, a forecasts is a probabilistic statement.
For example, a forecast might indicate a 30 percent chance of a magnitude 6.0 or greater earthquake occurring within the next 10 years in the Los Angeles region. This approach does not aim to specify the exact time or location of an event but instead provides a probability based on historical data, geological knowledge, and patterns of past seismicity. The spatial and temporal windows, as well as the magnitude range and number of earthquakes, can be tailored to different needs. For instance, a forecast could state the probability to observe at least three earthquakes of magnitude 4.0 or higher occurring between midnight today and midnight tomorrow within the official city boundaries of Tokyo, Japan. This flexibility allows forecasting models to be adapted for various scales and applications, from long-term regional assessments to short-term, localized hazard estimates.

The key distinction is that forecasting aims to provide useful information about earthquake risks in a manageable and realistic way, helping communities prepare for possible seismic hazards without attempting to specify the exact details of when and where the next big quake will strike. Because of its probabilistic nature, forecasting is seen as more reliable and scientifically grounded. The term "forecasting" is widely accepted in the statistical seismology community, as it reflects more modest and achievable objectives than the exacting demands of prediction, making it a more practical tool for disaster preparedness and risk mitigation.

\subsection{Testing the performance of earthquake forecasting models}\label{testing}

Despite the development of a large variety of methods, earthquake prediction remains controversial, particularly when it comes to major earthquakes. As noted by Huang et al. \cite{huang2015forecasting}, progress has been slow, with only a few notable cases of successful predictions. These rare successes are often overshadowed by the numerous failures, leading many to dismiss successful predictions as mere coincidences or “lucky guesses." Let us however mention the unique approach and the strength of the Chinese earthquake prediction program that developed over a decade from 1966 to 1976 \cite{MearnsSor2021}, but which has been abandoned for decades for complex political and organisational reasons and is now witnessing a recent renewal.
One should also not forget the Russian earthquake prediction program launched by Vladimir Keilis-Borok and his collaborators and now led by Vladimir Kossobokov focused on a pattern recognition approach to earthquake prediction invented by Gelfand et al \cite{Gelfandetal76}.
The routine real-time testing of the M8 and M8-MSc algorithms started in 1992 and continues until present with a six-months periodicity. The statistical significance of these on-going predictions can be found in Refs.\cite{HealyKovde92,KosspRoma99,IsmailKov2020,Kossosolv21}. 

Several factors contribute to the skepticism among scientists regarding the feasibility of developing reliable earthquake prediction methods.
\begin{itemize}
    \item Slow Progress: Research in earthquake prediction has been painstakingly slow, with very few verifiable successes.
    \item Inconsistent Results: The few cases of success are vastly outnumbered by instances of failure, reinforcing the notion that successful predictions are more due to chance than a reliable method. Many past claims of success have proven irreproducible \cite{hough2010predicting}.
    \item Lack of clarity on the underlying mechanism(s): As Freund et al. (2021)\cite{freund2021earthquake} pointed out, 
    the lack of a clear physical mechanism linking non-seismic precursors (such as atmospheric or ionospheric anomalies) to earthquakes has led to widespread criticism, even from within the seismology and physics communities.
    \item Lack of rigorous statistical testing methodology, in particular for works involving proposed non-seismic precursors \cite{Mignan-recom2021}. 
\end{itemize}
This history highlights the critical need for rigorous verification, testing, and validation of models before they can be deemed reliable for practical use \cite{Mignan-recom2021}. 

Leaving the realm of earthquake predictions to focus on the more scientifically robust approach to earthquake forecasting, seismologists have developed structured and systematic methods for assessing the effectiveness and reliability of forecasting models, providing a clearer path toward improving their real-world applicability.
Three key principles for testing earthquake forecasting models have emerged \cite{mizrahi2024developing}.
\begin{itemize}
    \item[(i)] Clear Parameterization: Forecasters must define clear parameters, including latitude, longitude, magnitude, and time range. Predictions may optionally include depth range, focal mechanisms, or other measurable earthquake characteristics.
    \item[(ii)] Objective Success Criteria: The success or failure of a forecast must be easily and objectively determined, without requiring further interpretation.
    \item[(iii)] Use of Prior Data: Models should be constructed using data available before the prediction time, to avoid introducing biases from data that would not be available in a real-world scenario. While this statement is a priori obvious, it can become subtly violated in statistical tests using past data that attempt to reproduce real-life real-time situations.
\end{itemize}
These principles are intended to ensure fair and objective evaluations of forecasting models, making it easier to assess their practical value without ambiguity. Rule (iii) is particularly important to trust that pseudo-prospective forecasting methods can be applied to real-world situations.

To rigorously evaluate earthquake forecasting models, prospective, retrospective and pseudo-prospective testing methods are employed \cite{mizrahi2024developing}.
\begin{itemize}
    \item[(1)] Prospective Testing: In prospective testing, forecasts are made in real-time and evaluated against earthquakes that occur after the forecast is issued. This approach is the gold standard because it eliminates biases that could arise from knowing future data. Delayed prospective testing, in which models are isolated from post-forecast data, is another method that helps ensure unbiased assessments.
     \item[(2)] Retrospective Testing: Retrospective testing compares forecasts against past earthquake data, which the modelers may already know. While valuable, this method can be prone to bias, as the modelers may unconsciously tailor their models based on the known data.
      \item[(3)] To bridge the gap between retrospective and prospective testing, a hybrid method called pseudo-prospective testing is often used. In this approach, models are calibrated using data only up to a certain time, $t_0$, and then forecasts are made for a period after $t_0$. Although this method mimics the conditions of prospective testing, it carries a risk of bias, as the modelers may be influenced by their knowledge of future data, even if indirectly.
\end{itemize}

In summary, rigorous prospective or pseudo-prospective testing is essential for evaluating the practical utility of earthquake forecasting models. While retrospective testing has its uses, only by testing models in a forward-looking, unbiased manner can we gain reliable insights into their true performance. Ensuring that these models can withstand such scrutiny is critical for their adoption in real-world disaster preparedness and risk mitigation. 

\subsection{Benchmarks for earthquake forecasting models}

"Benchmarks" refer to standards used to evaluate and compare performance, quality, or effectiveness. This topic originally belonged to the previous sub-section \ref{testing}. However, among AI experts unfamiliar with the intricacies of earthquake forecasting, many studies evaluating AI-based forecasting models suffer from notable methodological flaws and a lack of rigor, undermining their credibility and professional standards. Therefore, we believe it is necessary to dedicate a separate section to benchmarks to emphasize their importance. 

When assessing the performance of earthquake forecasting models, a significant knowledge gap exists between seismologists and AI experts\cite{R3}. AI experts often treat earthquake forecasting merely as a classification problem, overlooking key spatio-temporal features of earthquakes, such as the imbalance between positive and negative samples and the pronounced clustering of earthquakes in both space and time. On the other hand, statistical seismologists tend to adopt more specialized evaluation metrics, comparing their models with better benchmarks.

Let us review a longstanding debate concerning earthquake forecasting. DeVries et al.\cite{R1} built a deep learning network with 13,000 parameters to forecast the spatial distribution of aftershocks, and this work was published in the prestigious journal Nature. However, Mignan and Broccardo\cite{R2} reanalyzed these findings by applying a two-parameter logistic regression model (equivalent to a single neuron) and achieved results identical to those in the original study\cite{R1}. Furthermore, they found that the accuracy of this deep and complex neural network was no greater than that provided by a simple empirical law. We believe that the main reason for this phenomenon lies in the disconnect between AI and the field of geophysics. Due to a lack of understanding of specialized knowledge in earthquake forecasting, researchers often focus solely on high scores from evaluation metrics and hastily claim the success of their models, overlooking the deeper principles of seismology and the challenges present in practical applications.

Such issues are not isolated cases. The study by DeVries et al.\cite{R1} is frequently cited primarily because it was published in one of the world’s most prestigious journals, known for its highly selective peer review process. After reviewing 77 articles on the application of neural networks to earthquake prediction/forecasting from 1994 to 2019, Mignan and Broccardo\cite{R3} discovered that only 47\% of these studies included a comparison to a baseline. Of those that did, 22\% used a Poisson null hypothesis or randomized data as the baseline, while the remaining 78\% compared their results to other machine learning methods utilizing the same features and data. In our survey extending that of Mignan and Broccardo\cite{R3} to 141 papers, we arrive at similar conclusions. Only 66.9\% of works using AI to forecast earthquakes were compared to a baseline, most of which are other machine-learning methods or simple Poisson null hypothesis.  When we want to demonstrate that a newly developed model surpasses existing limitations, we should compare our model with the best current models, including both the best AI models and the most advanced geophysical or statistical seismological models. However, most previous studies evaluate their work primarily from the perspective of AI research, focusing on the use of state-of-the-art AI technologies or the development of improved artificial neural networks. However, they often neglect to benchmark their models against existing, high-performing geophysical models, which is critical for demonstrating their practical value in earthquake forecasting.

Due to this significant cognitive bias, Mignan and Broccardo\cite{R3} noted in their review that these artificial neural network (ANN) models do not appear to provide new insights into earthquake predictability, as they fail to offer convincing evidence that these models outperform simple empirical laws. In our extended investigation, we found that four studies from 2023 compared their models with reasonably strong geophysical models, six of which made comparisons with some versions of the general class of ETAS (epidemic type aftershock sequence) models. We will discuss these studies in detail later. 

In summary, one of the biggest issues with previous research is the lack of comparison between their models and powerful benchmarks, a point that has been repeatedly emphasized in several past specialized studies\cite{mizrahi2024developing,R19}.

\subsection{Recommended resources for non-experts in earthquake forecasting}

This section seeks to provide non-expert earthquake forecasters with essential resources to build a foundational understanding of statistical seismology and professional earthquake forecasting. Additionally, it introduces key Operational Earthquake Forecasting (OEF) platforms and relevant institutions, enabling newcomers to quickly integrate into the geophysics and statistical seismology communities.

Our first recommendation is to start with the Community Online Resource for Statistical Seismicity Analysis (CORSSA), a platform specifically designed for students, beginners, as well as researchers in statistical seismology. It is presently in hiatus but it has been organized and authored by globally renowned statistical seismologists and earthquake forecasting experts, including Andrew J. Michael, Stefan Wiemer, Jiancang Zhuang and several other dedicated scientists. The platform provides educational resources, software tools, and best practices aimed at improving the application of statistical methods in earthquake science. CORSSA offers tutorials, articles, and guidance on how to apply these methods in real-world research. On this platform, users can learn about statistical seismology models such as the ETAS model, the Reasenberg and Jones model, as well as various de-clustering methods for earthquake catalogs, earthquake catalog completeness algorithms, professional earthquake forecasting model evaluation metrics and so on. For more detailed information, visit \href{http://corssa.org/en/home/}{http://corssa.org/en/home/}. 

We also recommend the review paper titled "Developing, Testing, and Communicating Earthquake Forecasts: Current Practices and Future Directions" by Mizrahi et al.\cite{mizrahi2024developing} on Operational Earthquake Forecasting. This comprehensive review is an essential resource for anyone involved in earthquake forecasting, risk mitigation, and public communication. Based on discoveries in statistical seismology, scientists have proposed numerous models, many of which have been applied to operational earthquake forecasting (OEF). OEF refers to the authoritative, near-real-time application of earthquake forecasting, with countries like Italy, New Zealand, and the United States already having operational systems in place, and more nations expected to follow. Key models used in OEF include the Reasenberg and Jones model\cite{reasenberg1989earthquake}, various versions of the Epidemic-Type Aftershock Sequence (ETAS) model\cite{R6}, the Epidemic-Type Earthquake Sequence (ETES) model\cite{helmstetter2006comparison}, the Short-Term Earthquake Probability (STEP) model\cite{gerstenberger2005real,woessner2010building}, the Every Earthquake a Precursor According to Scale (EEPAS) model, and ensemble models that integrate multiple approaches. Among these, sophisticated versions of the ETAS model has consistently demonstrated the best performance, particularly in aftershock forecasting. The review\cite{mizrahi2024developing} serves as an invaluable guide for those seeking to engage in earthquake forecasting research. The review highlights the significance of transparency, benchmark comparisons, prospective testing, and reproducibility in developing effective earthquake forecast models. Furthermore, it emphasizes the critical importance of collaborating with end-users to ensure that earthquake forecast  outputs are both socially relevant and practically applicable. Providing authoritative guidance on the subject, this review strikes a balance between scientific rigor and practical applications for risk management, making it an excellent frontier guide for professional earthquake forecasters.

This section concludes with an introduction to several established and emerging earthquake forecasting platforms, as well as research organizations dedicated to advancing earthquake forecasting. As Mizrahi et al.\cite{mizrahi2024developing}  already provide a detailed overview of the official Operational Earthquake Forecasting platforms, we will only give a brief introduction. The Italian OEF system generates forecasts but currently limits public access to its detailed models and code. Information may be available through authorized channels, but specific resources are not openly shared. The New Zealand system offers publicly accessible forecasts and has made efforts to communicate its findings. However, the underlying models may not be fully open-source. More information can be found on GNS Science. The United States Geological Survey (USGS) provides a range of earthquake-related data and resources, including aftershock forecasts. They also share some models and methodologies openly, which can be found on their website https://www.usgs.gov/programs/earthquake-hazards. 

Since its inception in California in 2007, the Collaboratory for the Study of Earthquake Predictability (CSEP)\cite{CSEP1,CSEP2}
has been conducting forecast experiments in a variety of tectonic settings and at a global scale. It has been operating four testing centers on four continents to automatically and objectively evaluate models against prospective data.
This research organization has been focused on earthquake forecasting, dedicated to assessing and validating the performance of earthquake forecasting models. It has promoted collaboration among different research teams by providing a unified testing framework and datasets, enhancing the scientific rigor and reliability of earthquake forecasting. Forecasters have been able to submit their forecasts to this platform, thus completing prospective testing. Its operations are scaling down as new initiatives emerge to modernize and streamline its relatively cumbersome testing infrastructure. See https://cseptesting.org.

The RichterX platform\cite{kamer2021democratizing}  is a real-time earthquake forecasting system that utilizes a version of the ETAS model \cite{nandan-Richter2021} to construct scenarios and calculate forecast probabilities in adjustable space-time-magnitude windows. Designed for global accessibility, this platform integrates extensive historical seismic data with real-time monitoring to generate accurate forecasts of future earthquake events. It features a user-friendly interface that allows scientists, researchers, and the general public to access real-time predictions and alerts. Additionally, users can submit their own predictions to the Richter X platform and compete against the ETAS model; those who outperform the platform’s model will be rewarded. The creators of the platform aim to further explore the potential of earthquake prediction through this competitive approach. The website for the Richter X platform is \href{https://www.richterx.com/}{https://www.richterx.com}. 

Since 1992, Vladimir G. Kossobokov and collaborators have been sharing intermediate-term, middle- and narrow-range earthquake predictions on a global scale through biannual monthly emails shared with a select group of approximately 150 experts in earthquake prediction research worldwide, using algorithms called M8 and and M8-MSc\cite{Kossobokov1999,Kossobokov2021}. The focus is on predicting earthquakes with magnitudes of 7.5 or greater and 8.0 or greater within predefined large spatial regions.

AoyuX is an emerging online research platform focused on enhancing earthquake predictability in China. Developed by the Institute of Risk Analysis, Prediction, and Management (RisksX) at the Southern University of Science and Technology (SUSTech), in collaboration with multiple research institutions, including the China Earthquake Networks Center and Beijing Normal University, with support from the National Natural Science Foundation of China - Earthquake Science Joint Fund and the Special Fund of the China Seismic Experimental Site.
In Chinese mythology, "Aoyu" is a legendary creature with a dragon's head and a fish's body, believed by ancient Chinese to cause earthquakes. The "X" represents the unknown and infinite in Western culture, symbolizing goals and hopes. The AoyuX platform aims to use cutting-edge statistical seismology tools for forward earthquake forecasting research. It will also serve as an online research platform for forward earthquake forecasting competitions with other models. In the future, the AoyuX platform will integrate geophysical observation data, artificial intelligence (AI). AoyuX will also create a visualization platform to improve communication between experts and the public, aiding collaborative efforts in monitoring and assessing major earthquakes. At the time of writing, AoyuX shares with Chinese experts a monthly earthquake forecasting report for the China Seismic Experimental Site of Sichuan-Yunnan region.

\section{A quick review of AI algorithm used for earthquake forecasting}

In Section \ref{AIA}, we present a concise overview of the AI algorithms and the principles behind them, as applied in previous studies.
Section \ref{Seisdemand} explores the potential role of AI technology in earthquake forecasting, emphasizing its integration from a seismological perspective. 

\subsection{AI algorithms used in earthquake prediction}\label{AIA}

\subsubsection{Traditional Machine Learning Models}

\paragraph{Decision Tree}
A decision tree~\cite{song2015decision} is a non-parametric supervised learning algorithm used for classification and regression tasks. It is structured hierarchically, consisting of a root node, internal nodes, and leaf nodes. The root node contains the entire sample set, internal nodes correspond to feature attribute tests, and leaf nodes represent the outcomes of the decisions. The construction of a decision tree typically involves three steps: feature selection, decision tree generation, and decision tree pruning. Feature selection aims to identify features that are highly correlated with the classification results, typically based on the information gain criterion. Decision tree generation begins after selecting the features. Starting from the root node, the algorithm computes the information gain for all features at the node and selects the feature with the highest information gain as the node's feature. Sub-nodes are then created based on the different values of this feature. This process is recursively applied to each sub-node until the information gain becomes negligible or there are no more features to select. Decision tree pruning is primarily aimed at combating "overfitting" by proactively removing certain branches to reduce the risk of overfitting. Common decision tree algorithms include the Iterative Dichotomiser 3 (ID3), the C4.5, and the Classification and Regression Tree (CART) algorithms.

Building on the classical decision tree, many extensions have been developed. For instance, Random Forest, introduced by Leo Breiman in  2001 \cite{Breiman2001}, is an ensemble learning method that constructs multiple decision trees during training and outputs the mode of the classes (classification) or the mean prediction of the individual trees (regression). Another extension, Inverse Boosting Pruning Trees \cite{Tongetal2019}, proposed by Tong et al. in 2019, is an ensemble method that combines an improved AdaBoost algorithm with pruning decision trees for classification. Additionally, Boosted Decision Tree Regression, proposed by Jerome H. Friedman in 2001 \cite{Friedman2001}, is a variant of boosting where the model iteratively focuses on reducing the residual errors of the predictions by combining multiple weak learners (in this case, decision trees) into a robust predictive model for regression tasks.

\paragraph{Support Vector Machines (SVM)} SVM~\cite{hearst1998support} is a supervised learning model used for classification and regression tasks. SVM operate by finding the optimal hyperplane that maximizes the margin between different classes in a dataset. The core components of SVM include support vectors, the hyperplane, and the margin. Support vectors are the data points that are closest to the hyperplane and influence its position and orientation. The hyperplane is a decision boundary that separates different classes, while the margin is the distance between the hyperplane and the nearest support vectors. According to the margin, SVM can be divided into hard margin SVM and soft margin SVM. The hard margin SVM is applicable when the data are linearly separable. It finds the hyperplane that not only maximizes the margin but also ensures that all data points are correctly classified without any errors. This approach, however, is very sensitive to outliers and may not be suitable for datasets with overlapping classes or noisy data. The soft margin SVM, introduced to handle non-linearly separable data, allows for some misclassifications to achieve a better overall model fit. It introduces slack variables to permit certain data points to lie within the margin or on the wrong side of the hyperplane. The degree of tolerance for misclassifications is controlled by a regularization parameter $C$. A higher value of $C$ seeks to reduce misclassifications, potentially at the cost of a smaller margin, while a lower value of $C$ allows more flexibility and a larger margin, possibly tolerating more classification errors.

\paragraph{K-Nearest Neighbor (KNN)} The KNN algorithm~\cite{kramer2013k} is a simple yet powerful supervised learning method widely used in both classification and regression tasks. The fundamental principle of KNN is that similar data points tend to be close to each other in the feature space. When making predictions, KNN identifies the 'k' nearest neighbors to the query point and bases its prediction on these neighbors. In classification tasks, KNN assigns the query point to the most common class among the neighbors while, in regression tasks, it predicts the average value of the neighbors. The effectiveness of KNN is influenced by several key factors, such as the choice of distance metric, the choice of the 'k' value, and the distribution of the data. Common distance metrics include Euclidean, Manhattan, and Minkowski distances, each of which affects how similarity is assessed. The selection of 'k' is critical: a smaller 'k' may lead to overfitting, making the model overly sensitive to noise, while a larger 'k' can smooth predictions but may result in underfitting. Although KNN is straightforward and easy to implement, it can become computationally expensive when handling large datasets, particularly as the dimensionality increases. To address these challenges, dimensionality reduction techniques or data structures such as KD-Trees~\cite{zhou2008real} are commonly employed.

\paragraph{Naïve Bayes} Naïve Bayes~\cite{rish2001empirical} is a simple yet powerful family of probabilistic classifiers based on Bayes' theorem, which assumes that features are conditionally independent given the class label. Although this simplifying assumption often does not hold in reality, Naïve Bayes classifiers still perform exceptionally well in practical applications, especially in high-dimensional data environments. The algorithm works by calculating the posterior probability of each class given a set of features and assigning the data point to the class with the highest posterior probability. This process involves using Bayes' theorem to update the probability estimate for each class as more features are considered, with the estimates being gradually refined. Naïve Bayes is typically divided into three variants: Gaussian Naïve Bayes for continuous data, Multinomial Naïve Bayes for discrete data, and Bernoulli Naïve Bayes for binary data. The algorithm is particularly well-suited for tasks involving large feature spaces, performs well even with small training datasets, and is robust to irrelevant features. However, its performance may be negatively impacted if the independence assumption is severely violated or if there are strong interactions between features.

\paragraph{Polynomial Regression} Polynomial Regression~\cite{ostertagova2012modelling} is an advanced form of linear regression that models the relationship between independent and dependent variables using an $n$th-degree polynomial. Unlike linear regression, which fits a straight line, polynomial regression allows for the fitting of a curve, enabling the capture of more complex, non-linear relationships within the data. The model parameters are typically estimated using the least squares method, which minimizes the sum of the squared differences between the observed values and the predicted values. The flexibility of polynomial regression makes it particularly useful when the data exhibits non-linear trends that a straight line cannot accurately represent. However, as the degree of the polynomial increases, the risk of overfitting also rises, meaning the model may start to fit the noise in the data rather than the true underlying pattern. To mitigate this issue, regularization techniques such as Ridge regression or Lasso regression can be applied. Therefore, careful selection of the appropriate polynomial degree is crucial to balance the model's complexity and its ability to generalize to new data.

\paragraph{Mixture of regressions}  
Mixture regression~\cite{wedel2000mixture} is a probabilistic approach for modeling data generated by a combination of sub-populations within an overall population. It assumes that the dataset arises from $K$ distinct regression functions, each representing a specific cluster. Data points are probabilistically assigned to these clusters, and a corresponding regression function is fitted to each cluster. This method combines clustering and regression into a unified framework, with parameters and cluster assignments iteratively refined using algorithms such as Expectation-Maximization (EM) to maximize the likelihood of the data. This approach captures heterogeneity by identifying distinct relationships within subsets of the data while accounting for the overlap between clusters. The flexibility of this method makes it applicable to both linear and non-linear regression models, allowing it to effectively handle scenarios where multiple generating processes underlie the observed data.

\paragraph{Logistic Regression} Logistic Regression~\cite{menard2001applied} is a widely used statistical method for binary classification problems. The core idea is to use the logistic function to map predicted values to a probability range between 0 and 1, thereby estimating the probability that a given input point belongs to a specific class. The model calculates a probability value through a linear combination of input features, representing the likelihood that the input belongs to the positive class. In implementing logistic regression, the Maximum Likelihood Estimation (MLE) method is commonly used to determine the model's parameters, maximizing the likelihood of the observed data. Once trained, the logistic regression model can be used to classify new data and output a predicted probability based on the input features. To prevent overfitting, L1 or L2 regularization can be introduced, which adds a penalty term to the loss function to constrain the magnitude of the coefficients. 

Multinomial Logistic Regression is an extension of logistic regression that can handle classification tasks with more than two categories. The principle behind it is to calculate a logistic regression model for each class and then use the softmax function to convert the outputs of these models into a probability distribution over the classes. Each probability represents the likelihood that the input data belongs to a particular class, and the model ultimately selects the class with the highest probability as the prediction result.

\subsubsection{Deep Learning Models}

\paragraph{Feed-forward Neural Network (FNN) } A FNN~\cite{bebis1994feed} is a type of artificial neural network where information flows in a single direction, from the input layer, through one or more hidden layers, and finally to the output layer. Each neuron in one layer is connected to the neurons in the next layer via weights, without any backward connections or cycles, ensuring that information is not fed back into the network. Neurons in a FNN typically use activation functions, such as Sigmoid, Tanh, or ReLU, to introduce non-linearity, thereby enhancing the network's ability to model complex patterns. During training, the weights and biases within the network are adjusted using the backpropagation algorithm, aiming to minimize the loss between the predicted output and the true labels, with common loss functions being Mean Squared Error or Cross-Entropy Loss. Backpropagation is a gradient-based optimization method that computes the partial derivatives of the loss function with respect to the weights of connections between neuronal nodes and iteratively updates these weights to minimize the loss function. Depending on the depth of the network, FNNs can be classified into shallow networks with only one hidden layer and deep networks with multiple hidden layers, where deep networks are capable of capturing more complex patterns and features, forming the core foundation of deep learning.

Among the various implementations of FNNs, Multilayer Perceptrons (MLP), Radial Basis Function Neural Networks (RBFNN), and Extreme Learning Machines (ELM) are three significant variants, each optimized for different application scenarios and needs. The Multilayer Perceptron (MLP), one of the most classical architectures of feedforward neural networks (FNNs), features fully connected layers and is typically trained using the backpropagation algorithm. Its structured design and ability to model complex, non-linear relationships make it particularly well-suited for processing structured data. Radial Basis Function Neural Networks (RBFNN) leverage radial basis functions as activation functions, making them particularly effective for handling nonlinear classification and function approximation tasks. The Extreme Learning Machine (ELM) is notable for its exceptionally fast training process, made possible by randomly assigning and fixing the weights and biases of the hidden layers, rather than iteratively optimizing them. This approach reduces computational complexity significantly, as only the output weights need to be adjusted using a straightforward analytical solution, resulting in highly efficient training without sacrificing performance on many tasks. These three variants, with their unique strengths, have become effective tools for solving various practical problems in their respective application domains. In addition to these FNN variants, the Functional Link Artificial Neural Network (FLANN) is another significant model. FLANN enhances the traditional feedforward network architecture by incorporating a functional link layer that applies nonlinear transformations, such as polynomial or trigonometric expansions, to the input space. This augmentation significantly increases the network's ability to capture and model complex nonlinear relationships, making FLANN particularly well-suited for applications in nonlinear function approximation and time series prediction, while maintaining a streamlined and computationally efficient structure.

\paragraph{Convolutional Neural Network (CNN)} A CNN~\cite{li2021survey} is a deep learning model specifically designed for processing data with a grid-like topology, most commonly used for analyzing images and videos. CNNs effectively capture and extract spatial hierarchical features from data by utilizing a combination of convolutional layers, pooling layers, and fully connected layers. The core characteristics of CNNs include local connectivity, weight sharing, and downsampling. The convolutional layer is the central component of a CNN, and it uses kernels that slide over the input data to extract local features such as edges and corners. By employing weight sharing, CNNs reduce model complexity while retaining essential feature information. Pooling layers are used to downsample feature maps, reducing the spatial dimensions of the data, thereby lowering computational costs and preventing overfitting. The fully connected layer, located at the end of the network, maps the extracted features to the output space, performing the final classification or regression task. The training process of a CNN involves both forward propagation and backpropagation. During forward propagation, the input data passes sequentially through the convolutional, pooling, and fully connected layers, with features being extracted and predictions made layer by layer. In backpropagation, the network updates the parameters of each layer by using gradient descent to minimize the error between the predicted results and the actual labels, continuously optimizing the model. The strength of CNNs lies in their ability to automatically learn hierarchical feature representations of data and significantly reduce the number of parameters through the mechanism of weight sharing, making them particularly well-suited for processing image data. Overall, CNNs, with their unique architecture and powerful feature extraction capabilities, have achieved remarkable success in many complex tasks, becoming a crucial component of modern deep learning.

In addition to their success in processing grid-like data such as images, CNNs have been extended to handle temporal data through the development of the Dilated Causal Temporal Convolutional Network (Dilated Causal TCN). This variant of CNN is designed specifically for processing time-series data by capturing long-term dependencies while preserving causality, making it particularly effective in tasks where the sequence order is crucial. The Dilated Causal TCN achieves this by combining causal convolution, which ensures that each output only depends on the current and previous inputs, with dilated convolution, which expands the receptive field without increasing the computational load. This structure allows the network to efficiently model long-range temporal dependencies without requiring deep layers or excessive computational resources. This extension further underscores the critical role of CNNs in modern deep learning, offering powerful tools for a wide range of applications.

\paragraph{Recurrent Neural Network (RNN)} RNN~\cite{medsker2001recurrent} is a class of neural network models specifically designed for processing sequential data. Their defining feature is the use of a hidden state that acts as a memory, enabling the network to retain information from previous time steps and utilize it in subsequent computations. This memory capability is achieved through recurrent connections within the hidden layers, which update the hidden state at each time step based on both the current input and the previous hidden state. By sharing parameters across all time steps, RNNs capture temporal dependencies effectively, making them particularly well-suited for tasks involving time series or other sequential data. The recursive nature of their hidden state updates allows RNNs to dynamically model the evolving characteristics of sequences, making them indispensable for sequence-related applications such as natural language processing, speech recognition, and time series forecasting.

While standard Recurrent Neural Networks (RNNs) excel at capturing the dynamics of sequential data, they face significant challenges, including the vanishing and exploding gradient problems. These issues can severely impede their ability to learn and retain long-term dependencies in extended sequences, limiting their effectiveness in tasks requiring a deeper temporal understanding. To address these issues, researchers have developed several improved RNN architectures, with the most common being the Long Short-Term Memory (LSTM) networks and Gated Recurrent Units (GRU). LSTM networks introduce memory cells and gating mechanisms that effectively mitigate the vanishing gradient problem, ensuring that important information can be retained over long periods. GRU, a simplified version of LSTM, reduces network complexity while achieving comparable performance on certain tasks. Additionally, Bidirectional RNNs and Bidirectional Long Short-Term Memory (Bidirectional LSTM) cells consider both forward and backward input sequences, allowing for a more comprehensive capture of the overall context within sequential data. Building on these advancements, researchers have also proposed the Structural Recurrent Neural Network (SRNN), which incorporates structural information into the sequence modeling process, enabling the network to capture more complex temporal and spatial dependencies, thereby enhancing its ability to process intricate sequential data.

\paragraph{Graph Neural Network (GNN)} GNN~\cite{li2020topology} is a type of deep learning model specifically designed for processing graph-structured data, particularly suited for non-Euclidean structures. Unlike traditional neural networks, GNNs are characterized by their ability to directly handle the complex relationships between nodes and edges in graph data. The core concept of GNNs lies in the mechanism of message passing, where the features of each node are iteratively updated layer by layer through the exchange of information between neighboring nodes. This mechanism enables GNNs to effectively capture both local and global dependencies within the graph structure. GNNs typically involve two key steps: Aggregation and Update. During the aggregation step, each node gathers information from its neighbors and aggregates it using methods such as summation, averaging, or taking the maximum. In the update step, the aggregated information is combined with the node’s own features to update its representation. This process of aggregation and updating allows GNNs to effectively capture and leverage the complex dependencies inherent in graph structures, making them highly effective in tasks such as node classification, graph classification, and edge prediction. Additionally, GNNs exhibit locality and scalability, as the updates for each node primarily depend on its local neighborhood, enabling the model to efficiently operate on large-scale graphs. Through recursive propagation across multiple layers, GNNs can integrate information from distant nodes, further enhancing their performance in tasks like node classification, graph classification, and edge prediction. Owing to their robust capability in modeling graph structures, GNNs have been widely applied across various domains, demonstrating unique advantages in handling complex networked data.

\paragraph{Transformer} 
The Transformer~\cite{wen2025goal} is a groundbreaking deep learning model that relies entirely on the attention mechanism, eliminating the need for recurrent or convolutional structures. This innovative design enables efficient and effective handling of sequential data, particularly for tasks requiring long-range dependency modeling. The key components of the Transformer include the Multi-Head Self-Attention mechanism, the Feedforward Neural Network, and Positional Encoding. The self-attention mechanism allows the model to evaluate relationships between all elements in a sequence simultaneously, capturing global dependencies while processing each element in parallel. Multi-head attention enhances this capability by projecting the input into multiple subspaces, computing attention independently in each subspace, and aggregating the results, thereby improving the model’s capacity to capture complex relationships. The Feedforward Neural Network, a fully connected layer applied after the attention mechanism, further refines the representations by learning non-linear transformations for each position independently. Since the Transformer does not inherently preserve sequence order, Positional Encoding is introduced to embed explicit position information into the input vectors, maintaining the sequential context required for downstream tasks.

The Transformer's architecture consists of an Encoder and a Decoder, each composed of multiple identical layers. Encoder layers feature a multi-head self-attention mechanism and a feedforward neural network, while decoder layers include these components and an additional cross-attention mechanism. This cross-attention enables the decoder to integrate the encoder’s output while generating the target sequence step by step. The encoder transforms the input sequence into a set of rich representations, which the decoder leverages to produce the output sequence. By processing sequences in parallel and efficiently modeling long-range dependencies, the Transformer excels in tasks such as natural language processing, machine translation, and text generation, overcoming limitations of previous models reliant on recurrence or convolution.

Building upon the foundational Transformer, the Temporal Fusion Transformer (TFT) by Google extends its capabilities to time series forecasting. TFT dynamically selects relevant features at each time step using self-attention, making it particularly effective for multi-horizon predictions in complex time-series data. By learning both long-term temporal dependencies and context-specific patterns, TFT not only delivers high performance but also provides interpretability by quantifying the contributions of various input features to the predictions. This extension highlights the versatility of the Transformer architecture across diverse domains, from text processing to temporal data analysis.

\paragraph{Hybrid Models} The cross-application of hybrid models has become increasingly prevalent in the evolution of machine learning and deep learning, as these models leverage the strengths of different algorithms to address complex challenges more effectively. One such example is the integration of a Support Vector Regressor (SVR) with a Hybrid Neural Network (HNN), which combines the precise classification capabilities of SVM with the robust learning capacity of neural networks. This synergy is particularly effective in handling complex nonlinear relationships, leading to enhanced classification accuracy and improved generalization across various domains. Building on this concept of combining strengths, the integration of CNNs with LSTM networks, exemplified by the CNN-LSTM hybrid model or Convolutional LSTM (ConvLSTM), demonstrates significant advantages in processing data that contains both spatial and temporal features. By utilizing CNNs to extract spatial features from images or video sequences, and LSTMs to capture temporal dependencies, these models achieve efficient spatiotemporal modeling, making them highly effective in applications such as video analysis and speech recognition. Extending this approach further, the combination of RNNs with GNNs, as seen in models like the Physics-Informed Recurrent Graph Network or the Spatiotemporal Prior-Informed Deep Network, offers enhanced modeling capabilities for dynamic systems. These models effectively merge temporal sequence modeling with graph structure analysis, demonstrating superior performance in handling complex spatiotemporal data, particularly in the analysis of dynamic network structures. Moreover, the Attention-Based LSTM model enhances the processing of long sequences by incorporating attention mechanisms, allowing the model to focus on critical time steps within a sequence. This improvement is especially beneficial in tasks such as natural language processing and text generation, where capturing long-range dependencies is crucial. 

In addition to these hybrid models, the integration of fuzzy systems with neural networks has led to the development of a variety of neuro-fuzzy models. For example, the Hybrid Adaptive Neuro-Fuzzy Model combines the adaptability of neural networks with the interpretability of fuzzy systems, enabling the model to learn and adjust fuzzy rules dynamically based on data. The Adaptive Neural Fuzzy Inference System (ANFIS) further extends this concept by applying neural network learning algorithms to optimize the parameters of fuzzy inference systems, thus enhancing their accuracy and flexibility. Other models, such as Adaptive Neuro Fuzzy Modeling with Fuzzy Subtractive Clustering, introduce clustering techniques to improve the initialization and adaptation of fuzzy rules, allowing for more precise and context-aware decision-making. These neuro-fuzzy models are particularly effective in environments where data is uncertain or imprecise, as they leverage the strengths of both fuzzy logic and neural networks to create models that are both interpretable and powerful. In summary, these hybrid models represent a powerful evolution in machine learning, as they strategically combine different methodologies to address specific challenges, thereby expanding their applicability across a wide range of domains and providing robust solutions for complex real-world problems.

\subsubsection{Deep Learning Algorithms}

\paragraph{Meta Learning} Meta-learning~\cite{hospedales2021meta}, or ``learning to learn", is a machine learning paradigm that empowers models to adapt rapidly and effectively to new tasks by leveraging knowledge gained from a diverse set of related tasks. Its core idea is to train models on multiple tasks, enabling them to develop generalizable problem-solving strategies rather than excelling at a single task. This approach allows models to generalize across tasks and adjust quickly to unseen scenarios, making them particularly effective in dynamic environments. Meta-learning operates through a two-level optimization framework: task-specific learning, which fine-tunes the base model's parameters for individual tasks, and meta-optimization, which refines the meta-model to enhance adaptability across tasks. By learning to optimize the learning process itself, meta-learning achieves faster convergence and improved performance when addressing new challenges.

A key feature of meta-learning, especially in gradient-based methods like Model-Agnostic Meta-Learning (MAML), is its reliance on higher-order derivatives of the loss function. In these methods, the inner loop focuses on updating the base model’s parameters for each task using gradient descent, while the outer loop adjusts the meta-model by optimizing a meta-objective that evaluates the base model’s performance after fine-tuning. This requires second-order derivatives to capture how changes in meta-parameters influence the learning dynamics of the base model, providing insights into the curvature of the loss landscape. These insights allow the meta-model to refine initialization or learning strategies, ensuring efficient adaptation to new tasks with minimal data. This hierarchical optimization framework helps meta-learning generalize across diverse tasks, avoiding overfitting to specific ones, and has made it highly effective for applications such as few-shot learning, reinforcement learning, and hyperparameter tuning.

\paragraph{Federated Learning} Federated Learning~\cite{lin2025fedrsclip} is an innovative distributed machine learning paradigm that achieves data privacy protection and efficient model updates by training models across multiple devices or local data sources. The core idea is to perform local model training on each device and send the updated model parameters (such as gradients or weights) to a central server for aggregation, rather than centralizing the raw data. The central server receives model updates from all participating devices, aggregates these updates to generate a global model, and then sends the global model back to the devices for the next round of local training. This iterative process allows the model to be progressively optimized without compromising user privacy. Federated Learning excels in privacy protection by integrating techniques such as differential privacy and secure multi-party computation to further safeguard sensitive information. Differential privacy introduces noise into the model parameters to obscure individual data contributions, ensuring that personal data remains confidential. Secure multi-party computation enables multiple participants to collaboratively compute model updates securely without exposing their individual data. 

Moreover, Federated Learning effectively handles data distribution heterogeneity and computational capacity differences among devices, ensuring the global model's applicability across diverse scenarios. Communication efficiency is also a key consideration in Federated Learning; researchers have employed methods such as gradient compression and reduced communication frequency to lower communication costs and network bandwidth usage, thereby accelerating the training process. Federated Learning has demonstrated broad applicability in privacy-sensitive domains, providing a robust solution for deploying machine learning in environments where data privacy is a critical concern. By seamlessly integrating model training with data privacy protection, Federated Learning is emerging as a pivotal direction in the field of machine learning.

\subsection{An Introduction to Indirect Prediction Methods}

Using the literature search methods outlined in the introduction, we identified papers that diverge from the traditional focus on predicting earthquake magnitude, epicenter, occurrence time, and other seismic information. Instead, these studies aimed to detect physical or chemical precursors potentially linked to earthquakes, which we refer to as non-seismic alarms. While these studies are not the primary focus of our review, we provide a brief overview of this area.

There are two primary methods for extracting earthquake precursors using AI technologies:
\begin{enumerate}
    \item Background Field Method: This approach involves using AI to construct a background field that models the normal distribution and variation of signals in the absence of earthquake activity. When an observed signal deviates significantly from this background—exceeding a predefined threshold—it is flagged as potentially related to an earthquake. For example, Negarestani et al.\cite{R61} used a layered neural network to analyze variations in soil radon concentrations, determining whether these changes were linked to earthquakes or influenced by other environmental factors, such as rainfall.

    \item Direct Feature Extraction Method: This method leverages AI to directly extract specific precursors from raw data, particularly when signal characteristics associated with earthquakes are known. For instance, geomagnetic anomalies linked to earthquakes often manifest as square waveforms. Recognizing this, Xue et al.\cite{R62} developed a real-time automatic engine to extract square wave signals from background geomagnetic data, hypothesizing a correlation between these signals and earthquakes.
\end{enumerate}
While these approaches offer promising avenues for precursor detection, they share a critical challenge with earlier non-AI-based methods: the need for long-term statistical validation. Establishing a robust correlation between detected precursors and earthquake occurrences remains an essential next step. Additionally, researchers must build upon these precursor detections to develop reliable earthquake prediction or forecasting models, further integrating these findings into broader seismic research frameworks.

This area of study highlights the potential of AI not only to enhance traditional earthquake prediction models but also to advance the detection of subtle precursors that may hold the key to understanding earthquake mechanisms. However, rigorous testing and validation are necessary to ensure the reliability and applicability of these methods in practical forecasting scenarios.

\subsection{‘Seismologists’ demand for AI technology in earthquake forecasting}\label{Seisdemand}

Let us revisit a fundamental question: why incorporate AI technology into earthquake forecasting, and what specific role does it play? From a performance standpoint, we can categorize the contributions of AI models to the seismology and geophysics communities into three valuable types:
\begin{enumerate}
    \item {\bf Performance Parity with Speed}: AI models that achieve predictive accuracy comparable to leading geophysical models, such as some versions of the ETAS model, but operate significantly faster, thereby enhancing efficiency.
    \item {\bf Performance Supremacy}: AI models that surpass the predictive capabilities of the best existing geophysical models, offering improved accuracy and reliability.
    \item {\bf Complementary Innovation}: AI models that address limitations in geophysical models or uncover new physical laws and patterns, providing insights that were previously inaccessible to the geophysical community.
\end{enumerate}
While enhancing accuracy and speed is a fundamental expectation, the true value of AI lies in its ability to augment and expand the capabilities of traditional approaches, driving innovation and uncovering deeper insights in earthquake forecasting.

As outlined in Section \ref{funda}, it is essential to benchmark the AI model that one develops against the best available geophysical models. The research holds value for earthquake prediction if the model's performance is comparable to or exceeds that of the leading geophysical models, or if it uncovers previously unknown physical patterns. A review of 136 papers revealed that none had conducted prospective testing. However, six recent studies merit attention for comparing their AI models with geophysical or statistical seismology models, all of which were evaluated using pseudo-prospective testing. 

Chen et al. \cite{R4} developed a deep neural network that uses data from the Seismic Research Center Moment Tensor Catalog (SRCMOD) finite fault database and the International Seismological Centre (ISC) earthquake catalog to predict aftershocks following the 2008 Wenchuan and 2011 Tohoku earthquakes. The performance of this AI model was compared to the Coulomb Failure Stress Change ($\Delta$CFS) model. Their findings showed that the AI model achieved predictive accuracy comparable to the $\Delta$CFS model while offering significantly greater computational efficiency. 

Dascher-Cousineau et al. \cite{R5} developed a deep learning model to focus solely on the forecast of the occurrence times of earthquakes in small regions of Southern California, while magnitudes and spatial dependence were not considered. The benchmark is taken as a pure temporal ETAS-like model with the intensity proportional to an exponential function of the past earthquake history. This is different from the linear dependence in the standard ETAS model \cite{R6}. They claim a better performance than their “minimum performance benchmark” (in their own words). 

Zlydenko et al. \cite{R8} used an ANN encoder that imitates the mathematical structure of ETAS, with response functions learned by the ANN based on a number of features such as time intervals between present time and past earthquakes, distances, magnitudes, and so on. This ANN model performs on par with, or even surpasses, a standard ETAS model with isotropic spatial kernels
in terms of the average information gain per earthquake while achieving a 1000-fold reduction in computation time. Another valuable attribute of this ANN is that it is able to learn some anisotropic spatial dependence of the spatial kernel of earthquake triggering, in agreement with the known tendency of earthquakes to cluster along faults, thus improving on the ETAS versions that use isotropic spatial kernels. However, it is important to note that anisotropic spatial kernels for the ETAS model have been developed and applied in past statistical seismology research \cite{ZhOgVJ02,OgZh06}.

Stockman et al. \cite{R7} employed an extended temporal neural model to forecast short-term seismicity, formulating the neural point process in a manner similar to the ETAS model but with significantly greater flexibility in representing the intensity function. Their results demonstrated that the neural model not only outperformed their implementation of the ETAS model but also exhibited faster training times, highlighting its efficiency and adaptability.

Zhang et al.\cite{zhang/10.1093/gji/ggae373}  used a Fully Convolutional Network (FCN) model, taking as input the spatial map of the logarithm of past estimated released earthquake energies, to forecast future earthquakes. This model was applied to California and compared with an advanced version of the epidemic-type aftershock sequence (ETAS) model. Their pseudo-prospective experiments demonstrated that the FCN model performed similarly to the ETAS model in forecasting earthquakes with magnitudes $M \geq 3.0, 4.0$, and $5.0$, as assessed by the Molchan diagram. Furthermore, training and implementing the FCN model was 2000–4000 times faster than calibrating the ETAS model and generating its probabilistic forecasts.

Zhan et al. \cite{ZhanSTC} developed an ETAS-Inspired Spatio-Temporal Convolutional (STC) Model for next-day earthquake forecasting, building upon the structure of the Epidemic-Type Aftershock Sequence (ETAS) model. The model discretizes the time-space domain into bins. The continuous space-time convolution operation involved in the calculation of the intensity of the ETAS model that embodies the self-excited nature of earthquake triggering is replaced in the STC model by a discrete convolution operation. Since the structures of the spatio-temporal functions are copied from the ETAS model, only six neurons are sufficient to encode the six key parameters of the STC model. They encapsulate how past seismic activity contributes to the probability of future events in terms of their location, time, and magnitude. The convolution operation of the STC model provides the expected
number of triggered earthquakes for each spatio-temporal bin. In pseudo-prospective experiments conducted in California, the comparative performances of their reference ETAS model and of the STC model were evaluated using the receiver operating characteristic (ROC) curve, the precision–recall (PR) curve, and the parimutuel gambling score (PGS), all of which demonstrated that the STC model outperformed the ETAS model.

In summary, despite nearly 40 years of research and the identification of 140 studies on AI-based earthquake prediction in our review, a substantial number of these works have fallen short of the expectations of geophysicists. This shortfall is primarily due to a lack of expertise in geophysics and earthquake forecasting among AI researchers. They often prioritize the use of advanced AI techniques over improving the state of the art in earthquake forecasting.

However, we have highlighted six notable studies that offer valuable insights into building effective AI models. Incorporating the structure and domain knowledge of established models, such as the ETAS model, has proven to be a promising strategy, as demonstrated by works like \cite{R8, ZhanSTC}. Additionally, studies using RNN networks and encoder-decoder architectures \cite{R5, R7} show that these approaches can rival the ETAS model's performance while addressing some of its limitations. Furthermore, Zhang et al. \cite{zhang/10.1093/gji/ggae373} illustrate that AI models can autonomously learn the spatiotemporal distribution of earthquakes without requiring extensive knowledge of statistical seismology, achieving performance scores comparable to the ETAS model.

These findings underscore the potential of AI in earthquake prediction, particularly when AI techniques are integrated with geophysical insights and when models leverage the inherent strengths of both fields.

\textbf{Does this mean that the other 136 studies hold no value? Of course not!} This review systematically categorizes the inputs used in these studies, creating a valuable repository of resources for future researchers developing AI-based earthquake prediction models. Additionally, some studies have introduced innovative training methods that provide meaningful advancements in model optimization. Many AI researchers have also proposed ingenious approaches to reframe earthquake prediction as AI tasks, contributing fresh perspectives and methodologies to the field.

While previous reviews have primarily focused on cataloging AI techniques, our review takes a distinct approach by adopting a geophysical perspective. We prioritize the insights and practical experiences related to earthquake prediction derived from past studies, compiling them into the later sections of this review. This approach represents the key distinction and added value of our work—bridging the gap between AI technology and the specialized domains of geophysics and earthquake prediction. By emphasizing the integration and localization of AI within these fields, we aim to provide a framework that fosters deeper collaboration and more effective applications of AI in earthquake forecasting.

\section{An in-depth analysis of AI-based earthquake forecasting models from a seismological perspective}

\subsection{Background}

In the previous section, we introduced the AI algorithms employed in earthquake forecasting, an area of primary interest for AI experts. Previous review papers, such as Banna et al. \cite{R9} and others \cite{R10}, have also centered their analyses on AI algorithms, organizing comprehensive overviews around this theme. For instance, Banna et al. \cite{R9} categorizes AI approaches into rule-based methods, shallow machine learning, and deep learning techniques, and evaluates their performance in earthquake forecasting tasks.

However, providing a definitive assessment of the complex AI techniques discussed in these studies is challenging. Many do not systematically compare their AI models against professional geophysical or statistical seismology models, leaving their potential for breakthroughs in earthquake prediction uncertain. This gap highlights the need for deeper integration and benchmarking of AI methods within the established frameworks of geophysics and earthquake forecasting.

As observed by Mignan and Broccardo \cite{R3}, many of the studies we reviewed have not directly contributed new insights to improve earthquake forecasting. However, these studies possess significant potential value that warrants further exploration. It is well understood that a model’s performance depends not only on the AI algorithms employed but also on critical factors such as input data, the choice of loss functions, and evaluation metrics. Through careful analysis, we identified a wealth of valuable insights in the non-algorithmic components of these studies, which merit further attention.

To support future research, our review systematically organizes the inputs used in these studies, creating a repository of materials for developing AI-based earthquake forecasting models. Additionally, we highlight innovative training methods proposed in some studies, along with creative approaches by AI experts to reframe earthquake forecasting as an AI task. These contributions demonstrate the ingenuity and potential of integrating AI into this field.

Unlike previous reviews, which primarily focus on cataloging AI techniques, our work adopts a geophysical perspective. We prioritize the practical experiences and insights related to earthquake forecasting found in past studies, emphasizing how AI technology can be localized and effectively integrated within the specialized domains of geophysics and earthquake forecasting. This geophysics-centered approach represents the key distinction and added value of our review, aiming to bridge the gap between AI innovation and domain-specific expertise for meaningful advancements in earthquake prediction.

\subsection{Diversity in AI-Based Earthquake Forecasting Models: Approaches, Data, and Evaluation}\label{framework}

A comprehensive statistical analysis of the inputs, outputs, loss functions, and evaluation metrics from the 141 papers reviewed reveals significant diversity in AI-based earthquake forecasting models. These models utilize over 73 types of inputs, 20 types of outputs, and a wide range of evaluation metrics. While this diversity reflects the flexibility and creativity of researchers, it also hinders meaningful comparison and communication between models. Standardizing outputs, datasets, and evaluation metrics would facilitate a more straightforward assessment of model strengths and weaknesses, fostering better collaboration and progress in the field.

This approach is well-established in AI research, particularly in areas such as image recognition and computer vision, where models often rely on consistent datasets, evaluation metrics, and application objectives. By contrast, the lack of such standardization in AI-based earthquake forecasting models contributes to their perceived "heterogeneity." While this diversity enables tailored approaches, it also underscores the need for a unified framework to benchmark and advance these models effectively.

For instance, in the field of multi-object detection using AI models in real driving environments, researchers commonly use publicly available datasets like FLIR-ADAS and KAIST \cite{R11} to train their models. Performance is then evaluated using standardized metrics such as Average Precision (AP) and mean Average Precision (mAP). This consistency allows researchers to focus on developing better AI algorithms to improve object recognition accuracy. Here, the term "standardization" is not used negatively but rather highlights how it simplifies the comparison and evaluation of different models, driving advancements in the field.

Similarly, in earthquake prediction research, particularly in AI-based earthquake forecasting models, such standardization is essential for enabling meaningful comparisons between models. To address this need, this review introduces a model and benchmark framework based on our team’s prior research. This framework establishes a preliminary standard for future researchers to facilitate comparisons, discussions, and the collective advancement of the field.

\begin{figure}
\centering
\includegraphics[width=\linewidth]{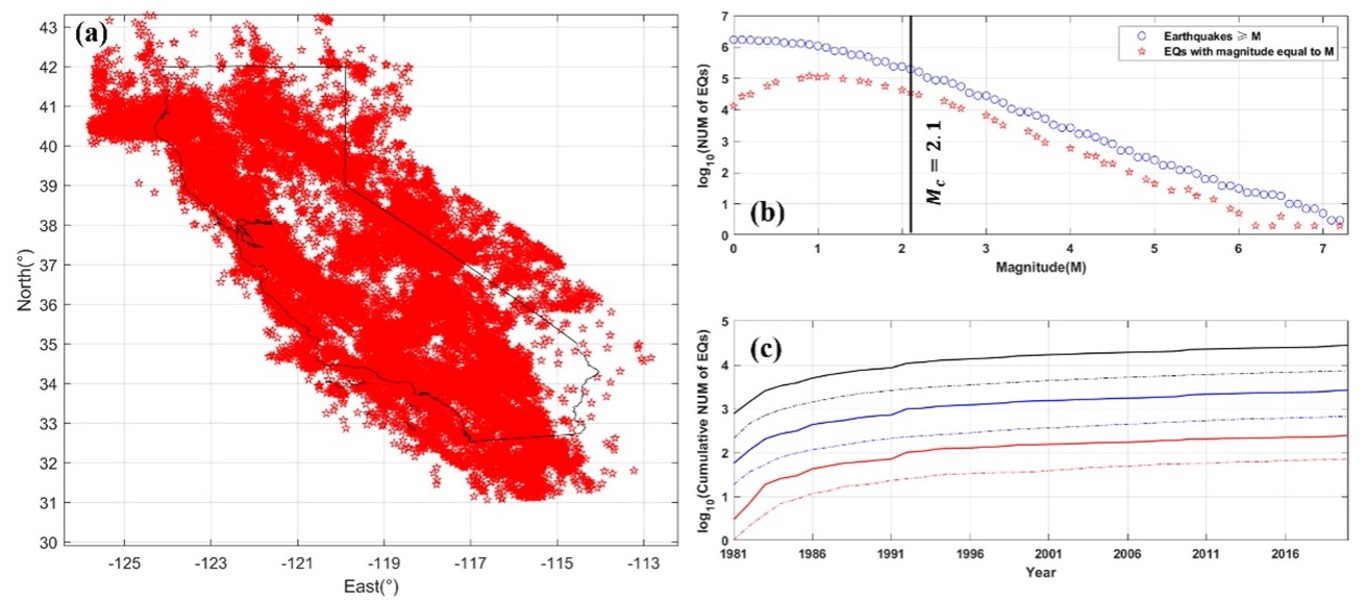}
\caption{(a) Spatial distribution of earthquakes with magnitude larger than 0 that occurred within the time period from 1 January 1981 to 31 December 2020 in the Regional Earthquake Likelihood Models (RELM) polygon. 
(b) Frequency-magnitude distribution of earthquakes. The black line indicates the magnitude $M_c$ of completeness over the whole time interval estimated using the method by Clauset et al.\cite{R43}(c) Black, blue and red lines are the logarithms of the cumulative numbers for events with magnitudes larger than 2, 3 and 4 respective. Solid lines are the numbers of events in the full sequence, while the dotted lines are the numbers of independent events. The sum of the independent probabilities of all events gives the number of independent events, where the independent probability refers to the probability that the event is independent (also called a background event), which is provided by the ETAS model\cite{nandan2021seismicity}. 
}
\label{es}
\end{figure}

In this framework, we define the spatial extent using the Regional Earthquake Likelihood Models (RELM) polygon for Southern California, as proposed by Schorlemmer and Gerstenberger \cite{schorlemmer2007relm}. The earthquake catalog spanning from January 1, 1981, to December 31, 2020, is sourced from the Advanced National Seismic System (ANSS), with the spatial distribution illustrated in Figure \ref{es}(a).
Southern California was selected for two main reasons: (i) it has been the focus of numerous earthquake prediction studies, providing a robust foundation for comparison, and (ii) the ANSS earthquake catalog for this region is of high quality and globally accessible. As illustrated in Figures \ref{es}(b) and (c), the completeness magnitude ($M_c$) for this region has varied over time but remains consistently below 3.0 throughout the study period \cite{nandan2017objective}.

We present two benchmark models for comparison: the ETAS model, a statistical geophysical approach, and our AI model, developed using a Fully Convolutional Network. Both models are employed to compute the probability of earthquake occurrence in each grid cell (0.1° $\times$ 0.1°) for various time and magnitude windows in California. In this context, $\mathbf{r}, t$ denote the location and time of the output associated with the spatiotemporal unit, which is represented by $P_r(\mathbf{r}, t, M, T)$ and denotes the probability that at least one earthquake with a magnitude $\geq M$ will occur in the grid cell at $\mathbf{r}$ within the time window $[t+1, t+T]$. In this study, the threshold magnitude $M$ for the target earthquakes is set to explore the values $\{3.0, 4.0, 5.0\}$ while the forecast time horizons $T$ are examined over $\{15, 30, 60, 90\}$ days. Therefore, we consider 12 different time-magnitude windows for forecasting earthquakes.

We consider a version of the ETAS model that account for the spatial inhomogeneity of background rates, as for instance developed by Nandan et al.\cite{nandan2017objective,nandan2021seismicity} The earthquake catalog with $M \geq 3.0$ from January 1, 1980, to December 31, 1989, is utilized to calibrate the initial model. The parameters are subsequently updated every 15 days, which also serves as our minimum horizon time for forecasting earthquakes. Through a total of 60,000 simulations, we generate ETAS earthquake forecasts for California from 1990 to 2020. More detail about the ETAS model used in this benchmark can be found in references\cite{nandan2017objective,nandan2021seismicity}.

The AI-based model utilizes the Fully Convolutional Networks (FCN) proposed by Long, Shelhamer and Darrell \cite{long2015fully} and adapted to earthquake forecasting by Zhang et al. \cite{zhang/10.1093/gji/ggae373}. Its input consists of the logarithm of past estimated released earthquake energies, and the loss function is the Balanced Mean Squared Error. Earthquake catalogs with $M \geq 3.0$ from January 1, 1990, to December 31, 1999, are used as the training dataset; from January 1, 2000, to December 31, 2009, as the validation dataset; and from January 1, 2010, to December 31, 2020, as the testing dataset.

Consequently, we finalize our selection of the testing dataset from the FCN model, specifically the data occurring from January 1, 2010, to December 31, 2020, for pseudo-prospective testing. We evaluate the model's performance using the Area Skill Score from the Molchan diagram, with a detailed explanation of this evaluation metric provided in the subsequent sections. The FCN model with inputs of earthquakes of magnitude 0 and above is labeled as FCN (M $\geq$ 0), while the FCN model with inputs of earthquakes of magnitude 3 and above is labeled as FCN (M $\geq$ 3.0). The performance metrics in the pseudo-prospective experiments, including speed and scores, are presented in Table \ref{tab:table_image}. Additionally, the results of the forecasts of both the ETAS model and our FCN model, along with the code of the FCN model, will be made publicly available at \footnote{https://github.com/AI-earthquake/Data-about-Seismically-informed-reference-models-enhance-AI-based-earthquake-prediction-systems}. The 11-year pseudo-prospective experiment shows that our proposed FCN model performs similarly to the well-established ETAS model, but is significantly faster. Our FCN model exemplifies the first type of valuable model we proposed earlier, offering performance comparable to the ETAS model while demonstrating a significant advantage in speed. Both the ETAS model and our FCN model serve as robust benchmarks for future research efforts.

Stockman et al.\cite{stockman2024earthquakenppbenchmarkdatasetsearthquake} also provide a platform and dataset to facilitate the comparison of AI-based earthquake forecasting models with their benchmark ETAS model. The datasets cover a range of target regions in California, from small to large, spanning from 1971 to 2021, and utilize various methodologies for dataset generation. In a benchmarking experiment, they compare three spatio-temporal Neural Point Processes (NPPs) with the ETAS model and find that none outperform ETAS in terms of spatial or temporal log-likelihood. These results suggest that current NPP implementations are not yet suitable for practical earthquake forecasting. However, the EarthquakeNPP platform will serve as a collaborative space for the seismology and machine learning communities, aimed at improving earthquake predictability. 

\begin{table}[H]
    \centering
    
    \caption{The numbers shown in this table are Area Skill Scores obtained for the ETAS with spatial inhomogeneity of background rates and for the fully convolutional network (FCN) model described in the text to forecast the number of events occurring within different time-magnitude windows. Time windows correspond to the four columns with headings 15, 30, 60, and 90 (days). Magnitude windows correspond to the three rows with headings 3, 4, and 5. Area Skill Scores are defined as the area between the Molchan curve and the diagonal line corresponding to random guessing. FCN (M$\geq$0) is the FCN model with earthquakes with M$\geq$0 as input, while FCN (M$\geq$3) is the FCN model with earthquakes with M$\geq$3 as input. The comparison of the running speeds of the two models is given in the last row of the table.
}
\includegraphics[width=0.7\textwidth]{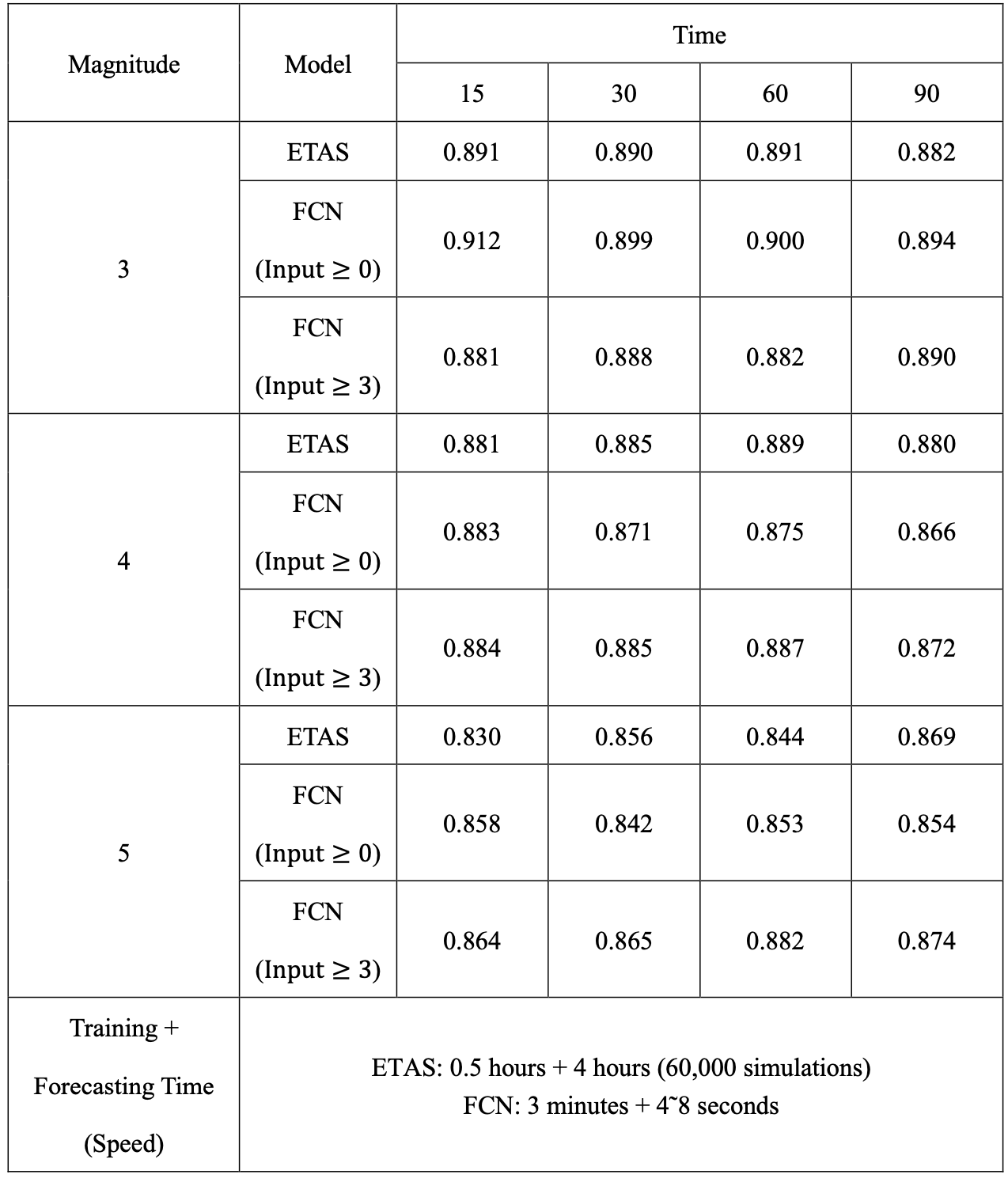} 
    \label{tab:table_image}
\end{table}

\subsection{Output of AI-based earthquake prediction/forecasting models}\label{output}

Since the inputs, loss function, and evaluation metrics in AI models are intrinsically tied to the model's output or prediction objectives, we begin by introducing the types of outputs generated by the model, followed by an in-depth exploration of the other components.

The key attributes of earthquakes—epicenter location, time of occurrence, and magnitude—are central to understanding and predicting seismic events. However, the underlying hypothesis about earthquake generation processes significantly influences how these attributes are forecasted. The debate between deterministic dynamics and stochastic processes plays a crucial role in shaping the types of outputs produced by earthquake prediction models.

On one hand, deterministic approaches suggest that earthquakes result from deterministic elastodynamic equations with strong nonlinearity. These models suggest that, with sufficient data and computational power, it might be possible to directly determine the epicenter, magnitude, and hypocenter of future earthquakes. Such outputs are highly specific and hold substantial value for hazard mitigation, but they are also beyond reach due to the inherent complexity of tectonic systems and the current limitations in observational data.

On the other hand, stochastic models view earthquakes as emergent phenomena of fundamentally random processes, influenced by factors such as stress accumulation and release, fault heterogeneity, and local geological conditions. In this context, forecasting becomes probabilistic, focusing on the likelihood of events within a predefined time-space-magnitude volume. This approach avoids the need for precise predictions of individual earthquake parameters and instead emphasizes statistical patterns and probabilities. For example, forecasting the probability of magnitude 5 or greater earthquakes occurring in California within the next 10 days relies on predefined parameters for time, space, and magnitude, simplifying the task to a probabilistic evaluation of such events.

From a technical standpoint, the probabilistic approach is more practical and easier to implement, as it reduces the dimensionality of the output space and aligns with the intrinsic uncertainties of seismic processes. While direct prediction of epicenters, magnitudes, and hypocenters is rare, probabilistic forecasts provide valuable insights into seismic hazard, enabling policymakers and communities to better prepare for potential risks.

Integrating technical and physical perspectives highlights the significance of different output types. Deterministic models aim for high-resolution, precise outputs, representing the ideal but currently unattainable goal of earthquake science. Meanwhile, probabilistic models offer a practical and scientifically grounded alternative, leveraging our understanding of stochastic processes to produce outputs that are both actionable and feasible. This dual approach underscores the importance of aligning forecasting methods with the underlying hypotheses of earthquake generation, ensuring that models are both scientifically valid and practically useful.

The definitions and commonly accepted descriptions of the four fundamental elements are as follows:

\textbf{Time span:} Typically describes a future time period, ranging from several hours to days, months, or even one year to several years. 

\textbf{Space area:} Describes the spatial extent of the target object. It can take various shapes and sizes, such as an irregular region like the entire region of Greece or California. It can also be artificially divided into grids, ranging in size from 0.1 degrees by 0.1 degrees to 10 degrees by 10 degrees.

\textbf{Magnitude range:} This refers to the magnitude information of potential events, typically expressed as exceeding a specified magnitude or falling within a defined magnitude range.

\textbf{Quantitative description of earthquake occurrence likelihood:} This can be a binary prediction (1/0) or a probability forecast. Additionally, some outputs may not represent the actual probability of earthquake occurrence but rather the relative likelihood, such as hazard level maps. 

Not all models predict the four elements mentioned above. For example, some models focus on forecasting the spatial distribution of aftershocks without providing magnitude information, while others predict the b-value of the Gutenberg-Richter distribution over a specified future time period. Through a detailed classification and analysis of the collected articles, we identified 20 distinct prediction methods or output types, which we grouped into nine major categories. The first eight categories pertain to predicting one or more of the four key elements of earthquakes, with some models focusing on a single element and others combining multiple predictions. The ninth category encompasses other types of predictions that do not address the four primary elements. Notably, some studies tackle multiple prediction tasks simultaneously, further diversifying their scope.
\subsubsection{Time}
\noindent\textbf{Output 1 (O1):} The lagged time between the given time point and the next earthquake meeting predefined conditions (within a spatial range and magnitude range).
\subsubsection{Magnitude}
\noindent\textbf{Output 2 (O2):} The maximum magnitude of earthquakes occurring within a predefined time-space window.

\noindent\textbf{Output 3 (O3):} The magnitudes of the next N earthquakes within a given area. 

\noindent\textbf{Output 4 (O4):} The magnitude of a given event. In this task, the epicenter location and occurrence time of earthquakes are assumed to be known. 

\noindent\textbf{Output 5 (O5):} The average magnitude of earthquakes occurring within a predefined time-space window. In this work\cite{R12}, the model also provides the number of events that will occur in the given time-space window. 

\noindent\textbf{Output 6 (O6):} The magnitudes of aftershocks greater than or equal to a specified magnitude, occurring between the \textit{i}\textsuperscript{th} and \textit{j}\textsuperscript{th} aftershock after a specific earthquake, with the predicted magnitudes given as specific values.
\subsubsection{Probability}
\noindent\textbf{Output 7 (O7):} Earthquake prediction is reduced to a binary classification, ie., the occurrence(1)/non-occurrence(0) of earthquakes in the given time-space-magnitude volume.

\noindent\textbf{Output 8 (O8): }Probability that earthquakes will occur in a predefined time-space-magnitude window.

\noindent\textbf{Output 9 (O9):} Probability that aftershocks will occur in a predefined time-space window. In these studies, the prediction of aftershocks is treated as a binary classification task, and these models do not provide information about the magnitude of future earthquakes. 

\subsubsection{Magnitude and Time}
\noindent\textbf{Output 10 (O10):} The model simultaneously predicts the magnitude and the occurrence time of future earthquakes within a given spatial range.
\subsubsection{Magnitude and Location}
\noindent\textbf{Output 11 (O11):} The model predicts the spatial locations and magnitudes of earthquakes within a certain time range in the future. Moreover, in these works\cite{R25,R26,R27}, they also predict the depth of future events. When predicting the location of future events, it is worth noting that not all relevant studies directly provide the latitude and longitude of these events. Some research divides the entire study area into multiple sub-regions, and the model predict in which specific region earthquakes will occur. 
\subsubsection{Magnitude and Probability}
\noindent\textbf{Output 12 (O12):} The probability and magnitude of future earthquakes that will occur in the pre-defined time-space window. Moreover, in this work\cite{R28}, the prediction of magnitude is also taken as a multi-classification task. 
\subsubsection{Time and Location}
\noindent\textbf{Output 13 (O13):} The model does not aim to predict the exact location and timing of the next earthquake. Instead, it focuses on forecasting relationships between events, such as estimating the time and distance at which a subsequent earthquake is likely to occur relative to a previous one.
\subsubsection{Magnitude, Time and Location}
\noindent\textbf{Output 14 (O14):} The model provides information on the magnitude, timing, and location of future earthquakes; however, the prediction format is not always explicitly presented as $(\mathbf{r}, t, M)$. Alves \cite{R29} predicts the time interval until the next earthquake from the time of the model's prediction, along with the earthquake's latitude, longitude, and magnitude. Similarly, Belouadha et al. \cite{R30} forecast the latitude, longitude, magnitude, and year of future earthquakes. Berhich et al. \cite{R31} focus on predicting both the timing and magnitude of upcoming earthquakes. However, their location predictions are limited to identifying the sub-region where an earthquake is most likely to occur. 
\subsubsection{Others}
\noindent\textbf{Output 15 (O15):} The model predicts the frequency of earthquakes occurring within a specific time-space window. 

\noindent\textbf{Output 16 (O16):} A seismogram is the graphical output of a seismograph, an instrument that detects and records vibrations in the ground caused by earthquakes or other seismic activities. Bahrami and Shafiee \cite{R32} use the first 560 seconds of a seismogram as input to predict the next 70 seconds of seismic activity. 

\noindent\textbf{Output 17 (O17):} The model predicts the b-value of the Gutenberg-Richter law, a key parameter in seismology that describes the frequency distribution of earthquakes \cite{R33}. The b-value represents the exponent in the relationship between earthquake magnitude and the number of seismic events, providing insights into the characteristics of seismic activity.

\noindent\textbf{Output 18 (O18):} The model predicts the acceleration, which in seismology refers to the rate of change of velocity of the ground caused by the propagation of seismic waves during seismic events.

\noindent\textbf{Output 19 (O19):} The model is used to recognize the preparatory phase preceding earthquakes, where small and stable ruptures progressively develop into an unstable and confined zone around the future hypocenter. In other words, the model is used to determine whether the current micro-seismicity consists of foreshocks that could trigger larger earthquakes, and it constructs earthquake prediction models based on foreshock activity. Iaccarino and Picozzi\cite{R34} take this work as a binary classification task, assigning '0' to stationary seismicity and aftershock sequences, while '1' refers to the preparatory phase.

\noindent\textbf{Output 20 (O20):} The model predicts the Mercalli intensity, a qualitative scale that measures the intensity of ground shaking during an earthquake as experienced at a specific location. Unlike magnitude, which quantifies the energy released at the source, Mercalli intensity focuses on the observed effects, such as structural damage, ground motion, and human perception of shaking. Wijaya et al.\cite{R35} predicted the Mercalli intensity map for Indonesia.

\subsubsection{Summary}
\begin{figure}[ht]
\centering
\includegraphics[width=\linewidth]{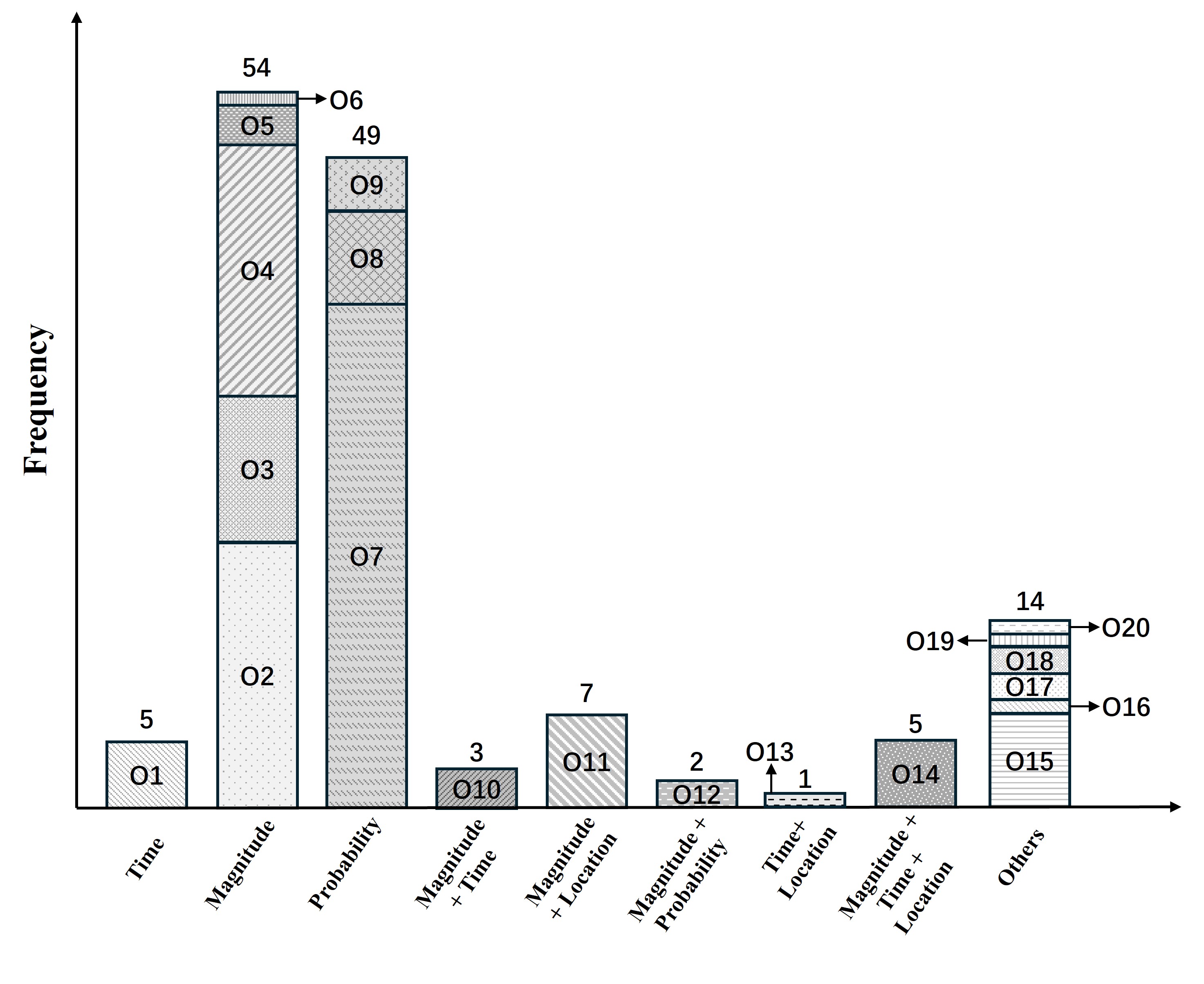}
\caption{Cumulative frequency of different types of outputs.
}
\label{figCul}
\end{figure}

\paragraph{Critique}
We have summarized nine major categories and twenty subcategories of AI-based earthquake prediction/forecasting model outputs. As shown in Figure \ref{figCul}, over two-thirds of these models focus solely on predicting earthquake magnitude or probability. However, not all output types provide complete information on the four key elements of an earthquake: magnitude, epicenter, occurrence time, and depth.

For O3 and O6-type models, predictions are limited to the magnitude of a certain number of upcoming earthquakes, without considering their time of occurrence. For O9-type models, while they estimate the probability of aftershocks occurring at different times and locations, they fail to provide information on the magnitude of these aftershocks. These models should establish a minimum magnitude threshold, such as predicting aftershocks with a magnitude of 3.0 or higher within a specific time-space window. Setting the threshold too low, for example at -3.0, renders the predictions meaningless, as very small earthquakes occur continuously across various locations. Thus, all studies must clearly specify the minimum magnitude of earthquakes being predicted.
For O15 to O20-type models, the outputs do not address core earthquake parameters such as magnitude, epicenter location, or occurrence time.

In the case of O4-type models, the epicenter and occurrence time of the target earthquake are assumed to be known and are sometimes used as inputs for predicting magnitude. However, in real-world earthquake forecasting, such information about future events is unattainable. This assumption violates the principles of pseudo-prospective testing, which aim to simulate realistic earthquake prediction conditions. Therefore, these models are better described as tools for analyzing the relationship between input data near the epicenter and earthquake magnitudes, rather than as genuine earthquake prediction or forecasting methods.

Finally, we emphasize the importance of rigorous pseudo-prospective or prospective testing to ensure that models are evaluated under conditions that accurately reflect real-world scenarios.

\paragraph{Suggestion}

Many forecasting models predefine the size of time-space volumes to manage the complexity of their outputs. The outputs of O7 and O8 types are among the predictive methods we highly value, serving as benchmark outputs presented in Section \ref{framework}. These models allow users to assess the likelihood of specific magnitudes occurring over varying time and spatial intervals, incorporating all essential elements for meaningful predictions.

The finer and smaller the time-space volume granularity in earthquake prediction models, the higher their application value for forecasting larger magnitude earthquakes. However, this also increases the technical complexity of the models. Regarding earthquake magnitude, most studies focus on events with a magnitude of 4.0 or higher.

On the time scale, research predominantly focuses on short- to medium-term predictions, ranging from 7 to 90 days, with some extending to annual forecasts. On the spatial scale, most studies concentrate on predicting earthquakes for entire regions, such as Greece, California, or Japan. A smaller number of studies explore fine-scale spatial predictions.

In Table \ref{tab:table_detail}, we highlight studies with a spatial resolution smaller than 1\textsuperscript{\textdegree} \ensuremath{\times} 1\textsuperscript{\textdegree}, along with their corresponding time windows and descriptions of the earthquakes they aim to predict. These finer-resolution studies demonstrate significant potential for advancing earthquake forecasting but also underscore the associated technical challenges.

\begin{table}[H] 
    \centering
    \caption{Detailed information for studies with spatial Resolution smaller than 1\textsuperscript{\textdegree} \ensuremath{\times} 1\textsuperscript{\textdegree}}
    \label{tab:table_detail}
    \small
    \begin{tabularx}{\textwidth}{|X|X|X|X|X|}
    \hline  
    Reference & Spatial Resolution & Time Window & Description of the targeted earthquakes & Region \\ 
    \hline  
    \cite{R14} & 0.1\textsuperscript{\textdegree} $\times$ 0.1\textsuperscript{\textdegree} & Next 1, 30, 90, 180, 365 days & Aftershocks & Sichuan Province (China) and Japan\\ 
    \hline  
    \cite{R1} & 5 km $\times$ 5 km & Next 365 days & Aftershocks & Taiwan Province (China), Japan, and Kashmir region (South Asia)\\ 
    \hline  
    \cite{R15} & 5 km $\times$ 5 km & / & Aftershocks (M>2.5)& Iran\\
    \hline  
    \cite{R16} & 0.1\textsuperscript{\textdegree} $\times$ 0.1\textsuperscript{\textdegree} & Next 2, 4, 8, 14, 26, 52, 104, 208 days & Events with magnitude $\geq M_{c}$ & United States\\ 
    \hline  
    \cite{R17} & 10 km $\times$ 10 km & The next 10 to 50 days & Events with magnitude $\geq$ 3.5 or 5.0 & Japan\\ 
    \hline  
    \cite{R18} & 72.92 km (longitude) $\times$ 67.71 km (latitude) & Next 3, 9, 15 days & Events with magnitude $\geq M_{c}$ & Worldwide\\ 
    \hline  
    \cite{R8} & 0.5\textsuperscript{\textdegree} $\times$ 0.5\textsuperscript{\textdegree} & Next 24 hours & Events with magnitude $\geq$ 4.0 or 5.0 & Japan\\ 
    \hline  
    \cite{R19} & 1\textsuperscript{\textdegree} $\times$ 1\textsuperscript{\textdegree} & Next 30, 60, 90 days & Events with magnitude $\geq$ 5.0 & China\\ 
    \hline  
        \cite{zhang/10.1093/gji/ggae373} & 0.1\textsuperscript{\textdegree} $\times$ 0.1\textsuperscript{\textdegree} & Next 15, 30, 60, 90 days & Events with magnitude $\geq$ 3.0, 4.0, 5.0 & California, United States\\ 
    \hline  
        \cite{ZhanSTC} & 0.1\textsuperscript{\textdegree} $\times$ 0.1\textsuperscript{\textdegree} & Next 1 day & Events with magnitude $\geq$ 3.0, 4.0, 5.0 & California, United States\\ 
    \hline  
    \end{tabularx}
\end{table}

In outputs (2) to (6) focused on predicting earthquake magnitude, most studies treat the task as a regression problem. However, a few studies \cite{R20,R21,R22,R23} approach it as a multi-class classification problem, where earthquake magnitudes are divided into discrete ranges, each assigned a unique label. This innovative approach leverages AI techniques originally developed for classification tasks, enabling them to provide valuable insights into the magnitude ranges of future earthquakes.

The accuracy and reliability of earthquake forecasting models are deeply influenced by the quality and completeness of the earthquake catalogs they rely on. These catalogs often vary significantly in terms of data quality and magnitude completeness across different geographical regions. For instance, Rojo Limón et al. \cite{R24} compared earthquake catalogs from the Kandilli Observatory (KOERI) and the National Earthquake Department (AFAD) in Turkey. Although there is significant overlap in the spatial and temporal coverage of  these agencies that operate independent seismic networks, notable differences were observed, particularly in the recording of large-magnitude events (M > 4). This underscores the variability and inconsistencies that can exist even in well-maintained earthquake catalogs.
Such discrepancies highlight a critical issue often overlooked by non-seismologists and data-mining experts applying AI techniques to earthquake prediction: the inherent variability and biases within earthquake catalogs. Assumptions about data quality that do not account for these limitations can lead to misleading results. A thorough understanding of the catalog's strengths and weaknesses is essential to ensure reliable and accurate predictions.

\subsection{Input of AI-based Earthquake Prediction/forecasting Model}
As noted earlier, AI-based earthquake prediction models utilize both seismic and non-seismic inputs. In this section, we will outline the inputs used in previous studies and provide a detailed explanation of the calculation methods for seismic inputs. It is important to highlight that some studies mention certain inputs without providing clear definitions or calculation methods. Such data will not be included or discussed in this section.

\subsubsection{Seismic types}
The seismic inputs include statistical seismology parameters calculated based on seismic catalogs and data directly related to earthquakes, such as seismographs. To introduce statistical seismological parameters, we first provide the following definitions related to earthquakes:
(\textit{$lat_i,long_i,t_i,m_i,d_i$}) - The latitude, longitude, occurrence time, magnitude, and depth of the \textit{$i^{th}$} event. The \textit{$i^{th}$} event is represented as \textit{$eq_i$}, and {\textit{$eq_1, eq_2, ..., eq_i, eq_{i+1},...,eq_n$}} represents the sequence or set of earthquakes from the first to the \textit{$n^{th}$}, arranged chronologically.

In some cases, model builders do not use all the earthquakes listed in the earthquake catalog as inputs for the model. Instead, they select a subset of \textit{N} specific events—{\textit{$EQ_1, EQ_2, ..., EQ_i, EQ_{i+1},...,EQ_N$}}—from the full set {\textit{$eq_1, eq_2, ..., eq_i, eq_{i+1},...,eq_n$}} according to specific criteria. For instance, they might choose earthquakes with magnitudes greater than or equal to the catalog's completeness magnitude or limit the selection to events occurring within a defined time period. When this process of manual selection is relevant to the input, we will provide a detailed explanation of the rules used for selecting earthquakes.

$M_{cutoff}$ - Minimum magnitude cut-off. There are several cut-off thresholds, such as the minimum recorded magnitude in the catalogs, the completeness magnitude ($M_c$) above which the catalog is considered complete, and the magnitude threshold above which earthquakes are included in the analysis and as input, and so on.

$t_{start}, t_{end}$ - Start and end time for the time window.
\[
I(\Omega)= \left\{
\begin{array}{ll}
1, & \text{if } \Omega \text{ is true} \\
0, & \text{otherwise}
\end{array} \right.
\parbox{8cm}{\centering This logical function takes the value 1 if the condition $\Omega$ is true, and 0 otherwise.}
\]

$t_{model}$ - The beginning time of the time prediction window.

$\mathfrak{V}_i$ - The $i^{th}$ time-space volume.

$\mathfrak{S}_i$ - The spatial range of $\mathfrak{V}_i$, and it also represents the spatial area of $\mathfrak{V}_i$ (in $m^2, km^2, $0.1\textsuperscript{\textdegree} $\times$ 0.1\textsuperscript{\textdegree}, ...).

$\mathfrak{T}_i$ - The time range of $\mathfrak{V}_i$, and it also represents the time span of $\mathfrak{V}_i$ (in $Seconds, Minutes, Days, ...$).

\noindent\textbf{Input 1 (I1):}  Time of the i-th earthquake - $t_i$.

\noindent\textbf{Input 2 (I2):}  Location of the i-th earthquake - $lat_i, long_i$. 

\noindent\textbf{Input 3 (I3):}  Magnitude of the i-th earthquake - $m_i$.

\noindent\textbf{Input 4 (I4):}  Depth of the i-th earthquake - $d_i$.

\noindent\textbf{Input 5 (I5):}  The elapsed time $\Delta t$ from  to a previous given time point ($t_{given}$), such as the occurrence time of the last earthquake. 
\begin{equation}\label{eq1}
    \Delta t=t_{model}-t_{given}
\end{equation}
\noindent\textbf{Input 6 (I6):}  Epicenter density - $EpDensity_i$. Divide the whole time-space volume into uniform sized time-space volume, denoted \[
{\mathfrak{V}_1,\mathfrak{V}_2,...,\mathfrak{V}_i,\mathfrak{V}_{i+1},...,\mathfrak{V}_{n-1},\mathfrak{V}_n}
\], and {$\mathfrak{S}_1=\mathfrak{S}_2=...=\mathfrak{S}_{i-1}=\mathfrak{T}_i,\mathfrak{T}_1=\mathfrak{T}_2=...=\mathfrak{T}_{i-1}=\mathfrak{T}_i$}. The Epicenter density of the i-th sub time-space element is the total number of events occurring this sub time-space unit divided by it volume.
\begin{equation}\label{eq2}
EpDensity_i=\frac{1}{\mathfrak{V}_i}\sum_{j=1}^{n}{I(eq_j\in\mathfrak{V}_i)} 
\end{equation}
\noindent\textbf{Input 7 (I7):}  Magnitude density - $MagDensity(m, \Delta m)$. Total number of $eq_i$ with $m\leq m_i <m+\Delta m$.
\begin{equation}\label{eq3}
MagDensity(m,\Delta m)=\frac{1}{\mathfrak{V}_i}\sum_{j=1}^{n}{I(m\leq m_i <m+\Delta m)}  
\end{equation}
\noindent\textbf{Input 8 (I8):}  Average magnitude of earthquakes in a moving time window - $\overline{M}_{t_{start},t_{end}}$.
\begin{equation}
    \overline{M}_{t_{start},t_{end}}=\frac{\sum_{i=1}^{n} {m_i*I(t_{start}\leq t_i\leq t_{end}})}{\sum_{i=1}^{n} {I(t_{start}\leq t_i\leq t_{end}})}
\end{equation}
\noindent\textbf{Input 9 (I9):}   Maximum magnitude of earthquakes in a moving time window - $M^{max}(t_{start},t_{end})$.

\noindent\textbf{Input  10 (I10):} Minimum magnitude of earthquakes in a moving time window - $M^{min}(t_{start},t_{end})$.

\noindent\textbf{Input 11 (I11):} Mean magnitude of the last $\mathfrak{N}$ events - $\overline{M}_{\mathfrak{N}}^{j}$.
\begin{equation}
    \overline{M}_{\mathfrak{N}}^{j}=\frac{1}{\mathfrak{N}}
    \sum_{i=j-\mathfrak{N}+1}^{j} {m_i}
\end{equation}
\noindent\textbf{Input 12 (I12):}  Number of earthquakes with a magnitude larger than a given threshold in a time window - $n(M_{cutoff,},t_{start},t_{end})$.
\begin{equation}
    n(M_{cutoff,},t_{start},t_{end})=\sum_{i=1}^{n} {I(m_i\geq M_{cutoff})*I(t_{start}\leq t_i \leq t_{end})}
\end{equation}
\noindent\textbf{Input 13 (I13):}  Mean seismicity rates in time intervals - $\overline{n}(M_{cutoff}, t_{start}, t_{end})$.
\begin{equation} 
    \scriptsize
    \overline{n}(M_{cutoff}, t_{start}, t_{end})=\frac{\sum_{i=1}^{n} {I(m_i\geq M_{cutoff})*I(t_{start}\leq t_i \leq t_{end})}}{t_{end}-t_{start}}=\frac{n(M_{cutoff,},t_{start},t_{end})}{t_{end}-t_{start}}
\end{equation}

\noindent\textbf{Input 14 (I14):} Duration of Seismic Belt - $\Delta T_{belt}$, defined as the time interval between the first and last earthquakes within the belt. Liu et al. (reference) defined the seismic belt as follows: 
\begin{enumerate}
  \renewcommand{\labelenumi}{(\arabic{enumi})}
  \item The seismic belt consists of at least 6 earthquakes with $m_i\geq M_{cutoff}$.
  \item When $M_{cutoff}({\rm for} ~M_L)<3$, the length of belt should exceed 200 $km$.
  \item The length-to-width ratio of the belt should be greater than or equal to 5.
  \item The empty segment length of the seismic belt should not exceed one-third of the total length.
  \item The number of earthquakes located within the seismic belt accounts for over 75\% of the total number of earthquakes both inside and outside the belt in that region.
\end{enumerate}
\noindent\textbf{Input 15 (I15):} Duration of the Seismic Gap - $\Delta T_{gap}$. Seismic gap refers to a segment along a fault line or within a tectonic plate boundary that has not experienced significant earthquakes for a considerable period, despite the surrounding areas having a history of seismic activity. Essentially, it is an area that has not ruptured in an earthquake when neighboring segments have. The duration $\Delta T_{gap}$ is supposed to be correlated with the magnitude of the future main earthquake. 

\noindent\textbf{Input 16 (I16):}  Velocity of seismic shear waves - $V_s$.

\noindent\textbf{Input 17 (I17):}  Velocity Ratio - ${V_p}/{V_s}$.  ${V_p}$ represents the velocity of seismic primary (or P) wave. P-waves are body waves, or compression waves.

\noindent\textbf{Input 18 (I18):}  Time elapsed over a predefined number of events with magnitude $\geq M_{cutoff}$ - $\Delta T_{elasped}^{j}(\mathfrak{N},M_{cutoff})$. In the earthquake sequence {\textit{$eq_1, eq_2, ..., eq_i, eq_{i+1},...,eq_n$}}, we select earthquakes with $m_i\geq M_{cutoff}$, forming a new sequence {\textit{$EQ_1, EQ_2, ..., EQ_i, EQ_{i+1},...,EQ_N$}}. Then, for the $EQ_j$, the time elasped over $\mathfrak{N}$ events is defined as: 
\begin{equation}
    \Delta T_{elasped}^{j}(\mathfrak{N},M_{cutoff})=T_j-T_{j-\mathfrak{N}+1}
\end{equation}
\noindent\textbf{Input 19 (I19):} Time elapsed between the $eq_i$ and $eq_j$ - $\Delta T_{i,j}$.
\begin{equation}
    \Delta T_{i,j}=t_j-t_i
\end{equation}
\noindent\textbf{Input 20 (I20):}  Mean time between selected events - $\mu$. This is the average time or gap between the last $N$ selected events, which are 
\[{{EQ_1, EQ_2, ..., EQ_i, EQ_{i+1},...,EQ_N}}\].  For example, earthquakes of magnitude 7 to 7.5 are grouped together. Then, $\mu$ is defined as,
\begin{equation}
    \mu = \frac{\sum_{i=1}^{N-1} {T_{i+1}-T_i}}{N-1}
\end{equation}
\noindent\textbf{Input 21 (I21):}  Distance between $eq_i$ and $eq_j$ - $r_{ij}$.

\noindent\textbf{Input 22 (I22):} Estimated energy released by the i-th earthquake - $e_i$. 

\noindent\textbf{Input 23 (I23):} Benioff strain - $dE_j^{\frac{1}{2}}(\mathfrak{N},M_{cutoff})$.  Cumulative sum of the square root of the seismic energy released over the time interval $\Delta T_{elasped}^{j}(\mathfrak{N},M_{cutoff})$ \cite{R36,R37,R38}. The Benioff strain has been used in various previous works on earthquake prediction.
\begin{equation}
    dE_j^{\frac{1}{2}}(\mathfrak{N},M_{cutoff})=\frac{\sum_{i-\mathfrak{N}+1}^{j}e_i^{\frac{1}{2}}}{\Delta T_{elasped}^{j}(\mathfrak{N},M_{cutoff})}
\end{equation}
\noindent\textbf{Input 24 (I24): } Slope of the curve formed by the logarithm of the earthquake frequency versus magnitude  - $b$. The frequency distribution of earthquake magnitudes is known as the Gutenberg-Richter Law\cite{R39}. The Gutenberg-Richter distribution of earthquakes magnitudes can be displayed in two equivalent forms: (i) as a frequency-magnitude distribution also called probability density function; (ii) as a complementary cumulative distribution function (CCDF) $N(m_i>m)$ defined as the number of $eq_i$ with $m_i\geq m$. Historically the latter has been preferred because it is very simple to construct from sorting magnitudes from the largest to the smallest: the CCDF is then just the normalized index of the sorting operation. Then, the GR law can be described as:
\begin{equation}\label{eqGR}
    logN(m_i>m)=-bm+a
\end{equation}
where $b$ is the coefficient of proportionality of this log-linear regression referred to as the  $b$ value. The constant a is equal to the total number of events in the sample with magnitudes larger than $m=0$. There are many methods to estimate the $b$ value\cite{R41,R42,R43}. The maximum likelihood method proposed by Aki\cite{R40} gives:
\begin{equation}
    \hat{b}=\frac{log(e)}{\overline{m}-(m_c-\frac{\Delta m}{2})}
\end{equation}
where $\hat{b}$ is the statistical  estimate of the $b$ value, $m_c$ is the minimum magnitude over which the earthquake catalog is deemed to be complete, $\overline{m}$  is the mean magnitude over all magnitudes larger than $m_c$, and $\Delta m$ is the finite magnitude binning width of the earthquake catalog. 

\noindent\textbf{Input 25 (I25):} Standard deviation of $b$ value - $\sigma _b$.
\begin{equation}
    \sigma_b=2.3\hat{b}^2\sqrt{\frac{\sum_{i=1}^n{(m_i-\overline{m})^2}}{n(n-1)}}
\end{equation}
\noindent\textbf{Input 26 (I26):} $b$ value variation - $\Delta b_{ki}(\mathfrak{N})$.
\begin{equation}
    \Delta b_{ki}(\mathfrak{N})=b_{i-(k-1)*\Delta n}(\mathfrak{N})- b_{i-k*\Delta n}(\mathfrak{N})
\end{equation}
where $b_i(\mathfrak{N})$ is the $b$-value estimated  for the set  {$eq_{i-\mathfrak{N}+1}, eq_{i-\mathfrak{N}+2},...,eq_{i-2},eq_{i-1},...,eq_i $}, and $\mathfrak{N}$ is usually set to be 50. $\Delta n$ can be varied in different implementations. For instance, it is equal to $4$ in the definitions of $b_{ki}(\mathfrak{N})$ given below. In the following parts, we abbreviate  $b_i(\mathfrak{N})$ as $b_i$, $b_{ki}(\mathfrak{N})$ as $b_{ki}$, and $k$ is usually set to be {$1,2,3,4,5$}. $\Delta b_{1i},\Delta b_{2i},\Delta b_{3i},\Delta b_{4i}$ and $\Delta b_{5i}$ are simultaneously used as inputs to the model, and they are defined as follows:
\begin{equation}
\begin{cases}
    \Delta b_{1i} = b_i - b_{i-4} \\
    \Delta b_{2i} = b_{i-4} - b_{i-8} \\
    \Delta b_{3i} = b_{i-8} - b_{i-12} \\
    \Delta b_{4i} = b_{i-12} - b_{i-16} \\
    \Delta b_{5i} = b_{i-16} - b_{i-20}
\end{cases}
\end{equation}
\noindent\textbf{Input 27 (I27):}  $a$ value of Gutenberg-Richter Law presented in equation.\ref{eqGR} - $a$.
\begin{equation}
    \begin{cases}
        \mathfrak{M}_j=m_c+j\Delta m\\
        a=\frac{\sum_{j=0}^{n-1}{(\log N(m_i>\mathfrak{M}_j)+\hat{b}\mathfrak{M}_j)}}{n}
    \end{cases}
\end{equation}
\noindent\textbf{Input 28 (I28):}  Summation of the mean square deviation about the regression line based on the Gutenberg-Richter Law - $\eta$.
\begin{equation}
\eta = \frac{1}{n-1} \sum_{j=0}^{n-1}(\log_{10}N(m_i>\mathfrak{M}_j)-(a-\hat{b}\mathfrak{M}_j))
\end{equation}
\noindent\textbf{Input 29 (I29):}  Magnitude surprise or the difference between the largest observed magnitude ($M_{max,observed}$) and the largest expected magnitude ($M_{max,expected}$) based on the G-R law - $\Delta{M}$.
\begin{equation}
    \begin{cases}
        \Delta M=M_{max,observed}-M_{max,expected}\\
        M_{max,expected}=\frac{a}{\hat{b}}
    \end{cases}
\end{equation}
Note that the largest observed magnitude is a random variable\cite{R44,R45,R46}. The typical value that is not exceeded with probability $1/e=37\%$ is given by the expression $ M_{max,expected}=\frac{a}{\hat{b}}$.

\noindent\textbf{Input 30 (I30):}  Coefficient of variation of the set {$T_N-T_{N-1},T_{N-1}-T_{N-2},...,T_2-T_1$} of waiting times between successive earthquakes - $c$. The coefficient of variation is calculated as the ratio of the standard deviation to the mean of a data set.
\begin{equation}
    c=\frac{\rm Standard~deviation~of~\{ T_N-T_{N-1},T_{N-1}-T_{N-2},...,T_2-T_1\}}{\mu}
\end{equation}
\noindent\textbf{Input 31 (I31):}  Probability of recording an earthquake with magnitude $\geq M$ - $Pr(M)$, and it is calculated from the following probability density function, 
\begin{equation}
    Pr(M)=10^{-b_i M}
\end{equation}
where $b_i$ is the estimated $b$ value for the set {$eq_{i-\mathfrak{N}+1}, eq_{i-\mathfrak{N}+2},...,eq_{i-2},eq_{i-1},...,eq_i $}.

\noindent\textbf{Input 32 (I32):}  The probabilistic recurrence time, $T_r(M)$, for a shock with magnitude equal to or greater than $M$, is given by
\begin{equation}
    T_r(M)=\frac{t_{end}-t_{start}}{10^{(a-\hat{b}M)}}
\end{equation}
where the $a$ and $\hat{b}$ value are derived from the earthquake catalog with time span ranging from $t_{start}$ to $t_{end}$.

\noindent\textbf{Input 33 (I33):}  Earthquake density - $EqDensity(x,y,R)$. This is the logarithm with base 10 of the number of earthquakes per unit area that occur within a circular region of radius $R$, and the center of the circle region is $(x,y)$,
\begin{equation}
    EqDensity=\log_{10}\left(\frac{N_{tot}}{\pi R^2}\right)
\end{equation}
where $N_{tot}$ is the total number of events ocurring in this circular region.

\noindent\textbf{Input 34 (I34):}  Standardized number of earthquakes \cite{matthews1988statistical}- $\beta_N(t,\delta)$. The raw earthquake catalog - {\textit{$eq_1, eq_2, ..., eq_i, eq_{i+1},...,eq_n$}} is de-clustered, and a new catalog is obtained that contains only background events - {\textit{$EQ_1, EQ_2, ..., EQ_i, EQ_{i+1},...,EQ_N$}}. There are many de-clustering methods, which will not be further discussed in this article. $N$ is the total number of events for the declustered catalog. The time span of the declustered catalog is normalized from $[T_1,T_N]$ to $[0,1]$, and $t\in [0,1]$ represents a certain moment in the normalized time span.  $M(t,\delta)$ is the number of earthquakes in the sub interval $[t-\delta,t]$. The standardized number of earthquakes is defined as, 
\begin{equation}
    \beta_N(t,\delta)=\frac{M(t,\delta)-n\delta}{\sqrt{n\delta(1-\delta)}}
\end{equation}
\noindent\textbf{Input 35 (I35):}  Standardize seismicity rate difference\cite{spassov2002spatial} - $Z$ value.
\begin{equation}
    Z=\frac{R_1-R_2}{\sqrt{\frac{S_1}{n_1}+\frac{S_2}{n_2}}}
\end{equation}
where $n_1$ and $n_2$ are the number of events in period 1 and period 2. $R_1$ and $R_2$ are the mean rates in period 1 and 2, with variance of $S_1$ and $S_2$.

\noindent\textbf{Input 36 (I36):}  The Omori-Utsu law \cite{utsu1995centenary} describes the rate of aftershocks decay as $1/(c+\Delta T_{Mainshock})^p$, where $\Delta T_{Mainshock}$ is the time counted from the occurrence time of the mainshock - $N_{aftershock}(\Delta T_{Mainshock})$,
\begin{equation}
    N_{aftershock}(\Delta T_{Maibshock})=\frac{K}{(c+\Delta T_{Mainshock})^p}
\end{equation}
where $K$ and $c$ are constants, and $p$ is the Omori-Utsu exponent using found in the range $0.8-1.2$ for large earthquakes.

\noindent\textbf{Input 37 (I37):}  Cumulative estimated energy released by earthquakes occurring in a given time-space volume $(V_i)$ - $CulE_i$.
\begin{equation}
    CulE_i=\sum_{j=1}^n {e_j*I(eq_j\in V_i)}
\end{equation}
\noindent\textbf{Input 38 (I38):}  Logarithm of the seismic moment of the i-th earthquake - $\log_{10}(\mathfrak{M}_{0i})$. It is estimated by the empirical relation \cite{R47,kanamori1977energy}
\begin{equation}
    \log_{10}(\mathfrak{M}_{0i})=1.5~m_i({\rm for}~ M_w)+16.1
\end{equation}
\noindent\textbf{Input 39 (I39):}  Average rate of earthquakes occurring in the time interval between two selected earthquakes $EQ_{k-1}$ and $EQ_k$ - $\overline{n}(EQ_{k-1},EQ_k)$.
\begin{equation}
    \overline{n}(EQ_{k-1},EQ_k)=\frac{\sum_{j=1}^n{I(T_{k-1}\leq t_j \leq T_k)}}{T_k-T_{k-1}}
\end{equation}
\noindent\textbf{Input 40 (I40):}  Average plate slip rate - $S_{plate}$.

\noindent\textbf{Input 41 (I41):}  Six independent components of the static elastic stress-change tensor, generated co-seismically and calculated at the centroid of a grid cell, along with their corresponding negative values - 
\[\sigma_{XX},\sigma_{XY},\sigma_{XZ},\sigma_{YY},\sigma_{YZ},\sigma_{XX},-\sigma_{XX},-\sigma_{XY},-\sigma_{XZ},\\-\sigma_{YY},-\sigma_{YZ},-\sigma_{XX}\]

\noindent\textbf{Input 42 (I42):}  Distance from mainshock rupture. 

\noindent\textbf{Input 43 (I43):}  Peak ground acceleration, which refers to the maximum acceleration experienced by the ground surface during an earthquake. 

\noindent\textbf{Input 44 (I44):}  Map of the slip distribution. In this work\cite{karimzadeh2019spatial}, the slip distribution map is obtained through inversion of ground deformation data observed after an earthquake. Specifically, Differential Interferometric Synthetic Aperture Radar (DInSAR) techniques are employed to measure surface displacements caused by the earthquakes using satellite imagery, such as ALOS-2 and Sentinel-1 data. These measurements, which include phase unwrapping results of surface displacements, reflect the changes in the ground due to the earthquake. Subsequently, these observations are inverted using a finite elastic half-space dislocation model. The inversion process consists of nonlinear and linear steps. In the nonlinear inversion, all the source parameters (i.e., the length, width, depth, longitude, latitude, strike, dip, rake, and slip) are initialized with the CMT solution with a uniform slip model. Then, to find the minimum value of the function, the Gauss-Newton iteration and the Levenberg-Marquardt least-squares algorithms were applied by considering multiple random restarts. In the linear inversion step, the fault geometry derived from the nonlinear inversion is fixed, and the slip distribution along the fault plane is estimated.  

\noindent\textbf{Input 45 (I45):}  The Coulomb Stress change - $\Delta CFF$ - is a measure used in seismology to assess the change in stress on a fault due to an earthquake or other tectonic activities to determine the likelihood of future earthquake triggering or inhibition. The Coulomb Stress change is defined in terms of the slip distribution model and the friction coefficient  $\mu_f$ of the fault. In addition, Skempton’s coefficient $\beta_S$ is used to modify the normal stress to account for pore pressure effects: 
\begin{equation}
    \Delta CFF=\Delta\tau + \mu_f(\delta \sigma-\beta_S\frac{Tr(T_S)}{3})
\end{equation}
where $\Delta \tau$ is the variation in the shear stress, $\Delta\sigma$ is the variation of normal stress and $Tr(T_S)$ is the trace of the stress matrix $T_S$ so that $Tr(T_S)/3$ is the change in pore pressure. An increase in the Coulomb stress change after an event indicates a high failure risk, and vice versa. 

\noindent\textbf{Input 46 (I46):} Distance from the fault. The inputs "I42. distance from the mainshock rupture" and "I46. distance from the fault" describe different spatial measurements used in seismology. Distance from the fault refers to the proximity to the fault trace, representing the general location of the fault on the Earth’s surface. In contrast, distance from the mainshock rupture specifically measures the distance to the actual rupture area of the mainshock, where fault slip occurred during the earthquake. 

\noindent\textbf{Input 47 (I47):}  Fault density using digitization obtained from Geological map.  

\noindent\textbf{Input 48 (I48):}  Mercalli Intensity - $MMI$.

\noindent\textbf{Input 49 (I49):}  Box-counting fractal dimension of a set of earthquake hypocenters - $D_c$\cite{R48}:
\begin{equation}
    D_c=\lim _{n\to 0}\frac{\log C_R}{\log(R)}
\end{equation}
where $R$ is the radius of the disk of analysis, $C_R$ is the correlation sum given by
\begin{equation}
    C_R=\lim_{n \to \infty} \frac{1}{n^2}\sum_{i=1}^{n}\sum_{j=1}^n{I(R-r_{ij}>0)}
\end{equation}
and $n$ is the number of events in the analysis window, which is often set to be $200$. $r_{ij}$ is the distance between $eq_i$ and $eq_j$. $C_R$ represents a function that defines the probability of two points being separated by a distance less than $R$.

\noindent\textbf{Input 50 (I50):}  Rescaled spatio-temporal distance between $eq_i$ and $eq_j$ - $\eta_{ij}$\cite{R49}. One first introduced the rescaled time interval between $eq_i$ and $eq_j$ as
\begin{equation}
    T_{ij}=t_{ij}10^{-\frac{bm_i}{2}}
\end{equation}
where $b$ is the $b$ value of the GR law and $t_{ij}=t_j-t_i$. Then, the rescaled distance between $eq_i$ and $eq_j$ is defined as
\begin{equation}
    R_{ij}=r_{ij}^{D_c}10^{-\frac{bm_i}{2}}
\end{equation}
Finally, $\eta_{ij}$ is defined as
\begin{equation}
    \eta_{ij}=T_{ij}R_{ij}=t_{ij}r_{ij}^{D_c}10^{-bm_i}
\end{equation}
\noindent\textbf{Input 51 (I51):}  Record of a seismogram in a given time window for a given seismograph as a specific location. 

\noindent\textbf{Input 52 (I52):}  Map of the ground motion due to an earthquake.

\noindent\textbf{Input 53 (I53):}  Seismic noise or ground shaking noise.

\subsubsection{Non-seismic types}

\noindent\textbf{Input 54 (I54):}  Seismic Electric signals. Noise is eliminated from electrotelluric field measurements by applying specific, well-defined criteria. 

\noindent\textbf{Input 55 (I55):}  Variation of radon concentration. 

\noindent\textbf{Input 56 (I56):}  Temperature at different elevation or isobaric surface. 

\noindent\textbf{Input 57 (I57):}  Rainfall.

\noindent\textbf{Input 58 (I58):}  Variation of geomagnetic field declination. 

\noindent\textbf{Input 59 (I59):}  Humidity. 

\noindent\textbf{Input 60 (I60):}  Electric and magnetic fields. 

\noindent\textbf{Input 61 (I61):} Geoacoustic data. 

\noindent\textbf{Input 62 (I62):}  Ionospheric variables recorded by satellites, including Electron density, Electron temperature, Ion temperature, Oxygen ion temperature, Hydrogen ion density, Helium ion density and so on.

\noindent\textbf{Input 63 (I63):}  Wind Speed. 

\noindent\textbf{Input 64 (I64):} Topographic altitude.  

\noindent\textbf{Input 65 (I65):}  Slope angle. It refers to the topographic gradient, representing the rate of elevation change over a given horizontal distance. It is derived from Digital Elevation Models (DEMs) and is commonly expressed in degrees or percentages.

\noindent\textbf{Input 66 (I66):}  Solid Earth Tide anomalies characterized by their dates, amplitudes, stress orientation, and so on.  

\noindent\textbf{Input 67 (I67):}  Outgoing Longwave Radiation anomalies that relate to earthquakes. 

\noindent\textbf{Input 68 (I68):}  Water level and water composition in wells.

\noindent\textbf{Input 69 (I69):}  Gas concentration observed by satellite, including Ozone, CO, CH4, and so on.   

\noindent\textbf{Input 70 (I70):}  Site properties (bedrock, rock, hard layer of soil, medium layer of soil, soft soil). 

\noindent\textbf{Input 71 (I71):}  Distance of the Moon from the Earth.  

\noindent\textbf{Input 72 (I72):}  Total Electron Content. 

\noindent\textbf{Input 73 (I73):}  Atmospheric pressure. 

\subsubsection{Summary}

\begin{figure}[ht]
\centering
\includegraphics[height=\textwidth]{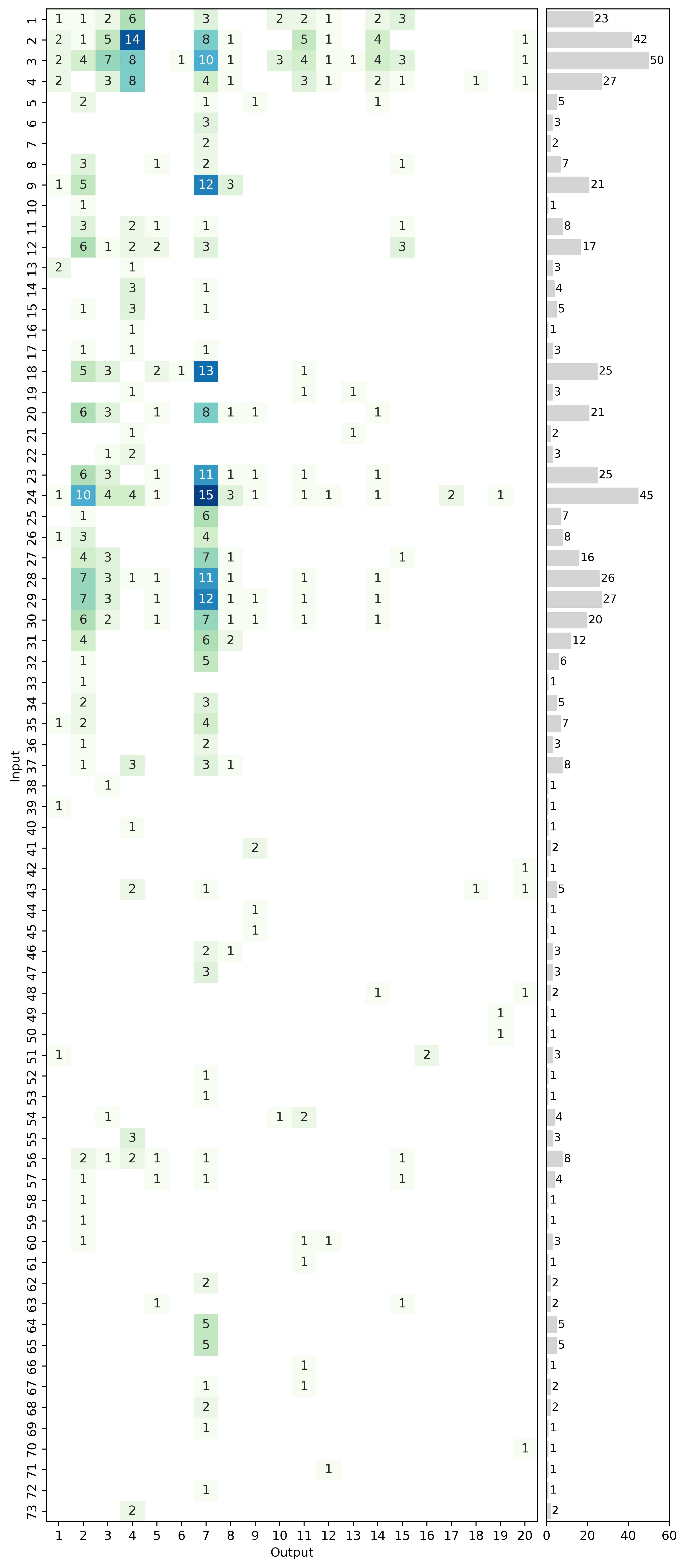}
\caption{Number of times different types of inputs have been used in published papers for different outputs. The histogram on the right shows the total number of papers that have used different types of inputs. 
}
\label{heatmap}
\end{figure}

\paragraph{Overview}
Figure \ref{heatmap} (left side) shows the number of times different inputs have been used to produce various outputs in the set of published articles. Figure \ref{heatmap} (right side) presents the total number of times each input has been used, independently of the outputs. The latter number (right side of Figure \ref{heatmap}) is therefore the sum over outputs of the former number (left side of Figure \ref{heatmap}).

Out of 141 investigated works, 113 of them used only seismic types as inputs, 8 works used only non-seismic types, and 19 works used both non-seismic and seismic types as inputs. Some studies incorporate a variety of inputs for earthquakes prediction, but whether all of these contribute to the prediction performance is not well-researched. 

There are significantly more seismic inputs than non-seismic ones. The average number of seismic inputs across the 129 published papers is 9.81, compared with an average of 2.4 non-seismic inputs. Among the seismic inputs, 11 parameters have been used more than 20 times, including the time ($t_i$, I1), location ([$\text{lat}i, \text{long}i$], I2), magnitude ($m_i$, I3), and depth ($d_i$, I4) of earthquakes, the maximum magnitude of earthquakes in a moving time window ($M^{\text{max}}(t{\text{start}}, t{\text{end}})$, I9), the time elapsed over a predefined number of events ($\Delta T_{\text{elasped}}^j(\mathfrak{N}, M_{\text{cutoff}})$, I18), the mean time between characteristic or typical events ($\mu$, I20), the Benioff strain ($dE_j^{\frac{1}{2}}(\mathfrak{N}, M_{\text{cutoff}})$, I23), the $b$ value of the GR law ($\hat{b}$, I24), and three other parameters related to $b$ values, which are $\eta$ (I28), $\Delta M$ (I29), and $c$ (I30).

This demonstrates that, apart from the time-location-magnitude-depth information of earthquakes, classical statistical seismology parameters used to describe the spatial and temporal variation of seismicity are more frequently applied in AI-based earthquake prediction models than non-seismic precursors.

\paragraph{Critique}

As inputs for AI-based earthquake prediction models, one should strictly adhere to the principles of simulating realistic earthquake prediction, as mentioned earlier. This involves using only historical data available up to the moment of prediction. However, in the case of the O4 task, some related studies incorporate the location and time of the target event as inputs, which is inconsistent with the requirements of a realistic earthquake prediction scenario.

\paragraph{Suggestion}

One common approach for evaluating the contributions of various features in earthquake prediction models is to compute feature importance. For example, Xiong et al.\cite{R50} introduced a machine learning algorithm known as Inverse Boosting Pruning Trees, which builds upon the Decision Tree framework. They utilized satellite observations of non-seismic precursors, including $CO$ column concentration, $CH_4$ column concentration, Outgoing Longwave Radiation (OLR), and other factors, as input features to predict earthquakes. To assess the relative importance of these precursors, they calculated the Gini importance, a metric that measures the contribution of each feature to the reduction of impurity in a decision tree. Their analysis revealed that OLR and $CO$ had significantly higher feature importance compared to $CH_4$, suggesting that these precursors might play a more prominent role in earthquake forecasting.

Another widely used approach for quantifying feature importance is the Shapley value, derived from cooperative game theory. The Shapley value offers a rigorous and equitable method for evaluating the individual contributions of features to the overall model output. In earthquake prediction models, where diverse input features are integrated, the Shapley value provides a robust framework for understanding feature relevance. Mathematically, for a set of \( N \) features and a model output function \( v(S) \) defined on subsets \( S \subseteq N \), the Shapley value for feature \( i \) is defined as:
\begin{equation}
\phi_i = \sum_{S \subseteq N \setminus \{i\}} \frac{|S|! \cdot (n - |S| - 1)!}{n!} \cdot \big[ v(S \cup \{i\}) - v(S) \big],
\end{equation}
where \( |S|! \) and \( (n - |S| - 1)! \) denote factorial weights that account for the permutations of subsets, and \( v(S \cup \{i\}) - v(S) \) represents the marginal contribution of feature \( i \) to the subset \( S \). By aggregating these marginal contributions across all possible subsets, the Shapley value ensures a fair allocation of feature importance. 

In practice, due to the exponential computational complexity of calculating Shapley values (\( O(2^n) \)), approximation techniques such as Monte Carlo sampling or SHAP (SHapley Additive exPlanations) are often employed. These methods make it feasible to apply the Shapley framework to large-scale models with numerous features. When applied to earthquake prediction, the Shapley value enhances model interpretability by identifying the most influential features, thereby guiding feature selection, improving model transparency, and refining predictive performance.

Additionally, ablation experiments, a common methodology in machine learning, can be employed to evaluate feature importance. This technique involves systematically removing specific inputs from the feature set and observing their impact on the model's performance. By comparing the model's accuracy or other performance metrics before and after feature removal, researchers can identify critical inputs and better understand their roles in the predictive process.

\subsection{Loss function}\label{loss function}
To determine optimal parameters, the AI community commonly relies on the loss function. The loss function quantitatively measures the disparity between predicted and target results within the training dataset. Artificial Neural Networks (ANNs) are designed to iteratively minimize this training loss to identify the optimal parameters.

Minimizing the loss typically involves gradient descent, which requires the loss function to be differentiable. The loss function adopts a penalizing approach, inspired by well-established techniques for solving constrained optimization problems, as described by Bertsekas\cite{R51}. In this framework, models incur penalties for inaccuracies in their predictions, encouraging improved performance over successive iterations.

In AI-based earthquake prediction and forecasting models, selecting an appropriate loss function depends on the model's output type and the specific objectives of the task. The most common task types in this context are regression and classification. This section provides a detailed explanation of the corresponding loss functions for each task.

The goal of a regression task is to learn the mapping relationship between inputs and outputs, enabling the model to accurately predict output values for new inputs. In contrast, the objective of a classification task is to determine the category to which the input data belongs, emphasizing the correct assignment of inputs to predefined classes.

Researchers often approach the task of predicting the probability of earthquake occurrence (O7) as a classification problem, aiming to determine whether an earthquake will occur within a specified time-space-magnitude window. This involves categorizing events into binary outcomes (e.g., earthquake or no earthquake) based on various input features.

In contrast, predicting parameters such as earthquake magnitude, time of occurrence, b-value, seismograms, and seismicity is typically treated as a regression task. However, as discussed in earlier sections, earthquake magnitude prediction can also be framed as a multi-class classification problem, where different magnitude ranges are assigned distinct labels.

\subsubsection{Loss function for regression}
Consider the set of predicted results $\hat{y_1}, \hat{y_2},...,\hat{y_n}$ and the corresponding true labels $y_1,y_2,...,y_n$ in the training dataset. For a pair of $\hat{y_i}$ and $y_i$, the error between the predicted result and the true label can be naturally represented as $y_i-\hat{y_i}$. Based on the error form $y_i-\hat{y_i}$, the loss functions used in 141 AI-based earthquake prediction/forecasting models include Sum of Squared Error (SSE), Mean Squared Error (MSE), Root Mean Squared Error (RMSE), and Mean Absolute Error (MAE), with specific definitions as follows:
\begin{equation}
    SSE=\sum_{i=1}^{n} {(y_i-\hat{y_i})^2}
\end{equation}
\begin{equation}
    MSE=\frac{1}{n}\sum_{i=1}^{n}(y_i-\hat{y_i})^2 = {SSE \over n}
\end{equation}
\begin{equation}
    RMSE=\sqrt{\frac{1}{n}\sum_{i=1}^{n}(y_i-\hat{y_i})^2} = \sqrt{MSE}
\end{equation}
\begin{equation}
    MAE=\frac{1}{n}\sum_{i=1}^{n}{|y_i-\hat{y_i}|} 
\end{equation}
In machine learning, to prevent overfitting, a penalty term is often added to the loss function, a technique known as regularized regression. Two common types of regularization used in regression are L1 regularization (Lasso) and L2 regularization (Ridge).
\begin{equation}
    L1=MSE+\Gamma\cdot\sum_{j=1}^{p}{|w_j|}
\end{equation}
\begin{equation}
    L2=MSE+\Gamma\cdot\sum_{j=1}^{p}{w_j^2}
\end{equation}
\noindent where $\Gamma$ is the regularization parameter that controls the strength of regularization, the $w_j$'s are the regression coefficients, and $p$ is the number of features. The regularization parameter $\Gamma$ controls the trade-off between fitting the training data well (minimizing MSE) and keeping the model simple (smaller coefficients). Selecting an appropriate value for $\Gamma$ is crucial and often relies on techniques such as cross-validation.

\subsubsection{Loss function for classification}\label{loss}
Cross Entropy and its derivatives are widely used in ANNs for dealing with binary classification problems. Cross Entropy (CE) is given by 
\begin{equation}
    CE(y_i,\hat{y_i})=\begin{cases}
        -\log(\hat{y_i}),~~ {\rm if}~ y_i=1 \\
        -\log(1-\hat{y_i}),~~{\rm if}~ y_i=0 ~,
    \end{cases}
\end{equation}
where $y_i\in \{0,1\}$ represents the true label of the $i^{th}$ sample: "1" denotes a positive sample, while "0" denotes a negative sample. $\hat{y_i}\in [0,1]$ is the probability estimated by the model that the $i^{th}$ sample is "positive", and $(1-\hat{y_i})$ is the predicted probability that it is "negative". $CE(y_i,\hat{y_i})$ can thus be equivalently expressed as 
\begin{equation}
    CE(y_i,\hat{y_i})= 
        -\log\left[\hat{y_i}^{y_i} (1-\hat{y_i})^{1-y_i}\right]~.
\end{equation}

Balanced Cross Entropy (BCE) is an improved loss function designed to address class imbalance in binary classification problems. In traditional binary classification tasks, when there is a significant difference in the number of positive and negative samples, models tend to be biased towards predicting the majority class, which can impact performance, accuracy as well as relevance. Balanced Cross Entropy aims to solve this issue by adjusting the weights of the loss function, prioritizing predictions for the minority class, thus improving model performance and relevance on imbalanced datasets. If the samples are dominated by positive samples, the BCE is given as,
\begin{equation}
    BCE(y_i,\hat{y_i},\alpha)=\begin{cases}
        -(1-\alpha)\log(\hat{y_i}), ~~{\rm if}~ y_i=1\\
        -\alpha \log(1-\hat{y_i}),~~{\rm if}~ y_i=0
    \end{cases}
    ~~~= -\log\left[\hat{y_i}^{y_i(1-\alpha)} (1-\hat{y_i})^{(1-y_i)\alpha}\right]
\end{equation}
\noindent where $\alpha$ is equal to the inverse of the proportion of positive samples.

To understand CE and BCE in the context of a specific application, let us consider a scenario related to earthquake forecasting within a certain time-space-magnitude volume, such as the Output 7 presented in section \ref{output}. In this scenario, if at least one earthquake occurs in the $i^{th}$ time-space-magnitude volume, then $y_i$ is set to be $1$ , indicating a “positive” sample. If no earthquakes occur in the $i^{th}$ time-space-magnitude volume, then $y_i$ is set to be 0, indicating a "negative" sample. $\hat{y_i}$ is the probability that at least one earthquake will occur in the  time-space-magnitude volume given by the model.

Lin et al.\cite{R53} proposed another loss function to deal with class imbalance, which is called Focal Loss (FL). Easily classified negatives contribute the majority of the loss and dominate the gradient, as BCE does not distinguish between easy and hard examples. To address this issue, Focal Loss (FL), compared to CE, assigns higher penalties to harder samples and lower penalties to easier ones, thereby focusing the model's learning on more challenging cases. Focal Loss is defined as
\begin{equation}
    FL(y_i,\hat{y_i},\gamma)=\begin{cases}
        -(1-\hat{y_i})^{\gamma}\log(\hat{y_i}),~~{\rm if}~ y_i=1\\
        -\alpha\hat{y_i}^{\gamma}\log(1-\hat{y_i}),~~{\rm if}~ y_i=0
    \end{cases}
    ~~=-\log\left[\hat{y_i}^{y_i(1-\hat{y_i})^\gamma} (1-\hat{y_i})^{(1-y_i)\alpha \hat{y_i}^\gamma}\right]
\end{equation}
The gradient harmonization mechanism classification (GHMC) loss function, which is a revised Focal Loss, is also used in AI-based earthquake prediction models. Both GHMC and FL aim to reduce the influence of easy samples during training, GHMC goes a step further by dynamically adjusting the loss weights based on sample gradients and margins, which effectively addresses both easy and difficult samples. Firstly, we define a gradient norm variable $\mathfrak{g}$,
\begin{equation}
    \mathfrak{g}_i=|y_i-\hat{y_i}|
\end{equation}
The smaller $\mathfrak{g}_i$, the more accurate the prediction is and the easier the sample is to classify, and vice versa. GHMC attenuates both easy and hard samples, with the degree of attenuation determined the number of samples. A gradient density (GD) is defined as
\begin{equation}
    GD(\mathfrak{g})=\frac{1}{l_{\epsilon}(\mathfrak{g})}\sum_{j=1}^{n}{\delta_{\epsilon}(\mathfrak{g}_j,\mathfrak{g})}
\end{equation}
\noindent $GD(\mathfrak{g})$ represents the calculation of the gradient density corresponding to the gradient modal length $\mathfrak{g}$. $\delta_{\epsilon}(\mathfrak{g}_i,\mathfrak{g})$ is a flag indicating whether the gradient of sample $i$ falls within the specified interval $[\mathfrak{g}-\frac{\epsilon}{2},\mathfrak{g}+\frac{\epsilon}{2}]$:
\begin{equation}
    \delta_{\epsilon}(g_i,\mathfrak{g})=\begin{cases}
        1, ~~{\rm if}~ \mathfrak{g}-\frac{\epsilon}{2}\leq g_i \leq \mathfrak{g}+\frac{\epsilon}{2}\\
        0,~~{\rm otherwise}~.
    \end{cases}
\end{equation}
$\epsilon$ represents a small positive constant that controls the width of the interval within which the gradient density is calculated. The value of $\epsilon$ determines the "neighborhood" around the gradient norm  $\mathfrak{g}$ that is considered for the density calculation. In the context of GHMC,  $\epsilon$ helps control the balance between the easy and hard samples by adjusting the sensitivity of the gradient density function.
The normalizing factor $l_{\epsilon}(\mathfrak{g})$ is defined as
\begin{equation}
    l_{\epsilon}(\mathfrak{g})=\min (\mathfrak{g}+\frac{\epsilon}{2},1)-\max (\mathfrak{g}-\frac{\epsilon}{2},0)
\end{equation}
This expression shows that $l_{\epsilon}(\mathfrak{g})$ is a linear function from $g=0$ to $g=\epsilon /2$ with slope $+1$ that attains the value $\epsilon$ at $g=\epsilon /2$. Then, for $\epsilon /2 \leq g \leq 1-\epsilon /2$, $l_{\epsilon}(\mathfrak{g}) = \epsilon$ is constant. For $g > 1-\epsilon /2$, $l_{\epsilon}(\mathfrak{g})$ is a decreasing function with slope $-1$ crossing $0$ at $g=1+\epsilon/2$. In other words, $l_{\epsilon}(\mathfrak{g})$ has the shape of a tent with flat roof.

The gradient harmonization mechanism classification (GHMC) loss can then be defined as
\begin{equation}
    GHMC(y_i,\hat{y_i})=\frac{CE(y_i,\hat{y_i})}{GD(g_i)}
\end{equation}
Apart from CE and its derivative, hinge loss function (HLF) and squared hinge loss function (SHLF) can also serve as loss functions for binary classification tasks. HLF and SHLF are defined as 
\begin{equation}
    HLF(y_i,\hat{y_i})=\max(0,1-y_i\cdot\hat{y_i})
\end{equation}
\begin{equation}
    SHLF(y_i,\hat{y_i})=\max(0,1-y_i\cdot \hat{y_i})^2
\end{equation}
The SLHF is essentially a squared version of the HLF. It penalizes misclassifications more severely than the HLF, making it less tolerant to misclassifications. It is important to note that the HLF and SLHF differ from the other loss functions in this section in that the labels of positive samples \( y_i \) are marked as 1, while the labels of negative samples are marked as -1. Meanwhile, the predicted result of $\hat{y_i}$ falls between -1 and 1.

Besides binary classification, there are also multi-class classification tasks, such as predicting earthquake magnitudes by assigning different magnitudes ranges to different labels. Popular loss functions for multi-class tasks include the standard softmax loss (SSL) function. 

Consider a set of $n$ training samples, where each sample $i$ has a true label $y_i$ from a set of $K$ classes ($y_i\in \{ 0,1,2,...,K\}$). In multi-class classification task, the output $\hat{y_i}$ of the $i^{th}$ sample is a vector $\hat{y_i}=(\hat{y_{i1}},\hat{y_{i2}},...,\hat{y_{iK}})$. One can use the softmax function to convert $\hat{y_i}$ into a probability distribution over each class,
\begin{equation}
    softmax(\hat{y_i})_k=\frac{e^{\hat{y_{ik}}}}{\sum_{j=1}^{k}{e^{\hat{y_{ij}}}}}
\end{equation}
\noindent where $softmax(\hat{y_i})_k$ is the probability that the sample $i$ belongs to class $k$. Then, the softmax loss or the cross entropy loss is used to measure the discrepancy between the predicted class probabilities and the true distribution of the labels. For a single sample , the standard softmax loss ($SSL$) is defined as
\begin{equation}
    SSL(y_i,\hat{y_i})=-\log(softmax(\hat{y_i})_{y_i})
\end{equation}
In decision three algorithms, the Gini coefficient is often used as a measure of data impurity and serves as the loss function or cost function of the decision tree. During the process of node splitting in decision trees, features with the smallest Gini impurity are chosen as the splitting criteria to achieve effective data partitioning. By minimizing the weighted average Gini impurity of the resulting child nodes after splitting, decision tree algorithms can gradually optimize the structure of the tree, enabling the final generated decision tree to achieve higher classification performance on the given dataset. The Gini impurity is typically represented as:
\begin{equation}
    Gini(T_{node})=1-\sum_{i=1}^{K}{(p_i)^2}
    \label{tbqgqd}
\end{equation}
\noindent where $T_{node}$ is a node (or dataset) of the tree, $K$ is the number of classes, and $p_i$ is the proportion of samples belonging to Class $i$ in node $T_{node}$. When all samples in the dataset belong to the same category, $Gini(T_{node})$ reaches its minimum value of 0, indicating the highest purity of the dataset; in contrast, when samples of each category are evenly distributed, $Gini(T_{node})$ reaches its maximum value of 1, indicating the highest impurity of the dataset. In decision tree algorithms, features with the smallest Gini coefficient are typically chosen for node splitting to achieve effective data classification.

It is noteworthy that the Gini impurity defined by expression (\ref{tbqgqd}) is used in other fields under various names. One minus the Gini impurity is called Herfindhal index in finance and quantifies the level of concentrations of financial porfolios. In the theory of spinglasses and of complex systems, one minus the Gini impurity is called the participation ratio and quantifies the extent to which the spin configurations or energy states are localized or delocalized, indicating how evenly the contributions are distributed across the system.

\subsubsection{Loss function designed for earthquake prediction/forecasting model}

Zhang and Wang\cite{R54} developed a spatiotemporal model for global earthquake prediction based on Convolutional LSTM. They divided the global map into spatial bins of size $72.92km\times 67.71km\ (longitude\times latitude)$. During training, they observed that the area over wihch earthquakes occur occupies only a small portion of the dataset, leading to a significant imbalance between earthquake and non-earthquake samples. Using traditional MSE or MAE as the loss function may lead to negative samples dominating the network gradients, thereby preventing the model from learning features of positive samples or earthquake regions. To solve this problem, they proposed a weighted MSE-MAE loss function. First, they extracted earthquakes with $Mw\geq 4$ from the training set onto the global map. Then, $m$ pixels around each earthquake in the map were marked as weight regions ($m$ depends on the map resolution). They used a weighted MSE-MAE loss function to assign greater weights to the earthquake regions and the surrounding weighted areas. The definition of this loss function is 
\begin{equation}
    Weighted\ MSE-MAE=w_i(y_i-\hat{y_i})^2+w_i|y_i-\hat{y_i}|
\end{equation}
\noindent where $w_i$ is the weight of each sample or spatial bin. The weights are determined by balancing the number of pixels in the three regions (area with no earthquakes, weighted map area, and earthquake area), and then adjusted after training on the training set.

Liu et al.\cite{R55} developed a spatiotemporal prior-informed deep network (STPiDN) to predict the epicenter and magnitude of future events. To address the spatial clustering in the distribution of earthquakes and the imbalance in the relative occurrence of earthquakes of different magnitudes according to the Gutenberg-Richter law $p(m)$, they proposed separate loss functions for predicting earthquake magnitude and epicenter location.

The loss for the magnitude prediction task is 
\begin{equation}
    L_{magnitude}=f^{m}(m_i)(m_i-\hat{m_i})^2
\end{equation}
\noindent where $m_i$ is the true magnitude of the $i^{th}$ event, and $\hat{m_i}$ is the predicted magnitude. $f^m(\cdot)$ is a mapping function to compute the balanced weight for sample based on $p(m_i)$ (probability that earthquakes with magnitude larger than $m_i$ will occur), and is computed by a magnitude probability function derived from the Gutenberg-Richter law,
\begin{equation}
    \begin{cases}
        f^m(m_i)=\frac{\mu_1}{p(m_i)+\mu_2}\\
        p(m_i)=b'\times 10^{-b'(m_i-m_c)}~.
    \end{cases}
\end{equation}
\noindent Here, $\mu_1,\mu_2>0$ are constants chosen by the model developer. $\mu_1$ is used to ensure that the loss weights are larger than 1 to amplify the loss of low-probability events. $\mu_2$ is a regularising constant used to avoid dividing by $0$. The constant $b'=\hat{b}/\log_{10}e$ is the estimated $b$ value and $m_c$ is the complete magnitude.

Similarly, the loss for the epicenter prediction task is 
\begin{equation}
    L_{epicenter}=f^r(r_i)|r_i-\hat{r_i}|
\end{equation}
\noindent where $r_i$ and $\hat{r_i}$ are true and predicted earthquake epicenter vector consisting of longitude and latitude information. $|r_i-\hat{r_i}|$ is the distance between the observed and predicted epicenter. $f^r(\cdot)$ is a mapping function that computes the balanced weight for the current sample based on distance to the nearest faults - $d_{r_i}$:
\begin{equation}
    f^r(r_i)=\mu_3\exp{\left[-\left(\frac{d_{r_i}}{\mu_4}\right)^2\right]}+\mu_5
\end{equation}
\noindent Here $\mu_3$, $\mu_4$ and $\mu_5$ are positive constants chosen by the model developer. $\mu_4$ is the bandwidth, and $\mu_5$ is used to ensure that the loss weights are larger than 1. 

The prediction loss is the sum of the magnitude prediction loss and the epicenter prediction loss, 
\begin{equation}
    L_{prediction}=L_{magnitude}+\exp{(-\omega)}L_{epicenter}+\lambda\omega
\end{equation}
\noindent where $\exp({-\omega})$ is the weight that controls the relative importance of the two loss contributions $L_{magnitude}$ and $L_{epicenter}$, $\omega$ is trained by the adversarial process between $\exp{(-\omega)}$ and $\lambda\omega$, and $\lambda$ is the weight of $\omega$.

Stockman et al.\cite{R7} constructed a new machine learning variant of point processes for short-term earthquake forecasting. The output of the model is the inter-event time $\tau$ and the magnitude $m$ above magnitude $M_d$ of the next event. The sequence is  
\begin{equation}
    \{ (t_j,m_j):m_j\geq M_d\}^k_{j=1}~.
\end{equation}
\noindent To introduce the loss function, The subscript $i$
is used to denote the model input events, while the subscript $j$ represents the predicted target events, providing a clear distinction between the two. The loss function is defined as 
\begin{equation}\label{LL}
\begin{split}
    \log L(t_j,m_j)=\sum_{i:m_i\geq m_c,t_i\leq t_k}[(\log \frac{\partial}{\partial\tau}\Phi(\tau_i|h_i)+\log\frac{\partial}{\partial m}\Psi(m_i|\tau_i,h_i))I(m_i\geq M_d)-\Phi(\tau_i|h_i)]
\end{split}
\end{equation}
\noindent where $h_i$ is the hidden state of the neural network with a two-dimensional input $\{t_i,m_i\}$. $\Phi(\tau|h_i)$ is a cumulative hazard function using the neural network defined by Omi et al. (2019), and $\tau=t-t_i$ is the elapsed time from the most recent event that occurred at $t_i$. $\Psi(m|\tau_i,h_i)$ is the cumulative distribution associated with the conditional density function of the magnitude mark at time $t$. More details about $\Phi(\tau|h_i)$ and $\Psi(m|\tau_i,h_i)$ can be found in \cite{R7}. Stockman et al. trained this model by maximizing the likelihood defined in Equation \ref{LL}.

Inspired by the concept of Focal Loss, Zhang et al.\cite{R52} propose a revised loss function based on MSE and CE to introduce some knowledge of statistical seismology. The revised MSE and CE are defined as
\begin{equation}\label{CE_zhang}
    CE^*(y_i,\hat{y_i},y_i^{ref})=-[y_i^{ref}*y_i\log(\hat{y_i})+(1-y_i^{ref})*(1-y_i)\log(1-\hat{y_i})]
\end{equation}
\begin{equation}\label{MSE_zhang}
    MSE^*(y_i,\hat{y_i},y_i^{ref})=(y_i^{ref}-y_i)^\gamma*(\hat{y_i}-y_i)^2
\end{equation}
\noindent where $y_i^{ref}$ is the predicted value provided by the reference model, and $\gamma$ is a modulating factor. In this way, the greater the difference between the predicted value of the introduced seismology model and the real label of the sample, the greater the effect of this sample on the gradient of the ANN, thus forcing the model to focus more on the difficult samples of the introduced seismology model. The revised loss function enables the ANN to uncover new insights that may not be captured by the introduced statistical seismological reference model.

\subsubsection{Summary}

\paragraph{Overview}
In recent decades, significant efforts have been dedicated to designing deeper and more complex neural networks to enhance the performance of AI-based earthquake prediction models, often overlooking the critical role of loss functions. The loss function not only determines the network's optimization strategy but also defines what the model is expected to learn. Different loss functions can result in vastly different training outcomes for the same network. Customizing the loss or objective function to align more closely with the specific problem addressed by the ANN can significantly boost model performance. However, our investigation reveals that only five studies have designed new loss functions or incorporated seismological loss functions to better suit the needs of earthquake forecasting applications. This highlights that the role and importance of loss functions in earthquake forecasting models have been largely underestimated.

The choice of the objective function to optimize is not unique to machine learning and AI; it is a universal challenge across various disciplines, including variational problems in physics, optimization in engineering, and economic modeling. The selection of this function inherently embeds all the operator's knowledge, assumptions, and biases --often implicitly or without their full realization. Ideally, the choice of the loss function would be informed by a comprehensive understanding of the underlying generative processes, real mechanisms, and true patterns of the system being studied. However, such an ideal scenario is rarely attainable. Consequently, the chosen loss function becomes a reflection of our limited and biased understanding, underscoring the need to remain aware of these limitations at all times.

\paragraph{Critique}
Existing AI-based earthquake forecasting models draw heavily from principles used in image classification. However, it is crucial to recognize the significant differences between image classification samples and earthquake samples. In image classification, the ratio of positive to negative samples is typically balanced, and the samples are independent of one another. In contrast, earthquake samples are highly imbalanced, with negative samples vastly outnumbering positive ones, and positive samples (earthquakes) exhibiting strong clustering in both space and time.

As a result, treating earthquake prediction as a simple binary classification problem is an oversimplification. Using Cross-Entropy as the loss function in this context, where false negatives and false positives are penalized equally, leads to an imbalance in the training process. The network's gradient becomes dominated by the overwhelming number of negative samples, hindering its ability to effectively learn the characteristics of the positive samples—those critical for accurate earthquake forecasting.

\paragraph{Suggestion}
The recognition of the critical role of loss function design, as discussed in {\bf Overview}, underscores the need for ongoing efforts to critically examine and refine the assumptions underlying loss functions. Developing innovative loss functions should be an iterative process, informed by the outcomes of training and testing, with the goal of progressively aligning these functions with the generative mechanisms we aim to understand. By bridging the gap between models and the complex realities they strive to predict, such advancements can drive improvements in both performance and understanding.

An exciting insight from related works is that revising the loss function offers a powerful way to incorporate domain-specific knowledge directly into the optimization process. For instance, Liu et al.\cite{R55} successfully integrated fault information into their model by embedding it within the loss function, rather than using it as an input feature. Similarly, Zhang et al.\cite{R52} enhanced their ANN by weighting the punishments and rewards for failed and successful predictions with prior probabilities derived from statistical seismology models, effectively embedding expert knowledge from seismology into machine learning techniques.

These examples highlight the transformative potential of tailored loss functions in enhancing model performance. By leveraging domain expertise within the optimization process, such approaches provide a novel pathway for improving the synergy between machine learning models and the underlying principles of seismology.

Future research should prioritize the development of advanced loss functions specifically designed for earthquake prediction applications. By continuously refining these functions and incorporating deeper domain insights, researchers can enable models to better capture the complex dynamics of seismic activity, unlocking new opportunities for progress at the intersection of seismology and AI.

To address the issue of data imbalance discussed in {\bf Critique}, the most straightforward idea is to increase the weight of minority samples, thereby increasing their influence on the network gradients. For example, the BCE has been proposed to address the issue of imbalanced positive and negative samples. Moreover, 
Liu et al.\cite{R55} tackled the challenge of imbalanced distributions between large and small earthquakes by increasing the weight assigned to the minority class of large earthquake samples. Some specialized loss functions are adaptations of common machine learning loss functions, such as Cross-Entropy (CE). Experimental results from their study demonstrate that this approach can significantly improve model performance. An alternative strategy involves directly adopting objective functions from statistical seismology, such as the likelihood function employed in Stockman et al.'s work\cite{R7}.

As previously mentioned, earthquake data exhibits not only an imbalance between positive and negative samples but also an imbalance between large and small earthquakes, governed by the Gutenberg-Richter (GR) law, which relates magnitude to frequency. To address this, the weight of large earthquakes can be artificially increased during training, as demonstrated with $L_{magnitude}$.
Another significant imbalance exists between background (independent) events and triggered events. When the full sequence of earthquakes is used as input and prediction targets, in general, the number of triggered events far exceeds that of background events. Consequently, the network’s gradient becomes dominated by triggered events, making the model more effective at predicting these events but less capable of accurately forecasting background events.
In other words, the model excels at forecasting aftershocks but performs poorly in predicting mainshocks. This represents a significant deficiency and limitation, particularly for the practical application of earthquake forecasts in societal and operational contexts where accurate mainshock prediction is critical.
It is important to emphasize that this same issue affects various versions of the ETAS model, as reflected in its very name, ``epidemic-type aftershock sequence.'' The ETAS model focuses on earthquake triggering and treats mainshocks as inherently unpredictable background events. Consequently, the predictability of large earthquakes within this framework is primarily determined by the fraction of such events that are triggered by preceding earthquakes \cite{HSor03}.

From the perspective of statistical seismology and the analysis of seismicity spatio-temporal-magnitude patterns, the limits of large earthquake predictability are fundamentally linked to the apparent lack of significant correlations between earthquake magnitudes. In simpler terms, once an earthquake nucleates, its magnitude (or size) appears to behave as a random variable, largely independent of, or only very weakly influenced by, prior seismic activity. This widely acknowledged lack of correlations between earthquake magnitudes represents a major obstacle to the reliable prediction of large earthquakes. It highlights the intrinsic difficulty in anticipating their size based on previous seismicity patterns alone.

To advance efforts in predicting background seismic events and addressing the challenges of large earthquake forecasting, a two-pronged approach is essential:
\begin{enumerate}
    \item Enhancing Detection of Magnitude Correlations: Efforts should focus on improving our ability to identify correlations not only between earthquake magnitudes but also between magnitudes and other seismic parameters, such as spatial clustering, fault interactions, and energy release patterns. Advanced machine learning techniques, coupled with large-scale, high-resolution datasets, could be leveraged to uncover subtle dependencies that might otherwise go unnoticed in traditional analyses.
    \item Incorporating Non-Seismic Variables: Expanding the predictive framework to include non-seismic and non-mechanical variables is critical. Geophysical, geochemical, and space-based observations—such as ground deformation, gas emissions, electromagnetic anomalies, and satellite-based measurements of Earth's surface—offer valuable complementary information that could provide insights into the underlying processes governing earthquake nucleation and magnitude determination. While such variables have been widely proposed in the past, they remain underexplored and underutilized, offering a promising avenue for innovation in earthquake prediction models.
\end{enumerate}
By pursuing these strategies, researchers can work toward overcoming the limitations imposed by the apparent randomness of earthquake magnitudes, refining predictive models, and advancing the broader goal of improving our understanding of the complex dynamics driving seismic activity.

To develop a model focused on predicting background events, one could leverage seismic quiescence theory and use parameters such as RTL or RTL+ETAS as input features. For the loss function, an Information Prior (IP) weighting approach can be applied, encouraging the model to concentrate on learning the spatial and temporal distribution of independent events. This targeted focus ensures a better balance in performance across different event types. 

To integrate the technologies and knowledge of Machine Learning and Seismology, "feature engineering" is a widely used and effective strategy that should be explored further.

\subsection{Evaluation metrics}
Evaluation metrics are used to quantitatively assess the performance and accuracy of a model's predictions. The choice of evaluation metrics depends on the type of output and is critical for understanding a model's effectiveness. In the field of machine learning research, a variety of evaluation metrics are available for AI-based earthquake prediction models, tailored to common regression and classification tasks. These metrics will be introduced separately in the following sections.

In addition to standard machine learning metrics, statistical seismologists have developed specialized evaluation metrics specifically designed to assess the performance of earthquake prediction models, further enriching the tools available for this unique application.

\subsubsection{Evaluation metrics for regression in AI applications}
Consider a set of predicted results $\hat{y_1},\hat{y_2},...,\hat{y_n}$ and the corresponding true labels $y_1,y_2,...,y_n$ in the testing dataset. Several loss functions introduced in Section \ref{loss}, including Mean Squared Error, Root Mean Squared Error, Mean Absolute Error, and Sum of Squared Error are commonly used as evaluation metrics. Sometimes, the norm of the error $||e||_p$ is used to evaluate the model performance:
\begin{equation}
    ||e||_p=(\sum_{i=1}^{n}{|y_i-\hat{y_i}|^p})^{1/p}~.
\end{equation}
\noindent For $p=1$, $||e||_1$ is the L1 Norm or Manhattan Norm defined as the sum of the absolute values of the errors. It is less sensitive to outliers and is suitable for applications where robustness to outliers is important. For $p=2$, $||e||_2$ is the L2 Norm or Euclidean Norm and is more sensitive than the L1 Norm to large errors. For $p\to +\infty$, $||e||_{\infty}(||e||_{\infty}=\max\limits_{i}|y_i-\hat{y_i}|)$ is the L$\infty$ Norm or Maximum Norm, and it is the most sensitive to large errors as it emphasises the largest discrepancies.

Relative error (RE) is calculated as the ratio of the absolute error to the true value, often expressed as a percentage. A small relative error indicates that the approximation is close to the true value, while a large relative error indicates a significant deviation from the true value, implying lower accuracy. $RE(y_i,\hat{y_i})$ is given as
\begin{equation}
    RE(y_i,\hat{y_i})=\frac{|y_i-\hat{y_i}|}{|y_i|}*100\%
\end{equation}
Mean Absolute Percentage Error (MAPE) is the average of Relative Error, and it is given as
\begin{equation}
    MAPE=\frac{1}{n}\sum_{i=1}^{n}{RE(y_i,\hat{y_i})}=\frac{1}{n}\sum_{i=1}^{n}{\frac{|y_i-\hat{y_i}|}{|y_i|}}*100\%
\end{equation}
Mean Squared Logarithmic Error (MSLE) quantifies the relative difference between the predicted and actual values after a logarithmic transformation, and it is given as
\begin{equation}
    MSLE(y_i,\hat{y_i})=(\log(y_i+1)-\log(\hat{y_i}+1))^2~.
\end{equation}
Log-Cosh Loss (LCL) combines the strengths of both Mean Squared Error (MSE) and Mean Absolute Error (MAE), offering sensitivity to small errors while maintaining robustness against outliers. The term “log-cosh” refers to the logarithm of the hyperbolic cosine function, which helps to mitigate the effect of outliers and improve robustness. The $LCL(y_i,\hat{y_i})$ is defined as
\begin{equation}
    LCL(y_i,\hat{y_i})=\log(\cosh(\hat{y_i}-y_i))~,
\end{equation}
where the hyperbolic cosine function is defined as
$\cosh{(\hat{y_i}-y_i)}=\frac{e^{\hat{y_i}-y_i}+e^{y_i-\hat{y_i}}}{2}$.

Similarity Ratio (SR) $SR(y_i,\hat{y_i})$ is another simple metric used to evaluate the deviation between predicted values and actual values defined as
\begin{equation}
    SR(y_i,\hat{y_i})=\frac{\hat{y_i}}{y_i}~.
\end{equation}
Obviously, the optimum Similarity Ratio is $ SR(y_i,\hat{y_i})=1$.

For a series of results obtained under continuous prediction, researchers also use the correlation coefficient to assess the model’s performance. The Pearson correlation coefficient $(\mathfrak{r})$, which quantifies the existence of linear dependence \cite{MalSor2006}, is defined as
\begin{equation}
    \mathfrak{r}=\frac{\sum_{i=1}^{n}{(y_i-\overline{y})(\hat{y_i}-\overline{\hat{y}}})}{\sqrt{\sum_{i=1}^{n}{(y_i-\overline{y})^2}\sum_{i=1}^{n}{(\hat{y_i}-\overline{\hat{y}})^2}}}
\end{equation}
\noindent where $\overline{y}$ is the average of ($y_1,y_2,...,y_n$), and $\overline{\hat{y}}$ is the average of ($\hat{y_1},\hat{y_2},...,\hat{y_n}$).

The Nash-Sutcliffe Efficiency (NSE) is a measure used to evaluate the performance of hydrological or environmental models, and it is sometimes also used for AI-based earthquake prediction models. The NSE is defined as
\begin{equation}
    NSE=1-\frac{\sum_{i=1}^{n}{(\hat{y_i}-y_i)^2}}{\sum_{i=1}^{n}{(y_i-\overline{y})^2}}
\end{equation}
\noindent The value $NSE=1$ corresponds to a perfect model efficiency, where model predictions match the observed values perfectly. When $NSE>0 ~ {\rm or} ~ <0$, the model performs better or worse than the mean of observed values, respectively. When $NSE=0$, the model performs as accurately as simply using the mean of the observed values.

The Wilmott Index (WI), also known as the Wilmott Q, provides a straightforward  intuitive measure of the forecasting model’s accuracy. It is defined as
\begin{equation}
    WI=1-\frac{\sum_{i=1}^{n}{(\hat{y_i}-y_i)^2}}{\sum_{i=1}^{n}{(y_i-\overline{y})^2}\sum_{i=1}^{n}{(\hat{y_i}-\overline{\hat{y}})^2}}
\end{equation}
\noindent The value $WI=1$ correspoinds to perfect model efficiency. For $WI<0$, the model is worse than the mean of observed values. 

\subsubsection{Evaluation metrics for classification used in AI communities}

Based on the true labels of the samples, they can be categorized as either positive or negative. For example, in output (7), this corresponds to the occurrence (1) or non-occurrence (0) of earthquakes within a specified time-space-magnitude volume. The terms in the confusion matrix are commonly used to evaluate the performance of a classification model. These terms represent the relationships between the model's predictions and the ground truth, providing insight into its accuracy and errors:
\begin{itemize}
    \item True Positive (TP): The number of samples correctly predicted as positive class or ‘1’.
    \item  True Negative (TN): The number of samples correctly predicted as negative class or ‘0’.
    \item False Positive (FP): The number of samples incorrectly predicted as positive class.
    \item False Negative (FN): The number of samples incorrectly predicted as negative class. 
\end{itemize}
False positive is also referred to as an error of type I. In statistical test theory, it corresponds to the rejection of the null hypothesis when it is actually true. A type II error, or a false negative, is the failure to reject a null hypothesis that is actually false. The null hypothesis corresponds to the negative class.

Building on the four statistical measures mentioned earlier, researchers have developed a series of evaluation metrics, which we will introduce individually. These metrics have been specifically applied to AI-based earthquake prediction models.

Percentage of correct prediction is the ratio of the number of correct predictions to the total number of predictions, and it is also known as accuracy (ACC):
\begin{equation}
    Acc=\frac{TP+TN}{TP+TN+FP+FN}
\end{equation}
False alarm rate (FAR) or False positive rate (FPR) or False Discovery Rate (FDR) is the proportion of actual negative samples that are incorrectly predicted as positive by the model:
\begin{equation}
    FAR=\frac{FP}{FP+TP}
\end{equation}
False negative rate (FNR), which is also known as missing rate, is the proportion of actual positive samples that are incorrectly predicted as negative by the model:
\begin{equation}
    FNR=\frac{FN}{FN+TP}
\end{equation}
Probability of detection (POD) or True Positive Rate (TPR) is the proportion of actual positive samples that are correctly detected by the model. It is also known as hit rate, recall, sensitivity ($S_n$):
\begin{equation}
    S_n=\frac{TP}{TP+FN}
\end{equation}
Specificity ($S_p$) is the proportion of actual negative samples that are correctly detected by the model:
\begin{equation}
    S_p=\frac{TN}{TN+FP}
\end{equation}
Negative predictive value (NPV) is the proportion of actual negative samples among the negative predictions made by the model, and it is denoted as $P_0$:
\begin{equation}
    P_0=\frac{TN}{TN+FN}
\end{equation}
Positive predictive value (PPV) is the proportion of actual positive samples among the positive samples made by the model, and it is denoted as $P_1$:
\begin{equation}
    P_1=\frac{TP}{TP+FP}
\end{equation}
F1-score is the harmonic mean of precision ($P_1$) and recall ($S_n$). F1-score ranges from 0 to 1, with higher values indicating better model performance:
\begin{equation}
    F1{-\rm score}=\frac{2*P_1*S_n}{S_n+P_1}
\end{equation}
R-score is the difference between $S_n$ and False Alarm Rate (FAR):
\begin{equation}
    R{\rm -score}=S_n-FAR
\end{equation}
Frequency Bias (FB) quantifies the ratio of the total number of observed events (TP+FN) to the total number of predicted events (TP+FP), indicating whether the model tends to overpredict (FB>1) or underpredict (FB<1) events:
\begin{equation}
    FB=\frac{TP+FN}{TP+FP}
\end{equation}
Geometric mean (GM) of $S_n$ and $S_p$ provides a balanced evaluation metric that reflects the model's ability to correctly identify both positive and negative samples. By combining 
$S_n$ (the proportion of actual positive samples correctly identified) and $S_p$ (the proportion of actual negative samples correctly identified) in a geometric mean, GM captures the trade-off between sensitivity and specificity, ensuring that both metrics contribute equally to the overall performance assessment:
\begin{equation}
    GM=\sqrt{S_n\cdot S_p}
\end{equation}
A higher GM indicates that the model performs well across both positive and negative sample classifications, while a lower GM
suggests imbalances or weaknesses in detecting one of the two classes.

Matthews Correlation Coefficient (MCC) takes into account true and false positives and negatives and ranges from -1 to 1, where 1 indicates perfect prediction, 0 indicates no better than random prediction, and -1 indicates total disagreement between prediction and observation:
\begin{equation}
    MCC=\frac{TP\times TN-FP\times FN}{\sqrt{(TP+FP)(TP+FN)(TN+FP)(TN+FN)}}
\end{equation}
Critical Success Index (CSI) measures the proportion of correct positive predictions out of all positive outcomes:
\begin{equation}
    CSI=\frac{TP}{TP+FP+FN}
\end{equation}
Average of $P_1,P_0,S_n,S_p$:
\begin{equation}
    AVG(P_1,P_0,S_n,S_p)=\frac{P_1+P_0+S_n+S_p}{4}
\end{equation}

\noindent Receiver Operating Characteristics (ROC) curve and Area Under Curve (AUC): 

The ROC diagram plots the FAR against the TPR at different detection thresholds. The x-axis is FAR, while the y-axis is TPR. Area Under Curve (AUC) is the area under a ROC curve, and the larger the AUC, the better the is performance of the model. A random model has a ROC curve that is a straight line with a slope of 1.0 and an AUC value of 0.5. The extent to which a model is better than random is then assessed by how much its area exceeds 0.5. The perfect model has an AUC of 1. 

\noindent Precsion-Recall Curve (PRC) and Average Precision (AP):

The Precision‐Recall Curve (PRC) gives the dependence of Precision against Recall. Similar to the ROC and AUC, Average Precision (AP) is the area under a PRC. The random model or the baseline of PRC is a straight line with a slope equal to 0 (i.e., the fraction of correct predictions is constant when varying the fraction of predicted earthquakes), and the intercept on the y‐axis is equal to the positive sample rate. If the difference between the AP and the baseline is larger than 0, the model is better than random guessing, and the larger the difference, the better the performance. The perfect model has Precision = 1 for all values of Recall and its AP is equal to 1.

\subsubsection{Evaluation metrics in statistical seismology}

Statistical seismology experts have designed a series of evaluation metrics for earthquake prediction models. Consider Output 15, which is the frequency or number of earthquakes occurring within a specific time-space window with magnitudes larger than or equal to a given threshold. Let $\omega_{i,j}$ represent the corresponding observed number of events in the earthquake catalog, where {$i\in T,j\in S$}. In this section, $T$ represents the set of indices corresponding to days while $S$ denotes the set of indices corresponding to locations. Let $o_{i,j}$ represent the predicted seismicity  at time $i$ and location $j$ provided by models. The labels and forecasts can be denoted as $\Omega=\Lambda(o_{i,j}|i\in T,j\in S)$ and {$\omega_{i,j}|i\in T,j\in S$}, respectively. We then utilize the L-test, S-test, N-test to evaluate the performance of the models from various perspectives.

\noindent L-test: it evaluates the log-likelihood of the observed catalog, assuming the model provides Poisson rates for each bin:
\begin{equation}
    L{\rm -test}(\Omega|\Lambda)=\sum\limits_{i\in T,\ j\in S}(\omega_{i,j}\cdot\log(o_{i,j})-o_{i,j})
\end{equation}
\noindent S-test: it assesses the log-likelihood of the spatial distribution of the catalog, assuming that the model's average output per spatial bin represents a Poisson rate. 
Let $N_{obs}$ be the total number of observed earthquakes, and $N_{fore}$ be the forecasted number of earthquakes. They are given by
\begin{equation}
    N_{obs}=\sum\limits_{i\in T,j\in S} {\omega_{i,j}}
\end{equation}
\begin{equation}
    N_{fore}=\sum\limits_{i \in T,\ j\in S} {o_{i,j}}
\end{equation}
\noindent Let the spatial distribution be written as $\Omega^S=\{\omega_j^S|j\in S\}$ and $\Lambda^S=\{o_j^S|j\in S\}$, where
\begin{equation}
    \omega^S_j=\sum\limits_{i\in T}\omega_{i,j}\ and\ o^S_j=\frac{N_{obs}}{N_{fore}}\sum\limits_{i\in T} o_{i,j}~.
\end{equation}
\noindent Then, the score is:
\begin{equation}
    S{\rm -test}(\Omega^S|\Lambda^S)=\sum\limits_{j\in T}(\omega^S_j\cdot\log(o_j^S)-o_j^S)
\end{equation}
\noindent N-test: it evaluates the probability of observing the actual number of earthquakes, assuming the model outputs Poisson rates for each bin. The cumulative distribution of the number of earthquakes can be calculated as
\begin{equation}
    F(x|N_{fore})=\exp(-N_{fore})\cdot\sum_{i=0}^x \frac{(N_{fore})^i}{i!}
\end{equation}
The N-test metrics are the probability $\delta_1$ of observing at least $N_{obs}-1$ earthquakes and the probability $\delta_2$ of observing at most $N_{obs}$ earthquakes:
\begin{equation}
    \delta_1=1-F(N_{obs}-1|N_{fore})~~~{\rm and}~~~ \delta_2=F(N_{obs}|N_{fore})~.
\end{equation}
For output (7), which predicts whether earthquakes will occur within a specified time-space-magnitude volume, statistical seismology employs the Molchan diagram as an evaluation tool.

\noindent Molchan Diagram: 

It compares the rate of missed events to the expected number of target events within the alerted time-space volume, which comprises all samples predicted as positive. The 
$y$-axis represents the missing rate or false negative rate (FNR). The interpretation of the $x$-axis depends on the chosen reference model:
\begin{itemize}
    \item Spatially Uniform Poisson Distribution: The $x$-axis corresponds to the fraction of the time-space volume occupied by alarms, calculated as TP+FP.
    \item Spatially Variable Poisson Distribution: Each space-time bin is weighted by the Relative Intensity index, which represents the rate of past earthquakes occurring in each spatial cell. This index is used to calculate the expected number of target events.
\end{itemize}
By adjusting thresholds for probabilities predicted by the earthquake prediction model, a trajectory is generated in the Molchan error diagram. The area above this trajectory represents the Area Skill Score (ASS), a measure of model performance.

In the Molchan diagram, a random guess produces a straight line starting at $(0,1)$ and ending at $(1,0)$ with a slope of $-1.0$, corresponding to an ASS of $0.5$. This provides a baseline for evaluating the skill of prediction models.

\subsubsection{Summary}

\paragraph{Overview}

It is important to note that this section focuses solely on the evaluation metrics used in the surveyed works, leaving many statistical seismology evaluation metrics undiscussed. A careful analysis of the information provided in the Supporting Information reveals that the choice of evaluation metrics is directly tied to the type of outputs. For instance, for output type (7), which predicts earthquake occurrence, studies typically select evaluation metrics designed for classification tasks. Conversely, for regression tasks such as predicting earthquake magnitudes, metrics like Mean Squared Error (MSE), Mean Absolute Error (MAE), and other regression-specific metrics are commonly used.

There are also some unique cases. For example, in reference \cite{R56}, the network's output is the maximum magnitude of earthquakes within a future time period—a regression task. However, this study further calculated the probability of earthquake occurrence using a danger function and ultimately employed classification metrics such as accuracy (Acc), False Positive Rate (FPR), and False Negative Rate (FNR) to evaluate the model's performance.

An interesting approach is presented in references \cite{R57, R58, R55}, where future earthquake magnitudes are predicted, typically a regression task. In these studies, a threshold (e.g., 0.5) is set, and the difference between the predicted and actual magnitudes is evaluated: if the difference is below the threshold, it is considered a True Positive; otherwise, it is treated as a False Negative. This transformation allows the use of classification metrics to assess performance, blending regression outputs with classification-style evaluation.

\paragraph{Critique}

When evaluating the performance of earthquake prediction models, there is a clear knowledge gap between seismologists and data mining experts\cite{R3}. In the AI community, classification tasks typically treat positive and negative samples as independent, with manual adjustments to ensure a balanced dataset. However, earthquakes exhibit distinct spatio-temporal characteristics, such as an inherent imbalance between positive and negative samples and the clustering of earthquakes in both space and time. These characteristics make many popular AI evaluation metrics unsuitable for earthquake prediction models.

For instance, when accuracy (ACC) is used to evaluate earthquake prediction models, the significantly larger number of negative samples (non-earthquake samples) compared to positive samples (earthquake samples) can lead to misleading results. A model that simply predicts all samples as negative (no earthquake occurrence) could achieve a high accuracy rate, but this does not indicate any practical utility for earthquake prediction.

To address data imbalance, the AI community has developed evaluation metrics such as Sensitivity, Specificity, ROC curves, and PRC curves, which are designed to be less affected by sample imbalance. However, further research by statistical seismologists has shown that even these advanced metrics are inadequate for evaluating earthquake prediction models. For example, Parsons\cite{R59} found that ROC tests are still impacted by data imbalance when applied to spatial earthquake forecasts. Similarly, Zhang et al.\cite{R52} observed that positive scores from PRC plots do not demonstrate superior performance of AI models compared to the Spatially Variable Poisson model, which is the simplest and most basic model in statistical seismology.

These findings underscore the need for evaluation metrics specifically tailored to the unique characteristics of earthquake prediction tasks, integrating insights from both seismology and AI to ensure meaningful performance assessments.

\paragraph{Suggestion}

It is evident that numerous evaluation metrics exist to assess the performance of models from different perspectives, with each focusing on specific aspects of model behavior. Relying solely on a single metric for evaluation leads to biased assessments and can overlook critical trade-offs. For instance, Zhang et al.\cite{R66} demonstrated that, when a model is evaluated exclusively based on the False Negative Rate (FNR) that prioritizes the minimization of missed events, the model's False Positive Rate (FPR) often becomes excessively high. Strategies that focus solely on FNR are suboptimal, as low FNR combined with numerous false alarms effectively leads to a “crying wolf” scenario, eroding the model's practical utility. Similarly, a group of alarms with a low False Discovery Rate (FDR) might suffer from a high FNR, failing to capture a substantial portion of true events. Therefore, it is crucial to evaluate AI-based earthquake prediction models using multiple metrics to achieve a more holistic and unbiased understanding of their performance.

Zhang et al.\cite{R52} further utilized the “bookmaker-gambler” analogy introduced by Zhuang\cite{R60} to explore the deeper reasons why some evaluation metrics fail. Their analysis revealed that these metrics often rely on the Spatially Uniform Poisson model as the reference, which assumes earthquakes are distributed uniformly across space and time. As a result, positive scores from such metrics merely indicate that the tested model outperforms this simplistic reference model; they do not necessarily signify that the model itself is effective. Many previous studies have overlooked this distinction, mistakenly attributing their "good" performance to the inherent strength of their models, when in reality, their success was measured against an inferior baseline vastly outperformed by the best prediction models in statistical seismology.

To address these shortcomings, we recommend using more specialized evaluation metrics rooted in statistical seismology to accurately assess the capabilities of AI-based earthquake prediction models. These metrics better reflect the unique characteristics of seismicity and provide more meaningful evaluations.

Moreover, to demonstrate that AI models have genuinely surpassed the performance bottlenecks of existing earthquake prediction models and contributed new insights to geophysics or statistical seismology, it is essential to benchmark AI models against the best existing geophysical or statistical seismology models, as discussed in Section \ref{Seisdemand}. Only through such rigorous comparisons can the true potential of AI-based earthquake prediction models be validated and their contributions to the field properly assessed.

\section{Perspective}\label{idea}
\subsection{Prioritizing the Earthquake Forecasting Problem Over AI Technology in AI-Based Prediction Model Research}

Over the past few decades, machine learning technology has advanced rapidly. AI experts have introduced revolutionary innovations by designing new neurons and network architectures. For example, the astonishing capabilities of the ChatGPT model have demonstrated the tremendous potential of the Transformer architecture. Many researchers are eager to develop large models specific to their own fields of study. The application of AI technology to diverse physical domains, including earthquake prediction, has become a prominent and rapidly growing trend.

In the realm of cross-disciplinary research between AI and earthquake prediction, as well as other intersections of AI with physical sciences, a concerning trend has emerged. Researchers often place excessive emphasis on the novelty or sophistication of the AI technologies they employ, while neglecting the core physical problems these technologies are meant to address.

In many cases, researchers in such interdisciplinary fields adopt a research methodology resembling that of the AI community: they develop or introduce new machine learning architectures and benchmark them against other state-of-the-art models. If the new model demonstrates improved performance, they claim significant success and herald their work as a breakthrough in the related field of physics.

While comparing new models with existing ones is a necessary step in validating advancements, the issue lies in the narrow scope of these comparisons. In physics, the benchmark should not be limited to other machine learning models. Instead, robust, well-established physical models—developed over decades of rigorous study and proven to be effective—must also be included. Ignoring these foundational models risks undermining the integrity and practical relevance of the research, as true progress in physics requires integrating AI advancements with the deep domain knowledge encapsulated in traditional physical models.

We need to critically reevaluate the role of AI technology in physical applications. Currently, AI serves as a powerful tool, enabling us to achieve results that were previously unattainable or highly inefficient using traditional methods. We identify two main contributions:
\begin{itemize}
    \item Accelerating Time-Intensive Processes: AI technology can dramatically speed up processes that are computationally expensive in traditional physical modeling or inversion methods, often achieving comparable or superior accuracy. For example, Wu et al.\cite{R63} utilized airborne electromagnetic (AEM) data as input for a Long Short-Term Memory (LSTM) network to generate resistivity models. Their approach completed the inversion of approximately 740,000 AEM soundings in under two seconds, enabling rapid geological inference with significant resistivity contrasts—a clear example of the successful application of AI technology to enhance both efficiency and practicality.
    \item Enhancing and Complementing Traditional Models: AI models can also address limitations in traditional physical models and, in some cases, uncover previously unknown patterns or laws. For instance, Zlydenko et al.\cite{R8} designed an AI model inspired by the Epidemic-Type Aftershock Sequence (ETAS) model. By leveraging its multi-layer structure and high degree of nonlinearity, the AI model improved upon the ETAS model’s isotropic spatial kernel assumption,.
\end{itemize}
When applying AI technology to physical problems, the focus should be on its contributions to advancing the relevant field of physics rather than solely on innovations for the AI community. The primary objective of integrating AI should be to solve physical problems more effectively, rather than pursuing trends or novelty for their own sake. By maintaining this focus, AI can serve as a transformative tool that bridges computational efficiency and scientific discovery, ultimately driving progress in both technology and physical sciences.

Based on our investigation, we have observed that, particularly in the field of AI-based earthquake prediction research, a substantial number of studies lack depth and rigor. These studies often mechanically integrate the latest AI technologies into earthquake prediction without meaningful consideration of the underlying physical problems, resulting in what can be described as "fast-food research." Such approaches fail to make valuable contributions to geophysics and statistical seismology. This superficial trend has also affected some readers, who now judge the quality of research solely by the use of state-of-the-art AI technologies or the development of more advanced neural networks, rather than by the research's actual relevance and impact on the field.

When applying AI technology to physics or geophysics, the focus should be on whether it effectively addresses the fundamental issues within these domains. The emphasis should lie in benchmarking AI models against already well-established physical models. If simpler AI networks can outperform these models or reveal previously unknown patterns, it represents true progress for physics and geophysics.

Thus, it is crucial to move away from an overemphasis on designing increasingly complex neural networks or blindly transplanting the latest AI technologies into physics problems without context or purpose. Instead, research efforts should be directed toward understanding and solving the underlying physics challenges. By prioritizing the physical problems themselves, AI can be harnessed as a tool for meaningful scientific advancement rather than a superficial trend.

\subsection{Guidelines for Recognizing and Addressing Key Characteristics of Earthquake Datasets.}

When training and evaluating machine learning models, it is essential to first understand the characteristics of the dataset and then select appropriate loss functions and evaluation metrics. For instance, in classification tasks with datasets that exhibit a significant imbalance between positive and negative samples, where negative samples far outnumber positive ones, it is necessary to increase the penalty for false negatives to amplify the influence of positive samples on the network gradient. This approach underpins the design of loss functions like Balanced Cross Entropy.

Earthquake datasets are characterized by two primary features: data imbalance and spatial-temporal clustering. These features must be carefully considered when designing loss functions and evaluation metrics. Detailed guidance on loss function design tailored to these characteristics can be found in the suggestions provided in Section \ref{loss function}.

When evaluating earthquake prediction models, some researchers rely on standard evaluation metrics like the ROC curve and PRC curve, interpreting positive scores as evidence of the model's effectiveness in predicting the time and location of earthquakes. However, these metrics often use a Spatially Uniform Poisson (SUP) distribution as the reference model, which assumes earthquakes are uniformly distributed in time and space and are independent of each other. This naïve reference model fails to account for the true spatial-temporal clustering of earthquakes, making it an inadequate benchmark. As previous studies have shown, spatial clustering can lead to overly optimistic evaluation results when the reference model is poorly designed.

For example, Zhang et al.\cite{R52} constructed an earthquake prediction model based solely on the monthly average of human-made Nighttime Light data. Due to the negative correlation between nighttime light intensity and the spatial distribution of earthquakes, their model performed better at identifying earthquake-prone locations than the SUP reference model, resulting in positive ROC and PRC scores. However, this outcome does not indicate that the model has genuine earthquake prediction capabilities—it merely highlights that the model outperformed an inadequate reference.  
The SUP reference model is so inadequate that just leveraging the negative correlation between human settlements and earthquake locations can result in outperforming it. However, there is no genuine physical interaction between human-made nighttime light and future earthquakes. This example serves as a clear illustration of the pitfalls in many AI-based studies, where reliance on flawed reference models can lead to misleading conclusions about a model’s predictive capabilities.

Researchers must recognize and clearly state that any evaluation of earthquake prediction models inherently involves an implicit or explicit reference model. A positive score or metric simply means the tested model is better than the reference model, not that it is effective in absolute terms.

Accurately evaluating earthquake prediction models requires specialized methods. The simplest and most reliable approach is to use established statistical seismology evaluation techniques, such as the Molchan diagram\cite{R64,R65}, 3D error diagram\cite{R66}, Gambling score\cite{R60}, parimutuel gambling score\cite{R67}, or RichterX Score\cite{R68}. Additionally, comparisons should be made against robust and powerful benchmarks, ensuring the tested models are evaluated in a meaningful and scientifically rigorous context.

\subsection{Further reflections on pseudo-prospective testing}

We have consistently emphasized the importance of aligning models with real-world conditions during their development. However, an often-overlooked yet critical factor is the timeliness and quality of earthquake catalogs. Most earthquake prediction models rely on pre-processed, high-quality catalogs, but obtaining such ideal datasets in practical, real-time scenarios is challenging. For instance, when predicting the probability of earthquakes occurring on day $T$, if we include data from the earthquake catalog of day 
$T-1$ as input, the near-real-time catalog available may be of lower quality. These near-real-time catalogs frequently suffer from missing events, inaccuracies in spatial locations, and even contamination with non-natural seismic information. It is well-known that quasi-instantaneous catalogs are prone to numerous errors due to the lack of expert curation or ``cleaning.''

Machine learning (ML) and artificial intelligence (AI) methods typically assume homogeneous inputs with stable statistical properties. However, earthquake catalog quality is inherently non-stationary: it is poorer in the recent past due to incomplete and rapid data collection, improves over an intermediate time scale as data is reviewed and refined, and deteriorates further back in time due to older seismometers, fewer monitoring stations, and other historical limitations.

Additionally, the problem of short-term incompleteness exacerbates these challenges\cite{short_term_incomplete}. In cases of saturation or insufficient coverage by seismic monitoring stations, significant events occurring in close temporal proximity may be missed. For instance, a magnitude 7 earthquake occurring shortly after a magnitude 8 event might be overlooked. 

Zhang et al.\cite{CL_ETAS} addressed this issue using a convolutional long short-term memory network. They used a simple ETAS model and earthquake catalog predictions as inputs to predict the frequency of aftershocks (magnitude 3 or higher) within 1-degree by 1-degree spatial cells in California at 7, 15, and 30 days following three major earthquakes of magnitude 7 or higher. Their results showed that the simple ETAS model predicted a much higher frequency of earthquakes than recorded in the seismic catalog, while the AI model's predictions were closer to the cataloged frequency and lower than the ETAS predictions. Based on this, they concluded that their AI model outperformed the ETAS model. However, they failed to account for the impact of short-term incompleteness, where the actual number of earthquakes is likely higher than the catalog records. This raises a critical question: is the true earthquake frequency closer to the ETAS model predictions or the AI model predictions? Without addressing short-term incompleteness, any conclusion is unreliable.

This highlights the importance of ensuring that AI researchers working on earthquake prediction possess foundational seismological knowledge and undergo rigorous peer review by statistical seismologists. Without such oversight, it is easy to overlook critical issues that compromise the validity of conclusions.

Discrepancies in data quality pose significant risks to model performance and reliability, yet this issue has been surprisingly neglected in both real-time and long-term simulation contexts. Despite frequent emphasis on simulating real-world conditions, the importance of data timeliness and quality has not received the attention it deserves. Addressing these challenges is essential to ensure that models are not only theoretically sound but also practically viable in real-world earthquake prediction and forecasting.

\subsection{Opportunities and Challenges of AI Technology in the Context of Next-Generation Catalog}\label{oandc}

Stockman et al.\cite{R7} found that the completeness of the earthquake catalog does not diminish the forecasting power of AI models. On the contrary, including events with magnitudes smaller than $M_c$ as input actually improves model performance. They argued that the most significant gains in their neural model stemmed from its ability to process incomplete data immediately following large earthquakes. Similarly, studies by Zhang et al.\cite{zhang/10.1093/gji/ggae373} and Zhan et al.\cite{ZhanSTC} also demonstrated that catalog completeness does not hinder the forecasting power of AI models. In fact, incorporating small-magnitude events ($M<M_c$) enhances the ability of AI models to predict smaller seismic events. These low-magnitude events carry valuable information about the dynamics of seismic activity, and their inclusion improves predictions for certain time-magnitude windows.

In contrast, traditional models like the ETAS model rely heavily on a complete training catalog for accurate predictions. This underscores one of the key advantages of machine learning over traditional statistical seismology models: the ability to effectively handle incomplete and noisy data.

The advent of machine learning has also enabled the creation of next-generation earthquake catalogs, leading to an exponential increase in the volume and richness of seismic data. Mancini et al.\cite{R69} demonstrated that combining the ETAS model with high-resolution seismic catalogs generated using deep learning significantly enhances the forecasting performance of the ETAS model. However, as highlighted by Beroza et al.\cite{R70}, these next-generation ``deeper'' catalogs, while rich in information, introduce significant challenges for exploration due to their sheer size and complexity. Traditional methods often fail to fully exploit the wealth of data encoded in these catalogs, leaving valuable insights untapped\cite{R70}.

Machine learning offers a promising solution for uncovering hidden relationships and patterns within these complex datasets. By leveraging AI technologies, researchers can more effectively analyze the intricacies of seismicity, paving the way for breakthroughs in earthquake forecasting and a deeper understanding of seismic processes.

\subsection{Methods that combine AI technology and knowledge of seismology}

AI technology possesses powerful data mining capabilities, and recent models have demonstrated performance on par with or exceeding that of some versions of the ETAS model, while achieving significantly faster processing speeds. This highlights the immense potential of AI technology in earthquake prediction. Looking ahead, the integration of seismological expertise with AI is expected to further elevate the accuracy, efficiency, and reliability of AI-based earthquake prediction models.

We have identified three effective strategies for achieving this integration:

\noindent(1) {\bf Feature engineering} is the process of using domain knowledge to extract features (characteristics, attributes, or properties) from raw data to improve the performance of AI-based models. It involves creating new features or modifying existing ones to better represent the underlying patterns in the data that model needs to learn. Feature engineering is a commonly used method in past AI-based earthquake prediction. For example, when using seismic types as inputs, instead of directly using earthquake catalogs as inputs, various indicators derived from the catalogs are used, such as b-value, estimated seismic release energy, and so on. By doing so, the features in the raw data are highlighted, making it easier for the AI model to learn underlying knowledge. 

\noindent(2) {\bf Choosing or designing better network structures based on the physical processes of earthquake.}
Zlydenko et al.\cite{R7} developed an Artificial Neural Network (ANN) encoder that emulates the mathematical structure of the ETAS model. Their framework consisted of three encoders: the recent earthquake encoder, the seismicity rate encoder, and the location encoder.
\begin{itemize}
    \item The recent earthquake encoder mimics the triggering function of the ETAS model, capturing the cumulative contributions of past earthquakes within a specified time frame.
    \item The seismicity rate encoder captures both long-term and short-term seismic activity in the study area, representing the average earthquake rate per day or month.
    \item The location encoder models the spatial correlation between the epicenters of historical events and the target output.
\end{itemize}
The model was trained on features similar to the core elements of the ETAS model, such as time intervals between past and present earthquakes, distances, and magnitudes. This ANN model achieved performance comparable to or exceeding that of a standard ETAS model in terms of average information gain per earthquake while providing a 1000-fold improvement in computational efficiency.

Since seismicity displays clear time-series characteristics and long-term dependencies between events, earthquake prediction requires models that can learn these dependencies and capture the temporal and spatial correlations inherent in seismic data. Recurrent Neural Networks (RNNs) and their derivatives are well-suited for this task due to their ability to retain information from previous time steps and model context effectively, which is essential for understanding long-term dependencies.

For example, Dascher-Cousineau et al.\cite{R5} utilized a general-purpose encoder-decoder neural network based on the Gated Recurrent Unit (GRU) to predict the timing of the next earthquake given the history of past events. At each time step, the network takes as input the previous hidden state $h_{i-1}$ along with the features of the previous event $(t_{i-1},m_{i-1})$. It then updates the hidden state $h_i$ using GRU update equations. An affine layer connects this hidden state to the parameters of a Weibull Mixture component, which determines the likelihood of the model given the event occurring at time $t_i$.

Similarly, Stockman et al. employed an RNN to create compact representations of event histories and simulate neural point processes. Their model took the inter-event times and magnitudes of the last $d$ events as input into a recurrent section with 64 recurrent units. The output from this section was processed through two fully connected layers. For the temporal network, the output was combined with the next inter-event time $\tau$, and for the magnitude network, it was combined with the next magnitude $m$. These outputs were then used to compute the log-likelihood of the next inter-event time and magnitude {$\tau,m$}

Both models demonstrated performance surpassing that of the temporal ETAS model, highlighting the potential of RNN-based architectures for capturing complex temporal and spatial dependencies in earthquake prediction tasks.

 \noindent(3) {\bf Designing Loss Functions or Optimization Objectives to Incorporate Prior Knowledge.}
 During the training stage, the influence of individual samples on the network's gradient can be adjusted by assigning different weights to samples. This approach allows for controlling the direction of network learning and optimization, ensuring it aligns more closely with specific application goals. Modifying the loss function or objective function to better suit the problem at hand and incorporating domain-specific knowledge can significantly enhance model performance.

For example, as previously introduced, Zhang et al.\cite{R52} introduced a method to embed specialized geophysical knowledge into machine learning models by revising the loss function (see Equations \ref{CE_zhang} and \ref{MSE_zhang}). They used estimated seismic energy as input to an LSTM network to predict earthquakes with $M\geq 5.0$ in Mainland China. The evaluation, based on time-space Molchan diagrams, demonstrated significant performance improvements when the ANN was trained using a more appropriate reference model, highlighting the value of integrating expert knowledge into AI-based earthquake prediction models.

Although the application of Physics-Informed Neural Networks (PINNs)\cite{raissi2018hidden,raissi2019physics} has not yet been explored in this context, they represent a promising approach for effectively integrating prior knowledge into AI models. Suppose there is a physical process described by a function - $f(x|\theta)$, where $x$ is the independent variable and the input of PINN, and $\theta$ is a set of parameters. For this physical process, there may be an additional set of partial differential equations that describe the boundary conditions, denoted as $h(x|\theta)=0$. PINN is designed to invert $\theta$. In other words, $f, h$, and the network share the parameter $\theta$. The network is designed to minimize the mean squared error loss to estimate:
\begin{equation}
    MSE_{PINN}=MSE_0+MSE_b+MSE_f
\end{equation}
\noindent where $MSE_0$ corresponds to the loss for the data, $MSE_b$enforces the boundary conditions, and $MSE_f$ penalizes the $f(x|\theta)$ not being satisfied on the collocation points. By incorporating the partial differential equations as constraints, PINNs can provide more accurate and physically consistent predictions even with limited data. PINNs can combine data-driven and model-driven approaches, leveraging both empirical data and theoretical models to enhance prediction accuracy and robustness, and this architecture has been successfully applied to many physical scenarios, such as inversion and surrogate modeling in solid mechanics, fluid mechanics and so on. PINNs also seem to have tremendous potential for application in earthquake prediction, making them worth trying out. 

\subsection{How to use AI technology to discover unknown seismicity patterns}
Researchers are increasingly exploring how AI technology can uncover new patterns in data, particularly in datasets containing previously unknown structures, thereby expanding human cognitive boundaries. For example, in earthquake prediction, the application of machine learning techniques and the continued development of seismic monitoring networks have significantly enriched earthquake catalogs, especially with information on minor earthquakes. These expanded datasets likely contain hidden patterns or rules yet to be discovered. It is worth investigating how AI’s advanced data processing and information mining capabilities can help reveal these potential new insights.

One promising approach is to treat the AI model as an independent "scientist," leveraging its computational and information processing power to extract unknown knowledge from vast, complex datasets that are difficult for humans to analyze. For instance, in rogue wave forecasting, Häfner et al.\cite{R71} constructed a causal graph mapping the relationships between various sea state parameters and the occurrence of rogue waves. Their dataset, derived from buoy measurements, included parameters such as significant wave height, spectral bandwidth, peak wave number, and directional spread. By applying symbolic regression and refining the model, they discovered that the resulting symbolic expressions resembled classical models but incorporated more factors and complex interactions. This allowed the model to more accurately capture the mechanisms of rogue wave generation in real ocean environments, offering new insights into rogue wave forecasting. Similarly, in earthquake forecasting, Zlydenko et al.\cite{R7} improved upon the isotropic spatial kernel assumption of the ETAS model by leveraging the multi-layer structural features and nonlinearity of their AI model.

Another effective strategy involves the use of revised loss functions, as proposed by Zhang et al.\cite{R52}. In their work, the loss function was further weighted using predictions from a reference model, enabling the network to focus on "difficult" samples where the reference model underperformed. This approach encourages the AI model to pursue knowledge that the reference model lacks, ultimately striving for superior performance. When the AI-based model surpasses the reference model, it suggests that the AI has likely learned new insights beyond the scope of the reference. However, due to the inherent black-box nature of neural networks, it remains challenging to pinpoint precisely what the model has learned.

A promising direction for future research is developing methods to visualize and interpret the knowledge acquired by neural networks in ways that are comprehensible to humans. Such efforts could bridge the gap between AI-generated insights and human understanding, enhancing the practical utility of AI models in expanding our knowledge of complex systems like seismicity and rogue waves.

\subsection{Beyond performance: AI as a tool for earthquake forecasting}

In Section \ref{Seisdemand}, we outlined the fundamental expectations for applying AI to earthquake forecasting, including improving forecast accuracy, enhancing model speed, and discovering previously unknown seismic activity patterns. In this subsection, we delve deeper into the broader and more ambitious expectations for the future application of AI technology in earthquake forecasting.

\noindent (1) Data-Driven Advantages:

Earthquake forecasting models often fail to fully utilize the wealth of information available, such as data on different magnitude categories of the same earthquake, fault-related information, and other underexplored sources. Additionally, the densification of global seismic observation networks has led to an exponential increase in both the volume and complexity of seismic data. As discussed in "Opportunities and Challenges of AI Technology in the Context of Next-Generation Catalogs," this data explosion poses challenges for traditional geophysicists and statistical seismologists, who struggle to process and interpret such extensive datasets.

AI, particularly deep learning, excels at extracting meaningful patterns and relationships from complex data, offering a way to overcome these challenges. For example, while models such as the ETAS model perform well in predicting triggered events, they face theoretical limitations, such as assuming that independent events follow a simple Poisson distribution. A promising avenue for improvement lies in integrating seismic data with non-seismic alarms (e.g., GPS and satellite observation data) to create multi-source, multi-modal forecasting models. AI provides an ideal platform for enabling this integration, offering the tools needed to handle and combine diverse data types effectively.

\noindent (2) Knowledge Integration Advantages:

Over decades, statistical seismology and geophysics have accumulated valuable theories and practical insights into earthquake forecasting. Experts are well aware of the limitations of their models and often have conceptual strategies for improvement. However, executing these ideas—such as integrating multiple models or enhancing already well-developed ones—is a significant challenge.

For instance:

The standard implementation of ETAS model with the assumption of isotropic earthquake triggering is known to be insufficient, and incorporating fault-related information is expected to improve its accuracy.
Predictive performance for independent events in the ETAS model could benefit from leveraging seismic quiescence theory.
The influence of tidal forces from celestial bodies on earthquake triggering, well-documented in studies like Cochran et al.\cite{cochran2004earth}, remains largely unincorporated into forecasting models.
While one solution is to have individual experts apply their models independently and combine the results, this approach is resource-intensive and lacks true integration. AI offers an alternative by enabling the simultaneous incorporation of diverse inputs. For example, an AI model could integrate parameters from the ETAS model (e.g., spatial distances and time intervals between predicted and past earthquakes), RTL-related seismic quiescence metrics, and computed tidal force potentials. By leveraging AI’s ability to synthesize and mine data, these models have the potential to overcome the performance limitations of traditional approaches.

\noindent (3) Independence from Traditional Model Assumptions:

Traditional earthquake forecasting models are often constrained by physical laws and assumptions, which may not always hold or be well-defined in certain contexts. AI, on the other hand, can learn directly from the data without relying on predefined physical assumptions. This flexibility allows AI to uncover novel patterns and knowledge that might remain hidden within the limitations of traditional models, offering new insights into seismic activity.

\noindent (4) Real-Time Processing Capabilities:

Rapid decision-making is critical in earthquake forecasting, particularly for early warning systems and aftershock predictions. AI’s ability to process and analyze large-scale data in real time makes it uniquely suited for these tasks. By delivering timely predictions, AI can help mitigate the impact of earthquakes, improve disaster response strategies, and enhance public safety.

In summary, the integration of AI into earthquake forecasting has the potential to transform the field by addressing data complexity, enhancing knowledge integration, breaking free from traditional constraints, and enabling real-time predictions. As AI technology continues to advance, it will play an increasingly pivotal role in both improving existing models and discovering new insights into earthquake phenomena.

\section{Epilogue}

In conclusion, we would like to offer sincere advice to geophysicists and AI experts aspiring to apply AI technologies in earthquake forecasting.

To geophysicists, we want to emphasize that failures in AI model training are a common occurrence. Sometimes, despite our best efforts to modify and refine the model, its performance may deteriorate, leaving us perplexed. It is important not to get discouraged, as this is a normal part of the process. We should leverage our strengths in geophysics and continuously experiment with and adjust the model.

A key point to consider is that there are many types of Artificial Neural Network (ANN) architectures, each with unique strengths and weaknesses, designed to address specific types of problems. Common architectures include Feedforward Neural Networks (FNNs), Convolutional Neural Networks (CNNs), Recurrent Neural Networks (RNNs), Long Short-Term Memory (LSTM) networks, Transformers, and Graph Neural Networks (GNNs). Each of these architectures is tailored for particular tasks—such as temporal sequence modeling, spatial feature extraction, or graph-structured data processing—and selecting the right architecture is critical for achieving effective results.

Our advice is to first understand the rationale behind the design of these architectures, including their intended strengths and limitations, and then combine this understanding with geophysical knowledge to choose the most suitable ANN for your specific application. Blindly relying on or copying advanced models from the AI community without considering the unique characteristics of geophysical problems can often lead to suboptimal results. As we have discussed earlier, directly adopting AI research methodologies without adaptation may result in failure.

Instead, focus on integrating domain-specific knowledge into critical areas such as feature engineering, network architecture, loss function design, and model evaluation. This thoughtful combination of AI capabilities and geophysical expertise will significantly improve the chances of achieving meaningful and successful outcomes in earthquake forecasting.

To AI experts, we urge you to look beyond the technology itself and prioritize its application to earthquake forecasting. Unlike simple tasks like classifying images of cats and dogs, earthquake forecasting is deeply complex and demands extensive background knowledge in statistical seismology and geophysics. Taking the time to understand this domain is essential. Additionally, avoid over-reliance on model scores. AI is sometimes humorously referred to as “modern alchemy,” highlighting the fact that even model builders may not fully understand why their models succeed. A high score may sometimes provide a false sense of accomplishment, as we have previously noted. Instead, focus on what the model has truly learned and how it contributes to advancing our understanding of seismic processes and geophysics.

Looking ahead, deep collaboration between geophysicists and AI scientists offers the most promising path forward. By fostering mutual understanding and working together to combine domain expertise with advanced computational tools, we can drive significant advancements in earthquake forecasting and prediction. This interdisciplinary partnership holds the potential to unlock new insights and improve the reliability and effectiveness of forecasting models, ultimately benefiting both science and society.

\vskip 1cm
{\bf Acknowledgements}: This work is partially supported by the National Natural Science Foundation of China (Grant no. T2350710802, U2039202), and the Center for Computational Science and Engineering at Southern University of Science and Technology. We would like to extend our gratitude to Dr. Li Min for her invaluable assistance in the preliminary stages, where she helped us organize the literature and streamline the information.

\bibliographystyle{elsarticle-num-names} 
\bibliography{reference}

\end{document}